\documentclass[review]{elsarticle}

\usepackage{hyperref}
\journal{Journal of \LaTeX\ Templates}
\usepackage{amssymb,amsmath,amsfonts,geometry,enumerate,float}
\usepackage{tikz}
\usepackage{tikz-qtree}
\usetikzlibrary{bayesnet}
\usetikzlibrary{trees,shapes,arrows}
\usepackage{graphicx}
\usepackage{amsmath}
\usepackage{bbm}

\journal{Artificial Intelligence}
\begin{document}

\begin{frontmatter}

\title{A Technical Survey on Statistical Modelling and Design Methods for Crowdsourcing Quality Control}

%\tnotetext[mytitlenote]{Fully documented templates are available in the elsarticle package on \href{http://www.ctan.org/tex-archive/macros/latex/contrib/elsarticle}{CTAN}.}

%% Group authors per affiliation:
%\author{Yuan Jin\fnref{myfootnote}}
%\address{}
%\fntext[myfootnote]{Since 1880.}

%% or include affiliations in footnotes:
\author[deakin]{Yuan Jin\corref{mycorrespondingauthor}}
\cortext[mycorrespondingauthor]{Corresponding author}
\ead{yuan.jin@deakin.edu.au}
\author[monash]{Mark Carman}
\ead{mark.carman@monash.edu}
\author[deakin]{Ye Zhu}
\ead{ye.zhu@ieee.org}
\author[deakin]{Yong Xiang}
\ead{yong.xiang@deakin.edu.au}
\address[deakin]{School of Information Technology, Deakin University, Victoria, Australia 3125}
\address[monash]{Faculty of Information Technology, Monash University, Victoria, Australia 3800}

\begin{abstract}
Online crowdsourcing provides a scalable and inexpensive means to collect knowledge (e.g. labels) about various types of data items (e.g. text, audio, video). However, it is also known to result in large variance in the quality of recorded responses which often cannot be directly used for training machine learning systems. To resolve this issue, a lot of work has been conducted to control the response quality such that low-quality responses cannot adversely affect the performance of the machine learning systems. Such work is referred to as the quality control for crowdsourcing. Past quality control research can be divided into two major branches: \textit{quality control mechanism design} and \textit{statistical models}. The first branch focuses on designing measures, thresholds, interfaces and workflows for payment, gamification, question assignment and other mechanisms that influence workers' behaviour. The second branch focuses on developing statistical models to perform effective aggregation of responses to infer correct responses. The two branches are connected as statistical models (i) provide parameter estimates to support the measure and threshold calculation, and (ii) encode modelling assumptions used to derive (theoretical) performance guarantees for the mechanisms. There are surveys regarding each branch but they lack technical details about the other branch. Our survey is the first to bridge the two branches by providing technical details on how they work together under frameworks that systematically unify crowdsourcing aspects modelled by both of them to determine the response quality. We are also the first to provide taxonomies of quality control papers based on the proposed frameworks. Finally, we specify the current limitations and the corresponding future directions for the quality control research. 
\end{abstract}


\end{frontmatter}

%\linenumbers

\section{Introduction}
With the advent of Web 2.0 functionality, users of the Web gained the ability to submit questions online and get answers from other users. Crowdsourcing provides a mechanism by which submitted questions are distributed and solved by generally large and anonymous online crowds. When answering questions, the online crowds are characterized by different and variable \textit{motivation} (e.g. being money-driven or enjoyment-driven), and different and varying degrees of \textit{expertise}. As a result, they exhibit more diverse and in general less accurate question-answering behaviour as compared to \textit{in-house workers} who are trained to work more professionally and specifically on internal platforms for tasks of particular companies \cite{DBS-044}. On the other hand, online crowds are more readily accessible and usually less expensive than in-house workers \cite{zaidan2011crowdsourcing, liu2012crowdsourcing, stol2014two}.

Over the past decade, online human intelligence marketplaces have been thriving, providing organized and billed crowdsourcing services to requesters all over the world. Online crowds are registered with the marketplaces as \textit{crowd-workers} who are autonomous and generally receive a small payment, typically a few cents \cite{ipeirotis2010analyzing}, for finishing each question. Two popular crowdsourcing marketplaces are Amazon Merchanical Turk\footnote{https://www.mturk.com/} (AMT) and CrowdFlower\footnote{https://www.crowdflower.com/}. 

In recent years, AMT and CrowdFlower have become very successful in providing human intelligence support to the machine learning and data mining communities. They allow the communities to perform large-scale label collection for data items used to train all kinds of machine learning systems such as learning-to-rank systems in Information Retrieval \cite{kumar2011learning}, machine translation systems \cite{ambati2012active} and general supervised learning systems \cite{brew2010interaction, deng2013fine, cheng2015flock}. 

As these platforms continue to grow by attracting more label collection tasks, they allow people with \textit{varied abilities} and \textit{motivations} to join them to share the labelling workloads. This results in a \textit{deterioration} in the \textit{quality of labels} and varieties of cheating behaviour prevalent on the platforms \cite{eickhoff2013increasing}.
\subsection{Crowdsourcing Quality Control}
To deal with the above issues, quality control for crowdsourcing (QCC) is required by which the influence of \textit{high-quality} labels is guaranteed to outweigh that of \textit{low-quality} ones.

\textit{Worker filtering} is the main QCC method used by crowdsourcing platforms to deal with label quality deterioration and cheating behaviour of crowd-workers. It removes two types of workers: \emph{unqualified} workers and \emph{low-performing} workers. The \emph{unqualified} workers are removed using a \textit{quiz} before a task commences. The quiz contains only \textit{control} questions (for which the true answer is known) and each worker must achieve a certain accuracy on the control questions to be admitted to the task. The \emph{low-performing} workers are removed from the worker pool during their participation in the task using unseen control questions embedded amongst the target questions. The worker filtering mechanism typically removes all the responses given by each low-performing crowd-worker. Other qualified workers then have to redo all the questions answered by these workers. This results in more budget and time consumption. 

\textit{The wisdom of the crowd} (WoC) \cite{surowiecki2005wisdom} is an alternative to worker filtering that does not discard any response from low-performing workers. It retains the influence of their correct responses and tries to \textit{smooth out} the influence of the incorrect ones. It centres on the observation that proper \textit{aggregation} of \textit{multiple answers} given by different people to the same question is able to yield a better answer. For example, websites, such as Rotten Tomatoes\footnote{https://www.rottentomatoes.com/}, Netflix\footnote{https://www.netflix.com/} and Last.fm\footnote{https://www.last.fm/}, utilize WoC methods (which aggregate reviews from their users) to provide summary reviews and overall recommendation scores about their items. These summary reviews and scores have turned out to be very accurate in depicting the underlying quality of the items.

The efficacy of the WoC approach relies on two crucial aspects. The first aspect is the \textit{redundancy} of the responses to the same question. Since one crowd-worker is usually not capable of consistently providing the correct answer, naturally more workers are needed to work on the same question. The second aspect is the \textit{aggregation} of the redundant responses for eliciting an accurate final answer. Two prerequisites need to be satisfied for the aggregation to work properly. First, the majority of the crowd are reliable in deriving their answers. Second, there are sufficient responses collected for each question. The WoC aggregation and the worker filtering can be applied together. The filtering is typically applied prior to or during the crowdsourcing, while the WoC aggregation is typically applied during or after the crowdsourcing.

The simplest WoC approach is the \textit{majority vote} (MV). It considers the correct answer to a question to be the one endorsed by the majority of the crowd-workers. It assumes that each response is independent to one another and has the same quality irrespective of workers' abilities. Consequently, its performance is usually limited, especially when responses collected for each question are scarce.

\subsection{Statistical Models for Quality Control}
Crowdsourced responses are fundamentally \textit{not independent} and \textit{vary in quality}. The former suggests that the quality of a response can be indicated by the quality of other responses to which it is related (e.g. by coming from the same crowd-worker). The latter implies that responses with higher quality should be modelled to have greater influence in the WoC aggregation and vice-versa.

\textit{Statistical models} provide a means of encoding the dependency of response quality on relevant aspects of crowdsourcing (e.g. crowd-workers who gave the responses). \textit{Inference} procedures can be applied to these models to estimate the response quality and important attributes (e.g. worker ability/expertise) of its dependent aspects. Aggregation for obtaining the correct response to each question is also carried out by the inference procedures.
\begin{table*}[t]
\label{table:chp1_survey_summary}
\scriptsize
\caption{A summary of current QCC surveys.}
\begin{tabular}{|l|l|l|l|l|l|} 
\hline
Survey &Areas&Details&Pros&Cons\\
\hline
& &Review of designs for tasks/&Diverse design problems&Lack of technical details about\\
&&measures/interfaces that allow: & and strategies related to&how statistical methods are\\
\cite{allahbakhsh2013quality} &Mechanism&- Assessment of response quality &measuring and controlling&used/integrated in the designs\\
\cite{chittilappilly2016survey}&Design&(e.g. by quiz, expert/peer review); &response quality are covered&for various crowdsourcing\\
\cite{daniel2018quality}&&- Assurance of response quality&with brief descriptions.&applications.\\
&&(e.g. by worker filtering, selection&&\\
&&,training, team work)&&\\
\hline
 & &Review of statistical models for&Technical details about a&Lack of design information on\\
 &&QCC which estimate:&variety of statistical models&crowdsourcing applications.\\
\cite{muhammadi2015unified}&Statistical&- Worker ability/expertise;&are specified. They include&\\
\cite{zhang2016learning}&Modelling&- Question difficulty;&model assumptions,variables&Ignoring aspects other than\\
\cite{zheng2017truth}&Methods&- Question true answer;&parameter estimation, etc.&worker and question which\\
&&- Response quality;&&also affect response quality:\\
&&&&- Context (in which workers\\
&&&& are situated, e.g. time, location);\\
&&&&- Answer options\\
&&&&(their semantic relationships).\\

\hline
\end{tabular}
\end{table*}
\subsection{Quality Control Mechanism Design}
A \textit{quality control mechanism} is a program which runs to (reactively or proactively) control modules of a crowdsourcing task which interact with crowd-workers to improve their performance. The \textit{design} of such a mechanism specifies how the control is conducted. For instance, a payment (and bonus) module is common to crowdsourcing tasks. In this case, a payment mechanism can be designed to manipulate this module on the time and amount workers get rewarded such that they are constantly motivated to do their best. 

Quality control mechanisms are usually based on statistical models which provide various estimates designed to trigger and direct the mechanisms' control of the task modules. For example, in payment mechanism designs, the ability estimate for a worker can be used to determine whether she needs to be rewarded or not. In addition, the modelling assumptions can also be used to derive theoretical guarantees for the mechanisms that they inspire desirable worker behaviour (e.g. being honest in their responses). We will discuss these in details later in the survey.  

\begin{figure*}[t]
\centering
\includegraphics[width=6in]{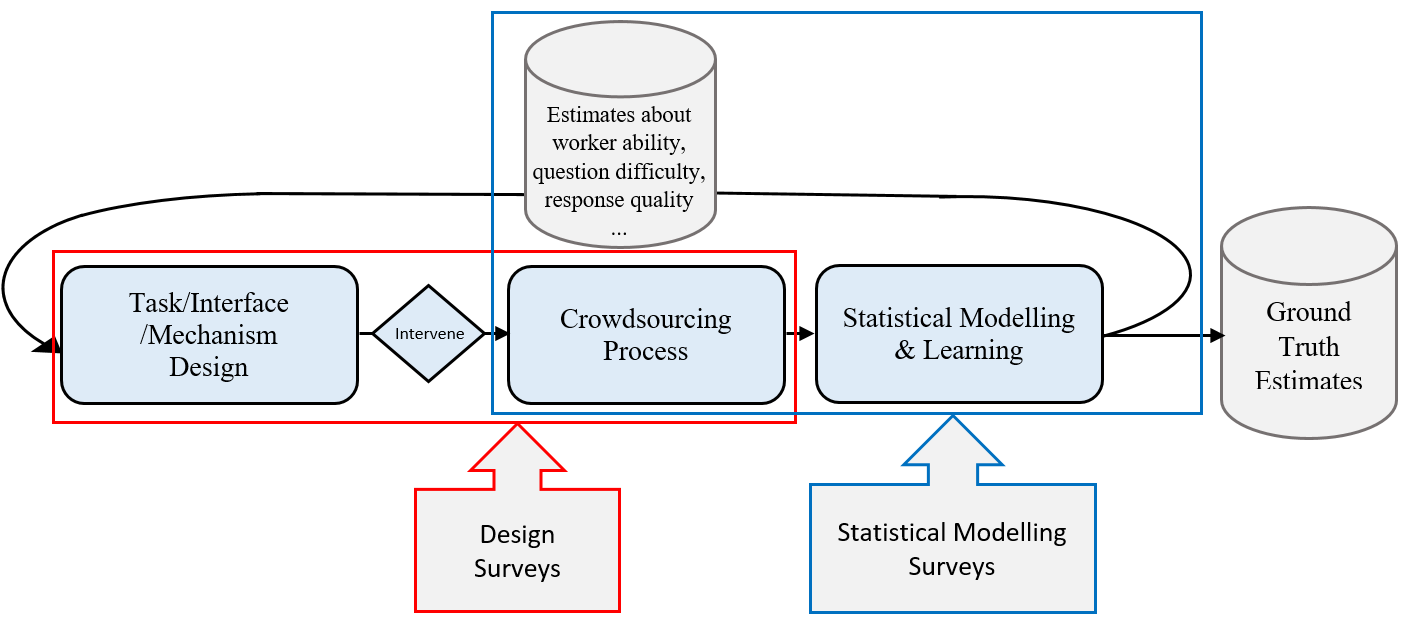}
  \caption{A diagram that shows a quality control process for crowdsourcing and its different parts reviewed by quality control design and statistical modelling surveys. There however lacks a current survey that links the statistical models with the designs.}
\label{chp2:design_stats_diagram}
\end{figure*}

\subsection{Related Surveys}
A variety of surveys in the area of crowdsourcing have been published in the past. The subjects of these reviews include general overviews of crowdsourcing \cite{kittur2013future, yin2014survey}, management of certain components of crowdsourcing platforms, such as the routing and recommendation of tasks \cite{geiger2014personalized, luz2015survey}, and different applications of crowdsourcing, such as information retrieval \cite{lease2013crowdsourcing}, software engineering \cite{mao2015survey}, data mining \cite{xintong2014brief}, health and medicine \cite{ranard2014crowdsourcing}, music \cite{gomes2012crowdsourcing} and neogeography \cite{heipke2010crowdsourcing}. 

With the development of sophisticated quality control mechanisms in recent years, surveys specifically regarding QCC research have started to appear. Existing surveys on quality control for crowdsourcing mainly fall into two areas: \textit{quality control mechanism design} and \textit{statistical (modelling) methods}, which is summarized in Table~\ref{table:chp1_survey_summary}. 

Current design surveys~\cite{allahbakhsh2013quality,chittilappilly2016survey,daniel2018quality} mainly review \textit{general-purpose} design strategies for building QCC mechanisms, response quality measures and user interfaces deployed on various task modules. As shown in Fig.~\ref{chp2:design_stats_diagram}, their main weakness is that they barely provide any technical details about the statistical methods (and vice versa) which provide the designs with various estimates, and how the designs utilize these estimates to achieve their goals.

Current statistical modelling surveys~\cite{muhammadi2015unified,zhang2016learning,zheng2017truth} focus on the models and inference procedures that learn attributes of two crowdsourcing aspects: the crowd-worker and the question. Table~\ref{table:chp1_survey_summary} shows their attributes reviewed by these surveys, namely: worker ability/expertise, question difficulty and true answer probability. There are other worker and question attributes that might affect or indicate response quality. For example, worker \textit{effort} and \textit{honesty} are attributes that have been frequently modelled by game-theoretic methods used for incentive designs~\cite{frongillo2015elicitation, ho2016eliciting}. Questions that might contain more than one correct answers (thereby suggesting \textit{subjectivity}) have also been studied and modelled in~\cite{Yuan2018Distinguishing}. Current surveys however ignore quality control methods that model and learn these attributes.

Apart from workers and questions, there are other aspects in crowdsourcing that also affect or indicate response quality. The \emph{context} in which each worker is situated is such an aspect. It is characterized by the time, location, labelling device, pay rate, Web-page, etc. It has been modelled and learned for quality control purposes in~\cite{jung2014predicting, venanzi2016time, jung2015modeling}. 

Another aspect is the \textit{response options} from which workers choose to answer questions. When the set of options is finite and large, their \textit{semantic relationships} might become very useful for indicating the responses according to recent QCC research~\cite{han2016incorporating,fang2017improving,Yuan2018LabelCategory}. Our survey reviews all the QCC research that deals with these aspects.
\subsection{Contributions}
In this work, we overcome the weaknesses of current QCC surveys with following contributions:
\begin{itemize}
    \item A unified taxonomy of the key aspects of crowdsourcing considered by the QCC research to determine the response quality, along with their important attributes.
    \item A graph framework of all the QCC research in which the nodes represent the crowdsourcing aspects and their attributes considered by the research. The graph contains paths indicating different lines of research.
    \item A systematic review of all the QCC research based on the graph. The review starts from the most basic work, which considered only the crowd-worker ability, and finishes with the most sophisticated work, which considered multiple aspects and attributes.
    \item Hierarchical categorization of QCC papers. The hierarchies are constructed according to features of the papers (e.g. considered aspects, modelling assumptions, parameter estimation techniques, design features, etc.). 
\item An in-depth discussion of the QCC research and identification of current research limitations along with the proposal of future research directions.
\end{itemize}

\section{Crowdsourcing Aspects}
\begin{figure}
\begin{tikzpicture}[scale=0.45, every node/.style={transform shape},grow'=right,level distance=2.5in,sibling distance=.2in]
\tikzset{edge from parent/.style= 
            {thick, draw, edge from parent fork right},
         every tree node/.style=
            {minimum width=1in,text width=1.5in,align=center}}
\Large            
\Tree
[.{\textbf{Crowdsourcing\\Aspects}} [.\node[draw,ellipse,text width=1in]{Question/Item}; 
[.\node[draw,ellipse,text width=1in]{Objective Question};  [.\node[draw,rectangle,rounded corners]{Difficulty}; ]] [.\node[draw,ellipse,text width=1in]{(Purely)\\Subjective Question}; [.\node[draw,rectangle,rounded corners]{Subjectivity}; ] ] [.\node[draw,ellipse,text width=1in]{(Partially)\\Subjective Question}; 
    [.\node[draw,rectangle,rounded corners]{Difficulty\\+\\Subjectivity}; ]]] 
    [.\node[draw,ellipse,text width=1in]{Crowd-worker};  
    [.\node[draw,rectangle,rounded corners]{Ability}; [.\node[draw,rectangle,rounded corners]{Motivation}; [.\node[draw,rectangle,rounded corners,text width=1in]{Effort}; ] [.\node[draw,rectangle,rounded corners,text width=1.2in]{Truthfulness}; ]] [.\node[draw,rectangle,rounded corners]{Expertise}; ]] 
    [.\node[draw,rectangle,rounded corners]{Preference}; ] ]
    [.\node[draw,ellipse,text width=1in]{Response Option};  
    [.\node[draw,rectangle,rounded corners]{Semantic Relationships}; ] ] [.\node[draw,ellipse,text width=1in]{Context};  
    [.\node[draw,ellipse,text width=1in]{Task-level\\Context}; [.\node[draw,rectangle,rounded corners,text width=2in]{Task features\\e.g. type of task (such as relevance judgement, image annotation), \\duration and location, \\pay rate and bonus,\\numbers of questions and workers, etc.}; ]]
    [.\node[draw,ellipse,text width=1in]{Session/Page-level\\Context}; [.\node[draw,rectangle,rounded corners,text width=2in]{Session features\\e.g. duration and location,\\pay rate and bonus,\\number of questions (in a session), etc.}; ]]
    [.\node[draw,ellipse,text width=1in]{Response-level\\Context}; [.\node[draw,rectangle,rounded corners,text width=2in]{Response features\\e.g. duration and location,\\pay rate and bonus,\\response order, etc.}; ]] ]]   
\end{tikzpicture}
\caption{Aspects of crowdsourcing considered by the QCC research. Each ellipse node denotes a particular crowdsourcing aspect. Question and context have finer definitions at level 2. Rectangle nodes denote attributes of aspects exploited by the QCC methods. }
\label{fig:crowdsourcing_aspects}
\end{figure}
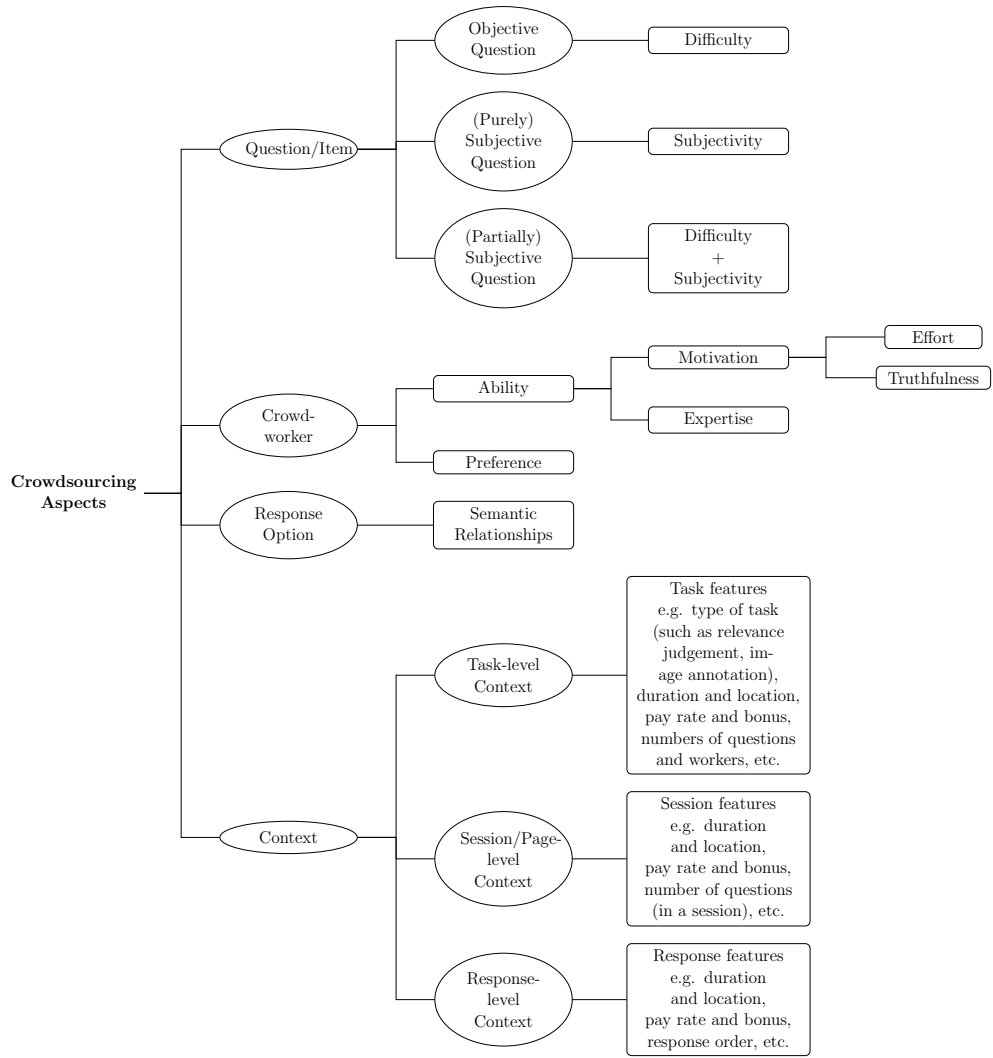

The QCC research makes (explicit or implicit) assumptions on how quality of responses is correlated with certain aspects of crowdsourcing. It proposes to encode these assumptions into quality control mechanisms and statistical models to effectively control the response quality. To conduct a systematic review on this research, we start by specifying four prominent aspects of crowdsourcing it has considered. These aspects (along with their key attributes) are shown in Fig.~\ref{fig:crowdsourcing_aspects}. 

\textit{Items/Questions}. A question is the smallest unit in a crowdsourcing task. It is \textit{objective} when it has a single correct answer. It is \textit{purely subjective} when every answer is correct (e.g. a demographic question). In between lies \textit{partially subjective} questions. Intuitively speaking, it has multiple correct answers but at least one incorrect answer. For instance, consider the task to judge if an image contains a person wearing fashionable clothes or not~\cite{Loni:2013,Loni:2014}. Workers can disagree on what the correct answers are for an image that contains a clothes-wearing person as they have different preferences for fashion styles. Nonetheless, they should agree on what the wrong answer is for that image (i.e. no person in it).

Both objective and partially subjective questions possess certain degrees of \textit{difficulty} which obscure their correct answers from crowd-workers to various extents. A difficult question leads to variation in the responses across crowd-workers. For extremely difficult questions, workers may have to resort to random guessing.

\textit{Crowd-workers}. A worker has certain \textit{motivation} for answering the questions, and a certain level of (\textit{domain}) \textit{expertise} required by the subject of a task. The motivation governs the levels of \textit{effort} exerted by the worker to answer each question, and also the \textit{truthfulness} of the worker's responses. When two workers possess the same level of expertise (in the same domain), the one motivated to exert more effort is more likely to yield high-quality responses. The truthfulness of the worker's response determines how likely she responds with the answer she believes to be correct (for a question). If the worker is malicious, she is more likely to give a different response.

When encountering a partially subjective question (e.g. judging whether an item is fashion-related or not), a worker would exhibit \textit{preferences} for certain \textit{features} of the item (e.g. vintage design and fabrics). Such preferences are independent from the ability of the worker, thereby having no effect on the correctness of a response.

\textit{Response options}. It is possible for a crowdsourcing task to have a large (but finite) set of response options for its questions. A typical example is the crowdsourcing for the ImageNet database \cite{deng2009imagenet} which stores millions of images according to tens of thousands of categories that are connected in a \textit{semantic relational} graph. In this case, the semantic relationships between these categories (as response options) would influence the workers' responses. For instance, consider the classification task of 120 dog breeds from the ImageNet~\cite{khosla2011novel}. A worker is more likely to confuse the correct breed (e.g. golden retriever) with a breed that is more related to it (e.g. Labrador) than with a less related one (e.g. Chihuahua).

\textit{Contexts}. A context in crowdsourcing is an \textit{environment} in which crowd-workers are situated. Changing or intervening in the context can affect the motivation of workers which further affects the quality of their responses. From the literature, we found that different QCC methods control the context at different \textit{levels of granularity}. We thus propose to refine the definition of the context according to the following three levels:
\begin{itemize}
    \item \textit{Task} level: a task-level context is characterized by features which distinguish multiple tasks on the same crowdsourcing platform. These features contain information about individual tasks such as the task durations, locations, domains, settings including the pay rates, instructions, minimum accuracy for quizzes and so on. They also contain information about the responses from workers who took multiple tasks such as micro-averaged and macro-averaged worker response time for each task.
    \item \textit{Session} level: a session-level context within a task corresponds to a question page of the task that a worker has completed and submitted. We call the process of the worker answering all the questions on that page a \textit{working session} of that worker. A session-level context can thus be described by features regarding a task page (e.g. the pay rate for answering each question on the page, their topics and total word counts, etc.). They also concern the worker's responses within a working session (e.g. the average response time on each question, the total response time, the device used for the response, etc.). 
    
    \item \textit{Response} level: a response-level context represents an even finer level of granularity for the within-task contexts. It corresponds to a single response given by a worker to a question, and is described by features regarding this response (e.g. its payment, duration, location and position in the sequence of all the responses given by the same worker).
\end{itemize}

\begin{figure}[t]
\centering
\includegraphics[width=5in]{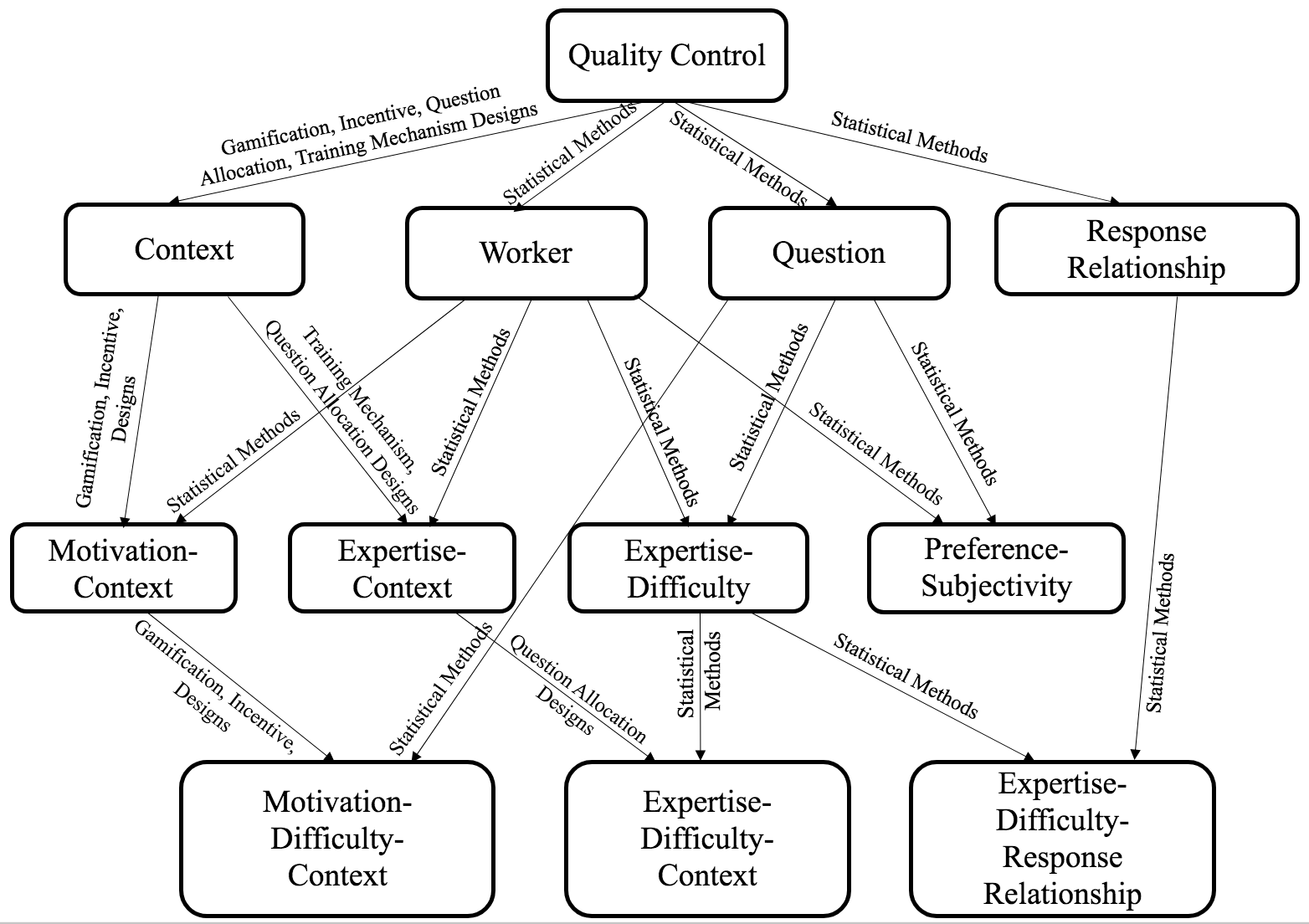}
  \caption{A graph framework of the past QCC research.}
\label{chp2:attribute_interact_diagram}
\end{figure}
\section{A Graph Framework of the QCC Research }
This survey is organized according to the graph shown in Fig.~\ref{chp2:attribute_interact_diagram}. This graph captures the crowdsourcing \textit{aspects}, their key \textit{attributes} and the past QCC research that has considered them jointly for developing control mechanisms and statistical models. The root node ``Quality Control'' has outgoing edges to the four major crowdsourcing aspects specified in Fig.~\ref{fig:crowdsourcing_aspects}. Nodes at the second level correspond to \textit{pairs} of attributes which have been jointly considered by some of the QCC research. Likewise, each node at the third level involves a \textit{triplet} of attributes. Research considering the triplets of attributes is usually the most sophisticated in terms of the assumptions made. 

Labels on the edges describe the methods developed by the QCC research that exploited the aspects. For example, \textit{gamification, payment, question allocation}, and \textit{training} mechanisms all make use of the worker context to control response quality. Edges labeled ``Statistical Methods", indicate that the corresponding research focuses on statistical modelling and inference of the attributes.

A path in the graph denotes a line of research. For example, the path from node ``Context'' to node ``Expertise-Difficulty-Context'' denote the line of research that designed question allocation mechanisms by considering more aspects (i.e. worker expertise and question difficulty).

Edges merging at a node indicate the combination of the corresponding QCC methods. For instance, the edges merging at node ``Motivation-Context'' indicate that the research considering worker motivation and contexts combines mechanism designs with statistical methods.

We carry out the rest of the survey according to the proposed graph. Each section corresponds to a node in the graph and the QCC methods reviewed in the section are given by the labels.

\section{Modelling Worker Ability}
Modelling the effect that individual crowd-workers have on the quality of responses (QoR) is most widely adopted by the QCC research. The effect is assumed to be determined by the ability/expertise of each worker.

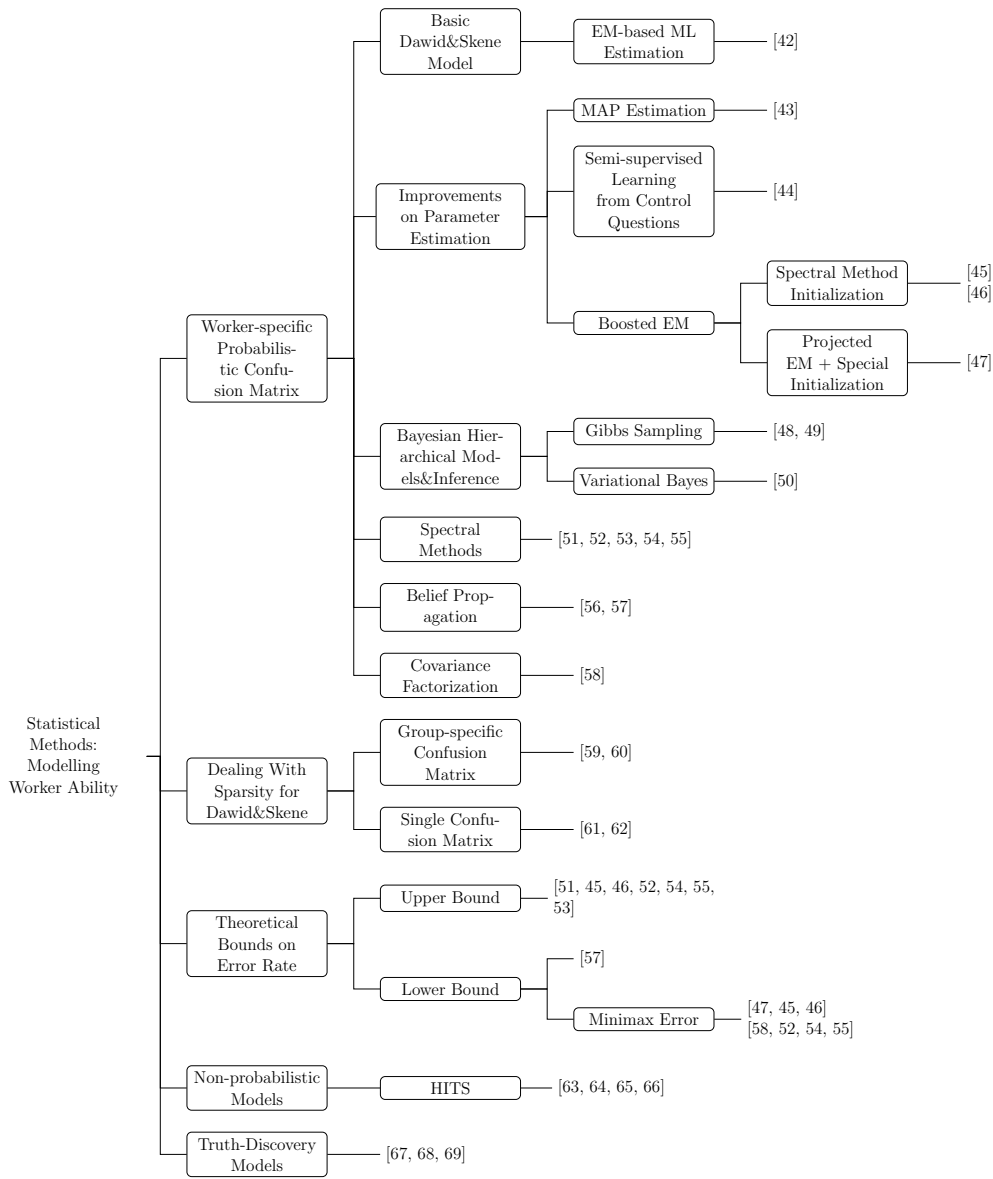
\begin{figure}[p]
\begin{tikzpicture}[scale=0.45, every node/.style={transform shape},grow'=right,level distance=2.25in,sibling distance=.25in]
\tikzset{edge from parent/.style= 
            {thick, draw, edge from parent fork right},
         every tree node/.style=
            {minimum width=1in,text width=1.5in,align=center}}
\Large            
\Tree
[.\node[text width=1.8in]{Statistical\\Methods:\\Modelling Worker Ability}; 
[.\node[draw,rectangle,rounded corners,text width=1.5in]{Worker-specific Probabilistic Confusion Matrix}; 
[.\node[draw,rectangle,rounded corners,text width=1.5in]{Basic\\Dawid\&Skene Model};  [.\node[draw,rectangle,rounded corners,text width=1.5in]{  EM-based ML Estimation }; [.\node[,align=left]{  \cite{dawid1979maximum} }; ] ]] [.\node[draw,rectangle,rounded corners,text width=1.6in]{Improvements on Parameter Estimation}; [.\node[draw,rectangle,rounded corners]{MAP Estimation}; [.\node[,align=left]{  \cite{snow2008cheap} }; ] ] 
[.\node[draw,rectangle,rounded corners]{Semi-supervised Learning from Control Questions}; [.\node[,align=left]{  \cite{tang2011semi} }; ] ]
[.\node[draw,rectangle,rounded corners]{Boosted EM}; [.\node[draw,rectangle,rounded corners]{Spectral Method Initialization}; [.\node[,align=left]{  \cite{zhang2014spectral}\\\cite{zhang2016spectral} }; ] ] [.\node[draw,rectangle,rounded corners,text width=1.5in]{Projected EM + Special Initialization }; [.\node[,align=left]{ \cite{gao2013minimax} }; ] ]   ] ]
[.\node[draw,rectangle,rounded corners]{Bayesian Hierarchical Models\&Inference}; [.\node[draw,rectangle,rounded corners]{Gibbs Sampling}; [.\node[,align=left]{  \cite{carpenter2011hierarchical,kim2012bayesian} }; ] ] [.\node[draw,rectangle,rounded corners]{Variational Bayes}; [.\node[,align=left]{  \cite{simpson2013dynamic} }; ] ] ]
[.\node[draw,rectangle,rounded corners]{Spectral Methods};
[.\node[text width=2in,align=left]{  \cite{ghosh2011moderates,karger2011budget,dalvi2013aggregating,karger2011iterative,karger2014budget} }; ]
]
[.\node[draw,rectangle,rounded corners,text width=1.5in]{  Belief Propagation }; [.\node[,align=left]{  \cite{liu2012variational,ok2016optimality} }; ] ]
[.\node[draw,rectangle,rounded corners]{Covariance Factorization}; [.\node[,align=left]{  \cite{bonald2017minimax} }; ] ]
]
[.\node[draw,rectangle,rounded corners,text width=1.5in]{Dealing With Sparsity for Dawid\&Skene};  [.\node[draw,rectangle,rounded corners,text width=1.5in]{  Group-specific Confusion Matrix }; [.\node[,align=left]{  \cite{venanzi2014community,moreno2015bayesian} }; ] ]
 [.\node[draw,rectangle,rounded corners,text width=1.5in]{  Single Confusion Matrix }; [.\node[,align=left]{  \cite{liu2012truelabel+,kamar2015identifying} }; ] ]]
[.\node[draw,rectangle,rounded corners,text width=1.5in]{Theoretical Bounds on Error Rate};  [.\node[draw,rectangle,rounded corners,text width=1.5in]{  Upper Bound }; [.\node[text width=2.1in,align=left]{  \cite{ghosh2011moderates,zhang2014spectral,zhang2016spectral,karger2011budget,karger2011iterative,karger2014budget,dalvi2013aggregating} }; ] ]
 [.\node[draw,rectangle,rounded corners,text width=1.5in]{  Lower Bound }; [.\node[text width=1.5in,align=left]{  \cite{ok2016optimality} }; ] [.\node[draw,rectangle,rounded corners,text width=1.5in]{  Minimax Error }; [.\node[text width=2.1in,align=left]{  \cite{gao2013minimax,zhang2014spectral,zhang2016spectral}\\\cite{bonald2017minimax,karger2011budget,karger2011iterative,karger2014budget} }; ] ] ]]  
[.\node[draw,rectangle,rounded corners,text width=1.5in]{Non-probabilistic Models}; [.\node[draw,rectangle,rounded corners,text width=1.5in]{HITS}; [.\node[text width=2in,align=left]{  \cite{aydin2014crowdsourcing,yan2015opportunities,kajimura2015quality,sunahase2017pairwise} }; ] ]]
[.\node[draw,rectangle,rounded corners,text width=1.5in]{Truth-Discovery Models}; [.\node[,align=left]{\cite{li2016survey,ouyang2015debiasing,ouyang2016aggregating}};  ]]
]   
\end{tikzpicture}
\caption{A taxonomy of QCC papers that only considered worker ability/expertise to determine response quality. They focused on statistical modelling and inference.}
\end{figure}

\subsection{The Dawid \& Skene Model}
The work that first introduced a model for estimating worker ability was that of Dawid \& Skene \cite{dawid1979maximum}, which we refer to as the DS model. It deals with the scenario where a worker $i$ reads a question $j$ which has underlying true answer $l_j$, and then gives her response $r_{ij}$ to the question. Both the true answer $l_j$ and the response $r_{ij}$ are members of a finite set of options $\mathcal{K}$ which is the same for each question. In this work, the crowd-worker $i$ is modelled by a $|\mathcal{K}|\times|\mathcal{K}|$ confusion matrix $\boldsymbol{\Pi}_{i}$ where $|\mathcal{K}|$ is the size of the option set $\mathcal{K}$. Each diagonal entry of the matrix $\pi_{ikk}$ records the probability of a response from worker $i$ being correct: $\pi_{ikk}=P(r_{ij}=k|l_j=k), k \in \mathcal{K}$. Each off-diagonal entry $\pi_{ikk'}$ records the probability of the response being incorrect: $\pi_{ikk'}=P(r_{ij}=k'|l_j=k), k\neq k'$. Since the $k$-th row of the confusion matrix stores conditional probabilities, the entries must sum to one $\sum_{k' \in \mathcal{K}}\pi_{ikk'} = 1$.

The DS model adopts the expectation-maximization (EM) algorithm to perform maximum likelihood (ML) estimation over all the worker responses. The ML estimation finds the locally optimal estimates for the model parameters: both the probabilities in the worker-specific confusion matrices and the true answers for the questions. In this case, the aggregation for inferring the true answers is essentially integrated into the EM estimation process.

The EM algorithm comprises two alternate steps which are iterated until the convergence of the likelihood. In the E-step, for each question $j$, the DS model estimates the probability of the true answer $l_j$ being equal to each category $k \in \mathcal{K}$ given the current estimates $\hat{\boldsymbol{\pi}}_{ik}$ for the entries in the $k$-th row of the confusion matrix $\hat{\boldsymbol{\Pi}}_{i}$ as:
\begin{eqnarray}
\hat{\rho}_{jk} = P(\hat{l}_j = k |\mathcal{R}_j,\{\hat{\boldsymbol{\pi}}_{ik}\}_{i \in \mathcal{I}_j, k \in \mathcal{K}})=\frac{\prod\limits_{i \in \mathcal{I}_j}\prod\limits_{k' \in \mathcal{K}}(\hat{\pi}_{ikk'})^{ \mathbbm{1}\{r_{ij} = k'\}  }P(\hat{l}_j = k)}{\sum\limits_{m \in \mathcal{K}}\prod\limits_{i \in \mathcal{I}_j}\prod\limits_{k' \in \mathcal{K}}(\hat{\pi}_{imk'})^{ \mathbbm{1}\{r_{ij} = k'\}  }P(\hat{l}_j = m)}
\label{eqn:correct_ans}
\end{eqnarray}In Eq.~\ref{eqn:correct_ans}, $\hat{l}_j$ is the estimate of the correct answer $l_j$; $\hat{\rho}_{jk}$ is the estimate of the probability $\rho_{jk}$ of the correct answer $l_j = k$; $P(\hat{l}_j = k)$ is the estimate of the prior probability of $l_j = k$; $\mathcal{I}_j = \{ i | (i \in \mathcal{I}) \land (r_{ij} \neq ?) \}$ is the set of workers who have answered the question $j$, with ``?'' denoting a missing value; $\mathcal{R}_j = \{ r_{ij} | i \in \mathcal{I}_j\}$ are their responses; $\mathbbm{1}\{...\}$ is the indicator function.
In the M-step, the algorithm estimates the rest of its parameters given the current estimates $ \{\hat{\rho}_{jk}\}_{k \in \mathcal{K}}$:
\begin{equation}
\begin{split}
\hat{\pi}_{ikk'} = \frac{\sum\limits_{j \in \mathcal{J}_i}\hat{\rho}_{jk}\mathbbm{1}{\{r_{ij} = k\}}}{\sum\limits_{k' \in \mathcal{K}}\sum\limits_{j \in \mathcal{J}_i}\hat{\rho}_{jk'}\mathbbm{1}{\{r_{ij} = k'\}}}
\label{eqn:worker_accu}
\end{split}
\end{equation}
\begin{equation}
\begin{split}
P(\hat{l}_j = k) = \frac{1}{|\mathcal{J}|}\sum\limits_{j \in \mathcal{J}}\hat{\rho}_{jk}
\end{split}
\label{eqn:correct_ans1}
\end{equation} In Eq. \ref{eqn:worker_accu}, $\mathcal{J}_i = \{ j | (j \in \mathcal{J}) \land (r_{ij} \neq ?) \}$ is the set of questions answered by worker $i$. In Eq. \ref{eqn:correct_ans1}, $|\mathcal{J}|$ is the total number of questions.

There has been simplification in some of the subsequent work to only consider binary response options. This means the confusion matrix specific to each worker is reduced to only two free parameters (i.e. the diagonal entries). Such a simplified model is called a \textit{two-coin} DS model \cite{raykar2010learning}. If the two diagonal entries in this model are assumed equal (i.e. the error probability is independent of the true answer), the model is further simplified into a \textit{one-coin} model \cite{zhang2014spectral}.

\subsection{Improvements on Parameter Estimation}
As an improvement on the maximum likelihood EM estimation for the DS model, Snow et al.~\cite{snow2008cheap} employed MAP estimation for the parameters. Later, Tang and Lease~\cite{tang2011semi} leveraged control questions for improving the ML estimation. This was achieved via semi-supervised learning based on the true answers of the control questions to refine the parameter estimation for the DS model. Zhang et al.~\cite{zhang2014spectral, zhang2016spectral} proposed to use spectral methods to initialize the EM algorithm to escape local optimum in the search for the optimal true answer probability estimates and confusion matrix estimates. Gao and Zhou~\cite{gao2013minimax} modified the M-step of the EM algorithm using a projection strategy which acts as an alternative to prior distributions over worker abilities to prevent EM estimation from over-fitting. The authors also customised the initialization procedures for the projected EM to avoid local optimum. 

Instead of point estimation based on EM, Kim and Ghahramani~\cite{kim2012bayesian} and Carpenter~\cite{carpenter2011hierarchical} applied their respective Bayesian treatments for building hierarchical DS models and used Gibbs sampling to infer posterior distributions for the models' parameters. Preserving the same Bayesian hierarchical frameworks, Simpson et al.~\cite{simpson2013dynamic} applied variational Bayesian inference to efficiently estimate the joint probabilities of the worker confusion matrices and the question true answers. The authors further extended the frameworks with dynamic worker confusion matrices and adapted the variational lower bound accordingly.

Ghosh et al.~\cite{ghosh2011moderates} proposed a spectral algorithm that decomposes a question-question matrix capturing response correlations across questions to learn workers' abilities and questions' true answers. The algorithm works with the one-coin DS formulation and requires the existence of one expert worker and that every worker answers every question. Removing the last two constraints, Dalvi et al.~\cite{dalvi2013aggregating} proposed spectral methods that focus on decomposing worker-worker matrices capturing response correlations across workers. Karger et al.~\cite{karger2011budget} proposed to apply spectral decomposition explicitly to the worker-question response matrix. Recently, Bonald and Combes~\cite{bonald2017minimax} factorized the response covariance matrix into one-coin worker abilities and developed a non-iterative algorithm that allows for real-time ability estimation. The algorithm leverages the current most informative pair of workers with the highest covariance and their respective covariance with a target worker to estimate the target worker's current ability.

Karger et al.~\cite{karger2011iterative,karger2014budget} combined the previous spectral analysis with belief propagation algorithms which achieved nearer optimality for true answer estimation. Liu et al.~\cite{liu2012variational} applied full belief propagation (BP) to both one-coin and two-coin DS models. They found that the efficacy of BP depends heavily on the choice of prior distributions over worker ability variables. In comparison, Ok et al.~\cite{ok2016optimality} proposed a practical belief propagation algorithm which works on the one-coin DS model and does not rely on choice of prior distributions over worker abilities.
\subsection{Dealing with Sparsity in DS}
The main drawback of the DS model is its vulnerability to sparsity in the responses from workers. When the number of responses per worker is small, the confusion matrix for each worker cannot be estimated reliably. If the number of response options is also large, each confusion matrix will be massive (quadratic in the number of response options), and estimating each matrix will most certainly result in overfitting to the sparse responses. To solve this problem, the sparse response information from individual workers needs to be combined and smoothed so that the overall response information is sufficient for reliable estimation of each confusion matrix. 

Venanzi et al.~\cite{venanzi2014community} applied Bayesian hierarchical modelling to infer clusters of workers, called ``communities''. The model allows for combining noisy response information across workers, such that the confusion matrix for each worker is smoothed based on the cluster to which the worker belongs. The model is parametric in the sense that it requires users to set up the number of communities in advance. 

In~\cite{moreno2015bayesian}, the authors proposed both the Bayesian non-parametric modelling alternative and its hierarchical extension to enable more flexible partitioning of the workers into communities. The number of clusters is learned jointly with the confusion matrices for the communities and the individual workers. 

Instead of determining a number of worker communities, Liu and Wang~\cite{liu2012truelabel+} and Kamar et al.~\cite{kamar2015identifying} have developed statistical models for an extreme case where the individual worker matrices is merged to form a single confusion matrix specific to the entire worker population. This confusion matrix is then balanced against the confusion matrix for each worker to smooth out its noisy information.

%There is a line of work that has modelled adaptive question assignment as changing the properties of a \textit{random bipartite} graph which links questions and workers to denote question assignment.

\subsection{Theoretical Bounds on Error-Rate of DS Estimation Techniques}
A line of work has investigated bounds on the error (convergence) rates of various parameter estimation algorithms employed for learning the DS model as the redundancy of responses increases. Among them, Ghosh et al.~\cite{ghosh2011moderates} first derived the upper bound for the error rate of a spectral inference method for true answer prediction. The technique considered binary responses under the one-coin DS model and assumed that each crowd-worker has answered a large number of questions. With the same setting, Gao and Zhou~\cite{gao2013minimax} showed the global maximum likelihood estimator follows a minimax lower bound with respect to the error rate, and their projected EM algorithm theoretically can achieve nearly that rate.

Changing the setting by allowing each worker to answer just a few (rather than many) questions, Zhang et al.~\cite{zhang2014spectral,zhang2016spectral} proved their proposed EM with spectral method initialization yielded a tighter upper bound than that of~\cite{ghosh2011moderates} and was faster to achieve the minimax error rate than~\cite{gao2013minimax}. Later, Bonald and Combes~\cite{bonald2017minimax} showed that their non-iterative algorithm can match an even stricter lower bound on the minimax error rate than the previous work. Karger et al.~\cite{karger2011budget} proved that when each worker provides only a few responses, their proposed framework based on low-rank spectral decomposition yielded a strict upper bound on the error rate. Meanwhile, they proved the framework matched a lower bound on the minimax error rate that could only be achieved by the best possible question
assignment with an optimal true answer inference algorithm. Later, they~\cite{karger2011iterative,karger2014budget} showed their framework based on belief propagation methods yields a tighter upper bound than~\cite{karger2011budget} and the same lower bound on the minimax error rate. The same strict upper bound was also achieved by the framework based on spectral methods proposed in~\cite{dalvi2013aggregating}. In~\cite{ok2016optimality}, the authors proved that their framework based on belief propagation is able to achieve the tightest possible error-rate lower bound under the same setting with an additional requirement that each worker is assigned at most two questions. Recently, Gao et al.~\cite{gao2016exact} established both the lower and the upper bounds of the error rates that match exactly the exponential rates under the setting in~\cite{karger2011iterative,karger2014budget}. 

Despite their theoretical soundness and empirical feasibility, current work in QCC error rate analysis has seldom relaxed the binary-response assumption and the one-coin DS modelling assumption. For works relaxing the binary-response assumption, we have only found that of~\cite{karger2013efficient} whose inference framework adopted the same setting as~\cite{karger2011iterative,karger2014budget} but extended the binary response options to multiple response options. They proved that a tight upper bound on the error rate can still be reached using proposed spectral methods. For works relaxing the one-coin DS assumption, we have only found that Liu et al.~\cite{liu2012variational} imposed a two-coin DS model and has done empirical error rate analysis based on belief propagation, EM and a mean field method with the conclusion that all of them can achieve nearly optimal rates with proper prior settings on worker confusion matrices. 

Despite these limitations, the current results of the QCC error rate analysis provide insights into setting up both the early stopping criteria and the response redundancy requirement for crowdsourcing provided that certain parameter estimation methods are used for true answer estimation.

%In \cite{karger2014budget} (\cite{karger2014budget}), a theoretical analysis was done based on the worker confusion matrix modelling between the expertise of workers and the difficulty of items for characterizing the trade-off between the total labelling cost and the prediction error on binary true labels. It proved that to achieve less than a certain amount of error, the budget needs to scale linearly with the number of items weighed by a logarithmic dependence on the target error.

\subsection{Non-Probabilistic Worker-Ability Models} 
A number of quality control methods that are not based on probabilistic inference have also been developed. In these methods, the abilities of crowd-workers and the quality of their answers are modelled to mutually support one another. The more workers who give the same response to a question, the higher the quality of that response will be. Likewise, the more high-quality responses provided by a worker, the higher the ability of this worker. The above mutually supportive relationship is analogous to the authority-hub relationship modelled by the HITS framework \cite{kleinberg1999authoritative}. In this case, the inference of the worker ability and the quality of responses has been conducted in similar ways to HITS in \cite{aydin2014crowdsourcing,yan2015opportunities,kajimura2015quality,sunahase2017pairwise}. Inferring a true answer is done by aggregating all the responses to a question weighed by the responses' respective quality estimates. Alternatively, the weights can be the difference between each response and the true answer estimate \cite{aydin2014crowdsourcing}.

\subsection{Truth-Discovery Worker-Ability Models} 
Research on \textit{truth discovery} from different (possibly unreliable) information sources \cite{li2016survey} shares similar modelling characteristics to the QCC methods. Each source of information (equivalently a crowd-worker) is associated with a reliability variable, called a weight, which measures the quality of the claim (i.e. a response) made by the source about an object (i.e. a data item). The general goal in truth discovery modelling is to minimize the sum of the weighted distance between each claim and the latent ground-truth of the corresponding object. This distance function can be any loss function depending on the data type of the claims and the ground truths. For example, the claims and the ground-truths can be observed as real-valued feature vectors, which do not often occur in QCC modelling, and their distances can be measured as the squared or absolute difference between these vectors. In comparison, QCC models mainly focus on minimizing the log-loss during the learning process. A comprehensive review on truth discovery models was provided by~\cite{li2016survey}. Thus, the details of these models will not be covered in our survey. A few QCC methods have adopted the idea of distances in truth discovery when handling tasks over ordinal or continuous responses such as object counting \cite{ouyang2016aggregating}, and percentage annotation \cite{ouyang2015debiasing}.

\section{Modelling Worker Expertise and Question Difficulty}\label{sec:worker_expertise_question_diff}
More sophisticated QCC methods consider not only the worker ability but also the question difficulty. The assumption is that some questions are intrinsically more difficult than others and thus are expected to receive less reliable responses.

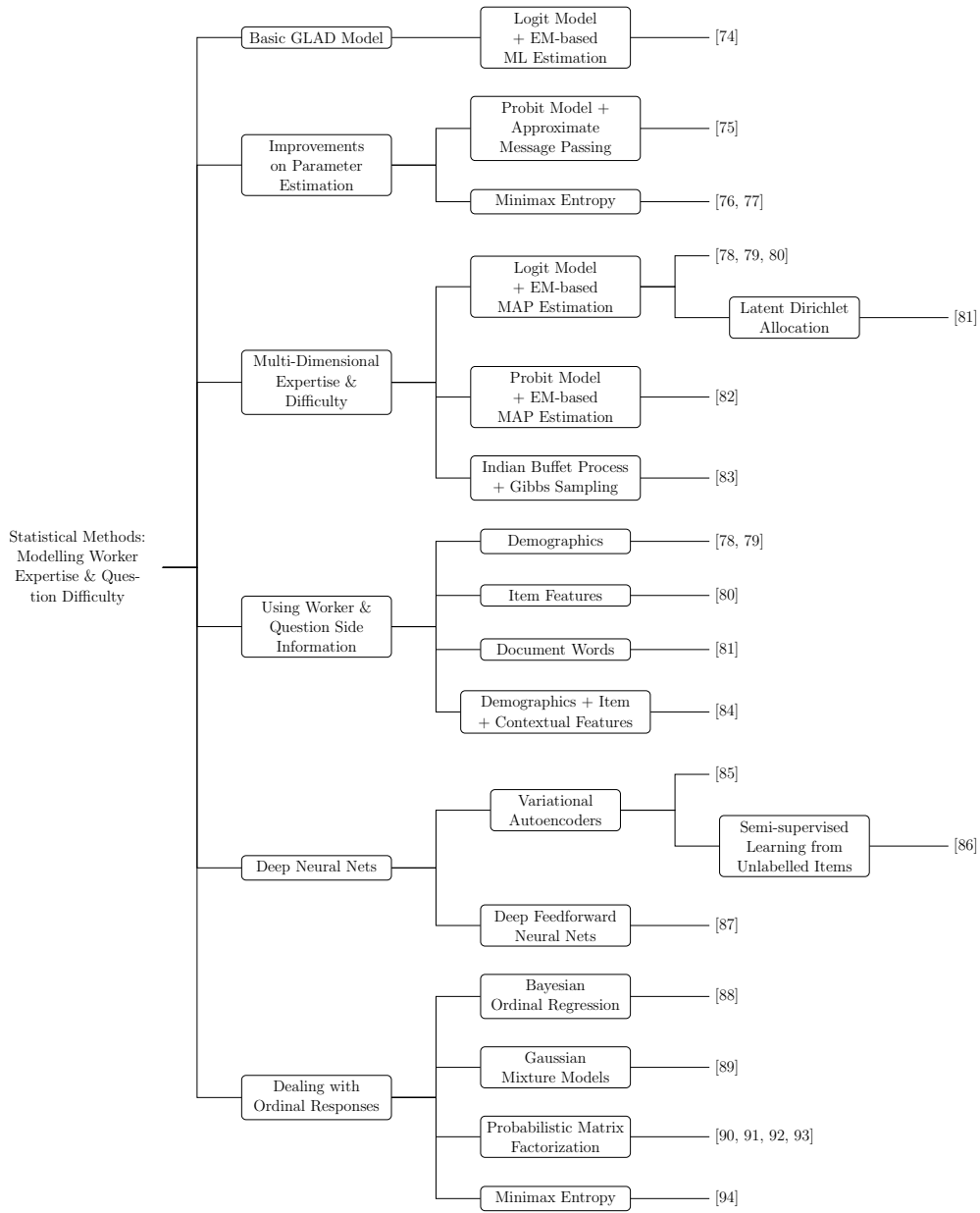
\begin{figure}[p]
\begin{tikzpicture}[scale=0.425, every node/.style={transform shape},grow'=right,level distance=3in,sibling distance=.35in]
\tikzset{edge from parent/.style= 
            {thick, draw, edge from parent fork right},
         every tree node/.style=
            {minimum width=1in,text width=2in,align=center}}
\Large            
\Tree
[.\node[text width=2in]{Statistical Methods:\\Modelling Worker Expertise \& Question Difficulty}; [.\node[draw,rectangle,rounded corners,text width=1.75in]{Basic GLAD Model}; [.\node[draw,rectangle,rounded corners,text width=1.75in]{Logit Model + EM-based ML Estimation}; [.\node[,align=left]{ \cite{whitehill2009whose} }; ] ] 
] 
[.\node[draw,rectangle,rounded corners,text width=1.75in]{Improvements on Parameter Estimation}; [.\node[draw,rectangle,rounded corners,text width=2in]{Probit Model + Approximate Message Passing}; [.\node[,align=left]{ \cite{bachrach2012grade} }; ] ] 
[.\node[draw,rectangle,rounded corners,text width=2in]{Minimax Entropy}; [.\node[,align=left]{ \cite{zhou2012learning,zhou2015regularized} }; ] ]
] 
[.\node[draw,rectangle,rounded corners,text width=1.75in]{Multi-Dimensional Expertise \& Difficulty}; [.\node[draw,rectangle,rounded corners,text width=2in]{ Logit Model + EM-based MAP Estimation }; [.\node[,align=left]{ \cite{ruvolo2010exploiting,ruvolo2013exploiting,kajino2012convex} }; ] [.\node[draw,rectangle,rounded corners,text width=1.5in]{Latent Dirichlet Allocation}; [.\node[,align=left]{ \cite{ma2015faitcrowd} }; ]  ] ] 
[.\node[draw,rectangle,rounded corners,text width=2in]{ Probit Model + EM-based MAP Estimation }; [.\node[,align=left]{ \cite{welinder2010multidimensional} }; ]  ] 
[.\node[draw,rectangle,rounded corners,text width=2in]{Indian Buffet Process + Gibbs Sampling}; [.\node[,align=left]{ \cite{wauthier2011bayesian} }; ]  ]
]
[.\node[draw,rectangle,rounded corners,text width=1.75in]{Using Worker \& Question Side Information}; [.\node[draw,rectangle,rounded corners,text width=1.75in]{Demographics}; [.\node[,align=left]{ \cite{ruvolo2010exploiting,ruvolo2013exploiting}  };  ]  ] [.\node[draw,rectangle,rounded corners,text width=1.75in]{Item Features}; [.\node[,align=left]{ \cite{kajino2012convex}  };  ]  ][.\node[draw,rectangle,rounded corners,text width=1.75in]{Document Words}; [.\node[,align=left]{ \cite{ma2015faitcrowd}  };  ]  ] [.\node[draw,rectangle,rounded corners,text width=2.25in]{Demographics + Item + Contextual Features}; [.\node[,align=left]{ \cite{Yuan2017sideinfo}  };  ]  ] ] 
[.\node[draw,rectangle,rounded corners,text width=1.75in]{Deep Neural Nets}; [.\node[draw,rectangle,rounded corners,text width=1.5in]{Variational Autoencoders}; [.\node[,align=left]{ \cite{yin2017aggregating} }; 
] [.\node[draw,rectangle,rounded corners,text width=1.75in]{ Semi-supervised Learning from Unlabelled Items }; [.\node[,align=left]{ \cite{atarashi2018semi} }; 
] ]
] [.\node[draw,rectangle,rounded corners,text width=1.75in]{Deep Feedforward Neural Nets}; [.\node[,align=left]{ \cite{gaunt2016training}  }; ] ] ]
[.\node[draw,rectangle,rounded corners,text width=1.75in]{Dealing with Ordinal Responses}; [.\node[draw,rectangle,rounded corners,text width=1.75in]{Bayesian\\Ordinal Regression};  [.\node[,align=left]{ \cite{lakshminarayanan2013inferring} }; 
] ] [.\node[draw,rectangle,rounded corners,text width=1.75in]{Gaussian\\Mixture Models};  [.\node[,align=left]{ \cite{metrikov2013modification}  }; 
] ] [.\node[draw,rectangle,rounded corners,text width=1.75in]{Probabilistic Matrix Factorization}; [.\node[,align=left]{\cite{jung2012inferring,jung2012improving,jung2013crowdsourced,jung2014quality} }; 
]  ] [.\node[draw,rectangle,rounded corners,text width=1.75in]{Minimax Entropy};  [.\node[,align=left]{\cite{zhou2014aggregating} }; 
] ] ]
]
\end{tikzpicture}
\caption{A taxonomy of QCC papers that considered both worker expertise and question difficulty to explain response quality. They focused on statistical modelling and inference.}
\label{}
\end{figure}

\subsection{The GLAD Model}\label{sec:expert_diff} 
Based on the above assumption, some QCC models have taken the question difficulty into account alongside the worker expertise for estimating the quality of each response \cite{whitehill2009whose,ruvolo2010exploiting,bachrach2012grade,ruvolo2013exploiting}. They model the quality of a response as the \textit{probability of it being correct}. The most fundamental work in this area is the \textit{GLAD} model, in which a logistic function $\delta_{ij}$ is used to represent the probability that response $r_{ij}$ is correct:
\begin{eqnarray}
\begin{split}
 \delta_{ij}=P(r_{ij} = l_j | e_i, d_j) = \frac{1}{1+\exp(-e_i/\exp(d_j))}
\end{split}
\label{eqn:correct_prob}
\end{eqnarray} where $e_i$ is a real-valued parameters that models the ability/expertise of the worker $i$, and (also real-valued) $d_j$ models the difficulty of the question $j$. The exponent transformation $\exp$($d_j$) serves to prevent negative difficulty. Compared to the DS model which considers the bias of a worker towards certain (possibly incorrect) responses, GLAD only models the probability of the correct response and ignores any biases by assuming their corresponding probabilities to be uniform as $\frac{\delta_{ij}}{|\mathcal{K}|-1}$. Here, $|\mathcal{K}|-1$ is the number of incorrect responses. 

Like the DS model, the GLAD model also adopts the EM algorithm for parameter estimation. More specifically, in the E-step, for each question $j$, the GLAD model estimates the probability of the true answer $l_j = k$ as:
\begin{eqnarray}
\hat{\rho}_{jk} = P(\hat{l}_j = k |\mathcal{R}_j,\{\hat{\boldsymbol{e}}_{i}\}_{i \in \mathcal{I}_j},\hat{d}_{j})=\frac{\prod\limits_{i \in \mathcal{I}_j}\delta_{ij}^{ \mathbbm{1}\{r_{ij} = k\}}\frac{1 - \delta_{ij}}{|\mathcal{K}| - 1}^{ \mathbbm{1}\{r_{ij} \neq k\}}P(\hat{l}_j = k)}{\sum\limits_{k' \in \mathcal{K}}\prod\limits_{i \in \mathcal{I}_j}\delta_{ij}^{ \mathbbm{1}\{r_{ij} = k'\}}\frac{1 - \delta_{ij}}{|\mathcal{K}| - 1}^{ \mathbbm{1}\{r_{ij} \neq k'\}}P(\hat{l}_j = k')}
\label{eqn:correct_ans2}
\end{eqnarray}
In the M-step, the expected joint likelihood over the observed responses and unobserved true answers with respect to $\hat{\rho}_{jk}$ is maximized over the rest of the parameters. The expected joint likelihood $Q$ is formulated as follows in GLAD:
\begin{align*}
\begin{split}
&Q(\{e_i\}_{i \in \mathcal{I}}, \{d_j\}_{j \in \mathcal{J}} \textbf{ ; } \mathcal{R},\{\hat{l}_j\}_{j \in \mathcal{J}})=\\& \sum\limits_{j \in \mathcal{J}}\sum\limits_{k \in \mathcal{K}}\hat{\rho}_{jk}\log\big(P(\hat{l}_j = k)\big) + \sum\limits_{j \in \mathcal{J}}\sum\limits_{i \in \mathcal{I}_j}\sum\limits_{k \in \mathcal{K}}\hat{\rho}_{jk}\log\Big(\delta_{ij}^{ \mathbbm{1}\{r_{ij} = k\}}\frac{1 - \delta_{ij}}{|\mathcal{K}| - 1}^{ \mathbbm{1}\{r_{ij} \neq k\}}\Big)
\end{split}
\label{eqn:em_opmtization}
\end{align*} where $\mathcal{R}$ is the set of all the responses. The sets of expertise $\{e_i\}_{i \in \mathcal{I}}$ and difficulty $\{d_j\}_{j \in \mathcal{J}}$ are estimated using gradient descent by taking partial derivatives of $Q$ with respect to each element.

\subsection{Improvements on Parameter Estimation}
Many later QCC methods have followed the main idea in GLAD \cite{whitehill2009whose}: the quality of a response being a probabilistic function over variables representing the worker ability and the question difficulty. In~\cite{bachrach2012grade}, the probit model is used where the logistic function is replaced by the Gaussian cumulative function with the mean being the expertise $e_i$ minus a question-specific bias term, and the variance being the difficulty $d_j$. Additionally, instead of using the EM algorithm for the ML estimation, this work employs approximate message passing inference on the model parameters. 

Zhou et al.~\cite{zhou2012learning} used a minimax entropy model to estimate the accuracy of each response. The entropy function is evaluated over all the responses with respect to their probabilities. The ability of worker $i$ and the difficulty of question $j$ are introduced as Lagrangian multipliers for the constraints derived from the $i$-th row and the $j$-th column of the response matrix. The authors maximized the constrained entropy function with respect to the response probabilities $P(r_{ij}=k)$, and the ability and difficulty multipliers. Then, the constrained maximization was minimized with respect to the latent true answer of each question, which was shown by the authors to be equivalent to minimizing the KL-divergence between the probability estimates of the true answers and their underlying distribution. Later, Zhou et al.~\cite{zhou2015regularized} extended their original work by regularizing the minimax optimization with relaxed constraints to prevent its response probability estimates from overfitting sparse responses.

\subsection{Multi-dimensional Worker Expertise and Side Information}
More recent work has extended the worker-question interaction to be \textit{multi-dimensional}. They argue that workers can have their own areas of expertise and questions can be associated with the different areas. The authors of GLAD exploited this idea by simply converting the worker expertise and the question difficulty into vectors. Correspondingly, they converted the original scalar product into a dot product between these vectors in \cite{ruvolo2010exploiting,ruvolo2013exploiting}. 

Ruvolo et al.~\cite{ruvolo2010exploiting} was also the first work to leverage side information of both crowd-workers (i.e. their demographics) and questions (i.e. the features of the data items) for further improving the parameter estimation. It incorporates the side information into the multi-dimensional GLAD model as the design matrices for the linear regressions that respectively determine the prior means of the expertise and difficulty vectors. Similar work was done in \cite{welinder2010multidimensional} which used the Gaussian cumulative function to represent the response quality with the mean being the dot product specified in~\cite{ruvolo2010exploiting} minus a worker bias term. 

In~\cite{wauthier2011bayesian}, rather than set up the dimension for the expertise and the difficulty vectors in advance as the previous work did, the proposed method modelled the dimension and the selection of the underlying latent expertise and difficulty components of the vectors as a Bayesian non-parametric Indian Buffet process \cite{griffiths2011indian}. This work applied Gibbs sampling to infer the model parameters including the true answers. 

In~\cite{kajino2012convex}, convex optimization techniques were proposed for training worker-specific binary classifiers which took side information features about questions (indicating question difficulty) into account. To allow for the multi-dimensionality of the question features, these classifiers are endowed with weight (expertise) vectors in the following logistic function: 
\begin{eqnarray}
\begin{split}
\delta_{ij}=P(r_{ij} = l_j | \boldsymbol{w}_i, \boldsymbol{x}_j) = \frac{1}{1+\exp(\boldsymbol{w}^{T}_i\boldsymbol{x}_j)}
\end{split}
\label{eqn:convex_logistic_fun}
\end{eqnarray}
In Eq.~\ref{eqn:convex_logistic_fun}, $\boldsymbol{w}_i$ is the real-valued weight vector specific to worker $i$ and $\boldsymbol{x}_j$ is the real-valued feature vector for question $j$. The weight vectors follow a Multivariate Gaussian and the maximum a posteriori (MAP) estimate of the mean vector serves as the weight vector of a base classifier to estimate question true answers using Eq. $\ref{eqn:correct_ans2}$.  

In~\cite{ma2015faitcrowd}, the multi-dimensionality of the expertise was estimated in a topic-wise manner for questions each associated with a text document. The difficulty of each question was modelled to be independent from their topics as a single variable. The model used Latent Dirichlet Allocation (LDA) \cite{blei2003latent} with a universal distribution of the topics across all the documents (assuming each of them to be short). It draws the topic of a question from that distribution and selects the corresponding topical expertise of each worker (from their topical expertise vectors) to calculate the correctness probabilities of their responses.

In~\cite{Yuan2017sideinfo}, the authors proposed to extend the GLAD model by incorporating various types of side information to improve the parameter estimation when responses are scarce. The side information features concern worker demographics, question content and contextual information such as devices (e.g. PC, mobile phone), browsers, time periods of each session, time duration and orders of each response and so on. The worker and question features are incorporated into the GLAD model as:
\begin{eqnarray}
\delta_{ij}=P(r_{ij} = l_j | e_i, d_j, \boldsymbol{\alpha}, \boldsymbol{\beta}, \boldsymbol{x}_i, \boldsymbol{x}_j) = \frac{1}{1+\exp\big(-(e_{i}+\boldsymbol{\alpha}^{T}\boldsymbol{x}_{i})/\exp(d_{j}+\boldsymbol{\beta}^{T}\boldsymbol{x}_{j})\big)}
\label{eqn:demographic}
\end{eqnarray}
The dot product $\boldsymbol{\alpha}^{T}\boldsymbol{x}_{i}$ forms expertise offsets across the workers with the global coefficients $\boldsymbol{\alpha}$ learned to bring the offsets of similar workers closer together. This helps to smooth the irregular expertise estimates that result from the sparse responses across workers. This effect is also applied to the dot product $\boldsymbol{\beta}^{T}\boldsymbol{x}_{j}$ for calibrating the difficulty $d_j$. The context features were further divided by the authors into the session-level features and the response-level features. The session-level features were incorporated into the expertise factor of the GLAD model same as the worker features but with different global coefficients. The response-level features are incorporated as $(\delta_{ij} + \boldsymbol{\eta}^{T}_i\boldsymbol{x}_{ij})$ where  $\delta_{ij}$ is derived from Eq.~\ref{eqn:demographic}, $\boldsymbol{\eta}_i$ are worker-specific coefficients, and $\boldsymbol{x}_{ij}$ are the feature values regarding response $r_{ij}$. The coefficients $\boldsymbol{\eta}_i$ form a local linear regression over the feature vectors of all the responses made by worker $i$. Such local regressions addressed worker-specific biases that the GLAD model failed to handle properly~\cite{welinder2010multidimensional}.  

\subsection{Neural Network Approaches}
Deep Learning approaches~\cite{lecun2015deep} have become very popular over the last few years in Machine Learning applications. Recently, Yin et al.~\cite{yin2017aggregating} used variational autoencoders \cite{kingma2013auto} to map responses to each question into latent true answer distributions. The inputs and outputs of the autoencoders are vectors corresponding to individual questions. Each vector is a concatenation of the \textit{one-hot} encoding of the response given by each worker to a question. Both the encoder and the decoder are implemented as single-layer networks and the global weight vector for each layer accounts for the biases across the workers towards different response options. In addition to the weight vectors, a question-specific scalar term is incorporated at each layer to account for question difficulty. It does this by scaling the layer outputs before fed into a softmax transformation. 

Atarashi et al.~\cite{atarashi2018semi} leveraged semi-supervised learning and variational autoencoders to facilitate true answer inference. They used features of unlabelled items to help distinguish the true answers from some (item-specific) latent factors, both of which are assumed to have generated the various feature values of the labelled and the unlabelled items. Instead of using responses and true answers as input-output pairs for the encoder part (as done by~\cite{yin2017aggregating}), they used the labelled and unlabelled item features as the inputs, and both true answers and latent factors as the outputs for the encoder part. The relationships between the responses and the true answers are captured using multi-class logistic regression.

In~\cite{gaunt2016training}, deep feedforward neural networks were trained at two consecutive stages. To train the network at the first stage, accuracy of workers and difficulty of questions were estimated based on degrees of response agreement. Then, the accuracy and difficulty estimates are segmented into different levels (e.g. ``low'' and ``high'' levels of accuracy/difficulty). The inputs to the network correspond to individual responses. Each input vector contains both the overall estimates of worker accuracy and question difficulty and their estimates across the different levels. The outputs of the network are the correctness probability estimates of individual responses. They are used to construct inputs to the neural networks at the second stage. Each of these networks corresponds to a response option. The input to the $k$-th network consists of the response probability of each worker. The response probability equals the output from the first stage, i.e. the correctness probability $P(r_{ij}=l_j)$, if workers' responses are the particular option $k$. Otherwise, the probabilities equal $\frac{1-P(r_{ij}=l_j)}{|\mathcal{K}|-1}$. The output from this network is normalized as $P(l_j=k)$ for each question.

\subsection{Dealing with Ordinal Response Data}
In crowdsourcing, the response options are sometimes not categorical but rather ordinal (e.g. the relevance level of a document to a query) or continuous (e.g. the count of an object in an image). In this case, it is natural to measure the distance between each worker's response and the corresponding correct answer. This distance directly reflects the quality of the response, which can be modelled by a Gaussian density function. In this function, the mean is set to be the latent true answer, and the precision (i.e. the inverse of the Gaussian variance) typically is set to be the ratio of worker expertise to question difficulty.

In~\cite{lakshminarayanan2013inferring}, the above framework was followed by one extra treatment that was to use global incremental intercepts of an ordinal regression to model the natural ordering existing in the responses (e.g. no/medium/high relevance of documents to queries). In~\cite{metrikov2015aggregation}, the framework was further extended to be a worker-specific Gaussian mixture with each Gaussian component acting as a ``soft'' intercept in the ordinal regression. This model deals with the situation where workers show their individual biases towards different response options. 

In~\cite{jung2012inferring} and their subsequent work \cite{jung2012improving, jung2013crowdsourced, jung2014quality}, a slightly different framework was adopted where each response is set to follow a Gaussian distribution with the mean being the dot product between the latent variable vectors of each worker and the target question. The variance is always set to be one. Sharing the same idea with~\cite{kajino2012convex}, these methods estimate the true answer of each question using the dot product between the question's latent vector and the MAP estimate of the mean vector over all the workers' latent vectors. Zhou et al.~\cite{zhou2014aggregating} adapted the minimax entropy principle from their previous work \cite{zhou2012learning} to make it compatible with ordinal responses by modifying the constraints to account for the natural ordering in the response options.

\section{Modelling Worker Preferences and Question Subjectivity}\label{chp2:worker_prefer_question_subj}
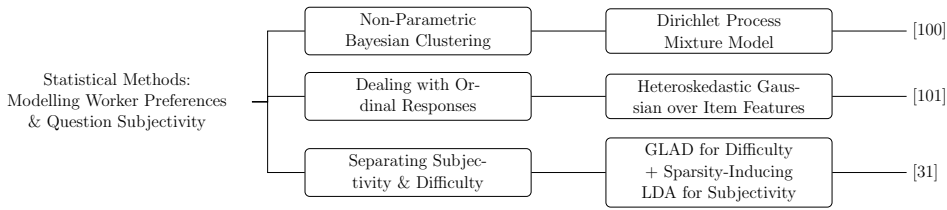
\begin{figure}[h!]
\begin{tikzpicture}[scale=0.45, every node/.style={transform shape},grow'=right,level distance=3.5in,sibling distance=0.2in]
\tikzset{edge from parent/.style= 
            {thick, draw, edge from parent fork right},
         every tree node/.style=
            {minimum width=1in,text width=2.5in,align=center}}
\Large            
\Tree
[.\node[text width=3in]{Statistical Methods:\\Modelling Worker Preferences \& Question Subjectivity}; [.\node[draw,rectangle,rounded corners,text width=2.5in]{Non-Parametric Bayesian Clustering}; [.\node[draw,rectangle,rounded corners,text width=2.5in]{Dirichlet Process Mixture Model}; [.\node[,align=left]{  \cite{tian2012learning} };  ] ]  ] [.\node[draw,rectangle,rounded corners,text width=2.5in]{Dealing with Ordinal Responses}; [.\node[draw,rectangle,rounded corners,text width=2.5in]{Heteroskedastic Gaussian over Item Features}; [.\node[,align=left]{  \cite{nguyen2016probabilistic} };  ] ]  ] [.\node[draw,rectangle,rounded corners,text width=2.5in]{Separating Subjectivity \& Difficulty}; [.\node[draw,rectangle,rounded corners,text width=2.5in]{GLAD for Difficulty + Sparsity-Inducing LDA for Subjectivity}; [.\node[,align=left]{  \cite{Yuan2018Distinguishing} };] ]]]
\end{tikzpicture}
\caption{A taxonomy of QCC papers that considered worker preferences and question subjectivity. They focused on statistical modelling and inference.}
\label{}
\end{figure}

In crowdsourcing, some tasks might contain partially subjective questions. These questions possess more than one correct answer and at least one incorrect answer. To answer such a question correctly, a worker needs to avoid any incorrect answers, which depends on her ability/expertise and the difficulty of the question. Meanwhile, her subjective preferences/opinions on different (subjective) features of the question will cause her to prefer one of the correct answer options over the others. %In the literature, \cite{welinder2010multidimensional} (\cite{welinder2010multidimensional}) has hinted from their modelling of the objective questions that the preference-feature interaction for the perception might precede the expertise-difficulty interaction which acts as the corruption of the perception.

To the best of our knowledge, very few QCC methods have endeavoured to distinguish between question difficulty and subjectivity in one unified model. In~\cite{tian2012learning}, the authors assumed that a higher joint degree of difficulty and subjectivity for an entire task can increase the number of underlying groupings of responses given to each question in the task, with the expected number of responses in each group becoming smaller. The authors proposed to infer the response groups using a Dirichlet Process Mixture Model~\cite{antoniak1974mixtures}. Despite attributing the response variation to both difficulty and subjectivity, the paper made no attempt to separately model the two even though they might induce different types of interactions with workers. 

In~\cite{nguyen2016probabilistic}, the authors focused on ordinal ratings given to partially subjective items with observed features. The rating $r_{ij}$ is assumed to follow a Gaussian distribution with the mean and the variance linearly regressed on the observed features $\boldsymbol{x}_j$ of item $j$ as $r_{ij} \sim \mathcal{N}(\boldsymbol{w}^{T}\boldsymbol{x}_j,\exp(\boldsymbol{v}^{T}\boldsymbol{x}_j))$. Here, $\boldsymbol{w}$ and $\boldsymbol{v}$ are global coefficient vectors. The variance encodes the mixing effects of the subjectivity and difficulty of the item. Thus, this method is also not intended to separate the modelling of the two properties.

Recently, Yuan et al.~\cite{Yuan2018Distinguishing} proposed the first QCC model that separates the difficulty and subjectivity. It replaces the single truth variable $l_j$ for a question $j$ with a subjective truth variable $l_{ij}$ specific to each response $r_{ij}$. The subjectivity of question $j$ is captured by factorizing $l_{ij}$ into another latent variables that represent worker $i$'s preferences and the question's features. The difficulty of the question is directly encoded as a variable $d_j$ which counteracts expertise $e_i$ in the following logistic function:
\begin{eqnarray}
\begin{split}
 \delta_{ij}=P(r_{ij} = l_{ij} | e_i, d_j) = \frac{1}{1+exp(-(e_i-d_j))}
\end{split}
\label{eqn:correct_prob1}
\end{eqnarray}
To prevent the preference-feature factorization of $l_{ij}$ from explaining all the response variation, the authors further replaced the preference vector of each worker with much sparser topic-preference vectors. These vectors are associated with each worker via an LDA. In this case, each worker has a distribution over a finite set of interested topics. A topic is associated with a topic-preference vector to be used in the preference-feature factorization for each $l_{ij}$ via topic-response assignment. This work is also the first to propose a measure of subjectivity in crowdsourcing which is the \textit{expected number of correct answers with respect to underlying groups of workers}. The worker groups were clustered using K-means with the Elbow method over the topic distribution of each worker.

\section{Modelling Worker Motivation and Worker Contexts}\label{sec:worker_motive_context}
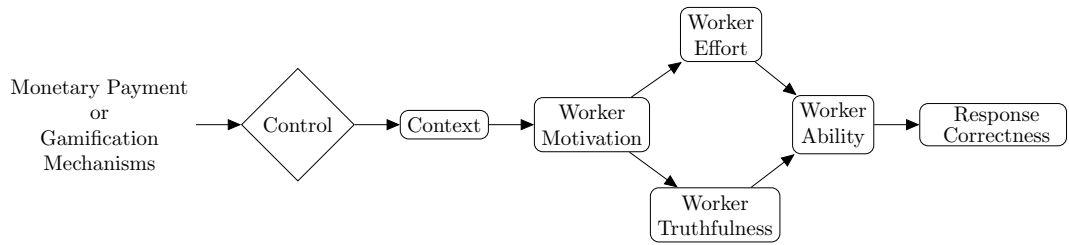
\begin{figure*}[ht]
  	\scalebox{.8} 		{\begin{tikzpicture}
\node[align=center] (im) {Monetary Payment\\or\\Gamification\\Mechanisms};
\node[draw, diamond, right=of im, xshift=-.25cm] (ic) {Control};
\node[draw,right=of ic,rectangle,rounded corners, xshift=-.25cm] (c) {Context};
\node[draw,align=center,right=of c,rectangle,rounded corners, xshift=-.25cm] (mi) {Worker\\Motivation};
\node[draw,align=center,right=of mi,rectangle,rounded corners, yshift=1.5cm, xshift=-0.5cm] (efi) {Worker\\Effort};
\node[draw,align=center,right=of mi,rectangle,rounded corners, yshift=-1.5cm, xshift=-1cm] (tfi) {Worker\\Truthfulness};
\node[draw,align=center,right=of efi,rectangle,rounded corners, yshift=-1.5cm, xshift=-0.5cm] (ei) {Worker\\Ability};
  \node[latent,align=center,right=of ei,rectangle,rounded corners, xshift=-.25cm] (qij) {~~~Response~~~\\~~~Correctness~~~};

%\node[draw,above=of qij,rectangle,rounded corners, xshift=1.5cm] (dj) {Question Difficulty};
%\edge{ei,dj}{qij};
\edge{ei}{qij};
\edge{efi,tfi}{ei};
\edge{mi}{efi,tfi};
\edge{c}{mi};
\edge{ic}{c};
\edge{im}{ic};

\end{tikzpicture}}
    %\caption{Motivation-Difficulty-Context Interaction}
    %\begin{subfigure}[b]{0.5\columnwidth}
  %\centering
  	%\scalebox{.8} 		{\input{models/ability_context_diagram}}
    %\caption{Expertise-Difficulty-Context Interaction}
    %\label{fig:chp1_ability_context_diagram}
  %\end{subfigure}%
  \caption{A diagram shows payment and gamification mechanisms control worker contexts to affect worker motivation, which further influence worker effort and truthfulness, and eventually the correctness of responses. %Figure~\ref{fig:chp1_ability_context_diagram} shows the \textit{expertise-difficulty-context} interaction. In this case, worker contexts can be intervened by training mechanisms to improve worker expertise and further response correctness.
  }
  \label{fig:chp1_motive_context_diagram}
\end{figure*}Different incentive mechanisms are designed to favour different characteristics of worker motivation. According to~\cite{ryan2000intrinsic}, motivations can be broadly characterized as being either \textit{extrinsic}, which is the desire to gain monetary payoffs and avoid costs, or \textit{intrinsic}, which is the desire to achieve fulfilment and enjoyment. Correspondingly, we categorize the past literature on incentive mechanisms into \textit{monetary payment} and \textit{gamification} approaches which are respectively responsible of motivating the workers extrinsically and intrinsically.

\begin{figure}[p]
\begin{tikzpicture}[scale=0.4, every node/.style={transform shape},grow'=right,level distance=1.5in,sibling distance=.25in]
\tikzset{edge from parent/.style= 
            {thick, draw, edge from parent fork right},
         every tree node/.style=
            {minimum width=1in,text width=1.25in,align=center}}
\Large            
\Tree
[.\node[text width=2in]{Statistical Method Based Design:\\Modelling Worker Motivation \&\\Intervened Context}; [.\node[draw,rectangle,rounded corners,text width=1.25in]{Monetary Payment Intervention}; [.\node[draw,rectangle,rounded corners,text width=1in]{Task-Level Context}; [.\node[draw,rectangle,rounded corners,text width=1.25in]{Worker Accuracy Evaluation over Control Questions}; [.\node[draw,rectangle,rounded corners,text width=1.25in]{Confidence-Aware Evaluation}; [.\node[,align=left]{ \cite{shah2015approval} };  ] ] [.\node[draw,rectangle,rounded corners,text width=1.25in]{Skip-Aware Evaluation}; [.\node[,align=left]{ \cite{shah2015double} };  ] ] [.\node[draw,rectangle,rounded corners,text width=1.25in]{Redo-Aware Evaluation}; [.\node[,align=left]{ \cite{shah2016no} };  ] ] ]   ]  [.\node[draw,rectangle,rounded corners,text width=1in]{Response-Level Context}; [.\node[draw,rectangle,rounded corners,text width=1.25in]{ Output Agreement };  [.\node[,align=left]{  \cite{von2008designing} };  ] ]  [.\node[draw,rectangle,rounded corners,text width=1.25in]{Modelling Worker Prior \& Posterior Beliefs of Correct Answers}; [.\node[draw,rectangle,rounded corners,text width=1.25in]{Homogeneous Priors \& Heterogeneous Posteriors}; [.\node[draw,rectangle,rounded corners,text width=1.25in]{Non-specific Prior \& Posterior Distributions}; [.\node[draw,rectangle,rounded corners,text width=1.25in]{Output Agreement};  [.\node[,align=left]{  \cite{waggoner2014output} };  ] ] ]  [.\node[draw,rectangle,rounded corners,text width=1.25in]{Assumed Prior \& Posterior Distributions}; [.\node[draw,rectangle,rounded corners,text width=1.25in]{Peer Prediction};  [.\node[,align=left]{  \cite{miller2005eliciting} };  ] [.\node[draw,rectangle,rounded corners,text width=1in]{Handling Worker Collusion};  [.\node[text width=1.25in,align=left]{  \cite{jurca2007collusion,jurca2009mechanisms,kong2016putting} };  ] ]  [.\node[draw,rectangle,rounded corners,text width=1.25in]{Requester-Provided Responses}; [.\node[,align=left]{ \cite{frongillo2015elicitation,ho2016eliciting} };   ]  ] ]    ]  [.\node[draw,rectangle,rounded corners,text width=1.3in]{Worker-Reported Posteriors};  [.\node[draw,rectangle,rounded corners,text width=1.25in]{Bayesian Truth Serum};  [.\node[,align=left]{  \cite{prelec2004bayesian} };  ]  [.\node[draw,rectangle,rounded corners,text width=1.25in]{Handling Few Reports};  [.\node[,align=left]{  \cite{parkes2012robust} };  ] ]   [.\node[draw,rectangle,rounded corners,text width=1.25in]{Handling Continuous Responses};  [.\node[,align=left]{  \cite{parkes2012robust} };  ] ]  ]  ] ]  ] [.\node[draw,rectangle,rounded corners,text width=1.25in]{Modelling Worker Prior Beliefs of Correct Answers};   [.\node[draw,rectangle,rounded corners,text width=1.25in]{Peer Truth Serum};  [.\node[draw,rectangle,rounded corners,text width=1.25in]{Assumed Homogeneous Priors}; [.\node[,align=left]{\cite{faltings2014incentives} };   ]  ] [.\node[draw,rectangle,rounded corners,text width=1.25in]{Estimated Homogeneous Priors};  [.\node[,align=left]{\cite{radanovic2016incentives} };   ] ]  ]   ]    [.\node[draw,rectangle,rounded corners,text width=1.25in]{Modelling Effort-Aware Response Correctness};  [.\node[draw,rectangle,rounded corners,text width=1.25in]{One-Coin Dawid\&Skene};   [.\node[draw,rectangle,rounded corners,text width=1.25in]{Constant Cost of Efforts}; [.\node[draw,rectangle,rounded corners,text width=1.25in]{Output Agreement};  [.\node[,align=left]{ \cite{dasgupta2013crowdsourced} };   ] ]  [.\node[draw,rectangle,rounded corners,text width=1.25in]{Peer Prediction};  [.\node[,align=left]{ \cite{witkowski2013dwelling} };   ] ]  ] [.\node[draw,rectangle,rounded corners,text width=1.25in]{Estimated Costs of Efforts}; [.\node[draw,rectangle,rounded corners,text width=1.25in]{Output Agreement};  [.\node[,align=left]{ \cite{liu2016learning} };   ] ]  [.\node[draw,rectangle,rounded corners,text width=1.25in]{Peer Prediction};  [.\node[draw,rectangle,rounded corners,text width=1.25in]{Sequential Learning};  [.\node[,align=left]{ \cite{liu2017sequential} };   ]  ] ]  ]  ]  [.\node[draw,rectangle,rounded corners,text width=1.25in]{Two-Coin Dawid\&Skene};   [.\node[draw,rectangle,rounded corners,text width=1.25in]{Constant Cost of Efforts};  [.\node[draw,rectangle,rounded corners,text width=1.25in]{Peer Prediction};  [.\node[draw,rectangle,rounded corners,text width=1.25in]{Classifier-Based Peer Responses}; [.\node[,align=left]{ \cite{liu2017machine} };   ]  ]  ]  ]  ]  ]     [.\node[draw,rectangle,rounded corners,text width=1.25in]{Empirical Evaluation on Peer Prediction Mechanisms}; [.\node[,align=left]{ \cite{gao2014trick} };   ]  ]     ] ]  ]
\end{tikzpicture}
\caption{A taxonomy of QCC papers that considered the interaction between worker context and (extrinsic) motivation. They focused on designing monetary payment mechanisms which rely on statistical modelling (and possibly estimation) of worker attributes.}
\label{}
\end{figure}
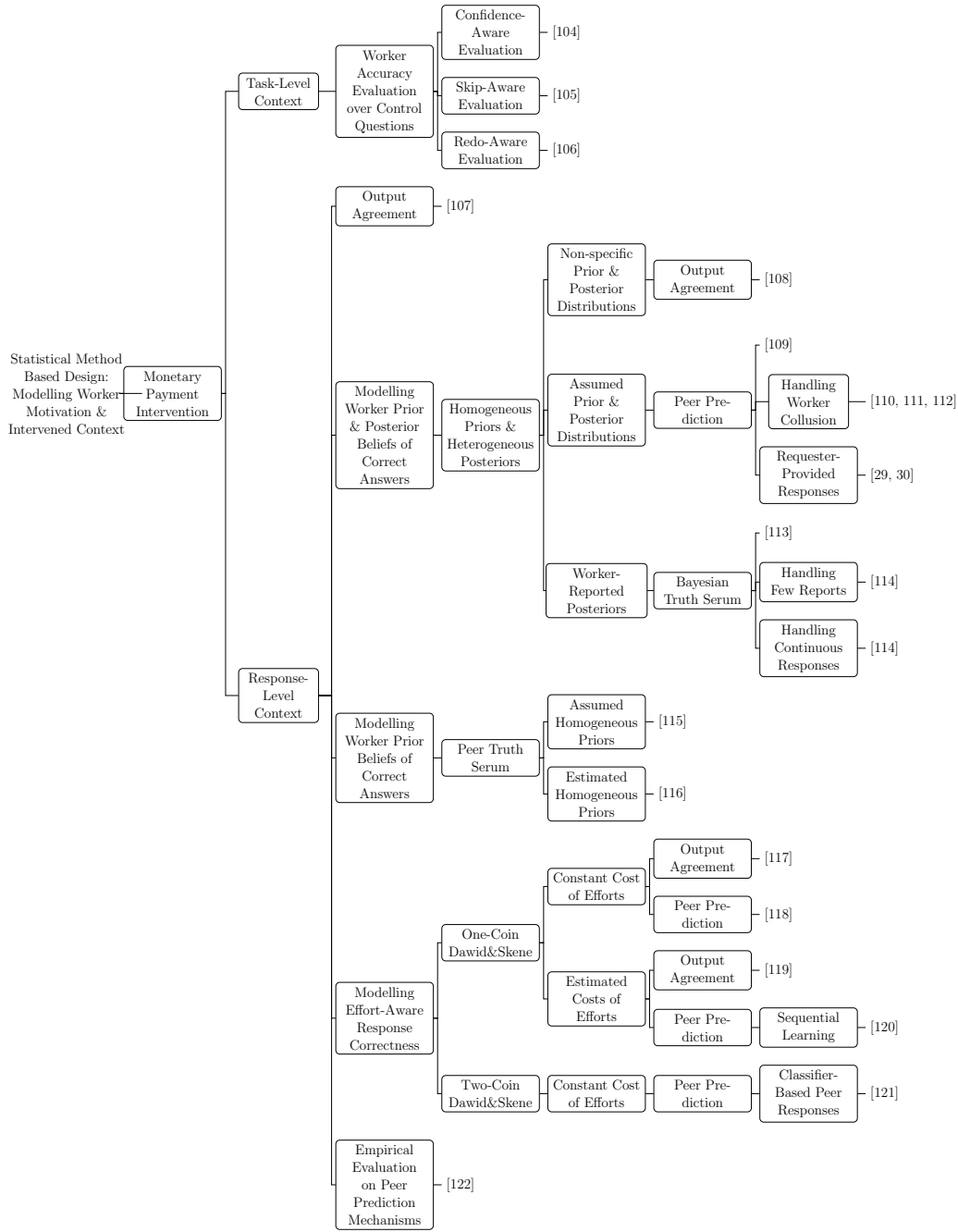

\subsection{Incentive Mechanisms Based on Monetary Payment}\label{chp2:monetary_payment_summary}
Monetary payments in crowdsourcing involve two types: \textit{base} payments and \textit{bonus} payments. The extrinsic motivation of rational workers is to choose answer options that maximize their monetary payoffs \cite{singer2013pricing,ho2015incentivizing,faltings2017game}. %With any uncertainty involved, the motivation is changed to maximize \textit{expected} payoffs. 
The payoffs are usually formulated as the difference between the monetary rewards given to the workers for the responses they provide and the costs incurred (the effort exerted) to generate these responses \cite{dasgupta2013crowdsourced,faltings2017game}. Maximizing the payoffs means minimizing the (costs of the) effort exerted. Consequently, one expects crowd-workers to minimize their effort, which generally leads to a deterioration in the quality of their responses. Moreover, some workers could even deliberately choose to not \textit{truthfully} report the responses that they elicited with effort (e.g. by flipping their responses) if they believe that doing so could result in higher payoffs \cite{radanovic2016incentives}. 

To address the above issues, specialized incentive mechanisms have been devised that alter either the fixed base payment or the fixed bonus payment, causing the payment to become \textit{adaptive} at the task level (or some lower levels of contexts). The aim of the adaptiveness is to ensure that the expected payoffs of the workers are maximized only when they exert sufficient effort to produce high-quality responses and report these responses truthfully.

\subsection{Monetary Payment in Task-Level Contexts}

In a task-level context, control questions are sometimes available and can be inserted randomly into each question page (such that crowd-workers do not know their whereabouts). Once the workers finish the task, the monetary payments to them can be adapted to be different (i.e. either the base or bonus payments) based on their accuracy on the control questions. The accuracy of the responses to those control questions serve as the inputs to some carefully designed \textit{payment function} which outputs a final reward for the worker and guarantees that this reward be maximized only when she has exerted sufficient effort and reported all her answers truthfully. %The design and the proof on the guarantee of such a payment function has first been investigated in \cite{lambert2009eliciting} for eliciting both categorical and ordinal true labels from expert workers who however might not truthfully report those labels. 

In~\cite{shah2015approval}, the task allows workers to express their confidence in different response options as the correct answer for each question. The confidence scores of a worker for all the control questions are then taken into account when estimating the worker's accuracy. The accuracy estimate is then used to calculate the payment to the worker. In~\cite{shah2015double}, the task additionally provides a ``skip'' option for each question and the final payment is decreased for each skipped control question and increased (multiplicatively) for each correctly answered one. In later work~\cite{shah2016no}, the authors proposed a two-stage design where a worker first answers all questions, and then is provided with the opportunity to change her answers after viewing a reference response from another worker to each of the same questions. At both stages, the payment function evaluated over the worker's responses to the control questions ensures her truthful reporting.
\subsection{Monetary Payment in Response-Level Contexts}\label{chp2:monetary_payment_final}
Most state-of-the-art incentive mechanisms that are based on monetary payments were developed under the assumption that the control questions are unavailable in crowdsourcing tasks or too scarce to be used reliably. Certain mechanisms were developed for making the payments adaptive to the response-level context. This means that a worker is not paid the same amount for every question and that different workers may be paid differently for their answers to the same question. The aim is to make the payments dependent on the quality of the response, where a response of higher quality deserves a higher base or bonus payment.

Since the quality of a response is unknown without control questions, the incentive mechanisms in this case resort to a strategy called \textit{peer consistency} to assess the quality of a worker's response. In this case, a response from another random worker, called a \textit{peer worker}, is selected for comparison with the target response. If these two responses are the same, then the target worker will be rewarded according to a payment function that is carefully designed to induce a \textit{game-theoretic equilibrium} among all the workers \cite{dasgupta2013crowdsourced, waggoner2014output, faltings2014incentives, radanovic2016incentives}. In such an equilibrium, no worker can improve their expected payoff by acting differently from what is required by the mechanism, namely that they truthfully report their answers to the data requester. Such an equilibrium is thus also referred to as a \textit{truthful} one. 

The game-theoretic incentive mechanisms usually model the belief systems of crowd-workers, which consist of their \textit{prior} and \textit{posterior} beliefs about the correct answers to questions. The belief systems are assumed by these mechanisms to be either \textit{homogeneous}, which means they are identical across all workers, or \textit{heterogeneous}, which means that different workers possess different prior and posterior beliefs as well as different ways of updating the beliefs.

Output agreement \cite{von2008designing} is the most basic peer consistency mechanism which does not assume any form of belief system (i.e. neither specific distributions over correct answers nor whether the distributions are shared) among workers. It only involves paying a worker for her response to a question when the response is the same as the one given by a randomly selected peer to the same question. Based on output agreement, Waggoner and Chen~\cite{waggoner2014output} assumed a homogeneous prior belief across workers (i.e. a shared non-specific prior distribution over correct answers) and heterogeneous posterior beliefs (i.e. private non-specific distributions) according to workers' individual understanding after reading the question. They defined a broader payment function by replacing the 0/1 error function in output agreement with the Euclidean distance between the response and the peer's response. They showed that output agreement based on this general payment scheme at best results in a strict equilibrium where workers report the correct answer according to the common part of their understanding.

Peer prediction \cite{miller2005eliciting} is another early work based on peer consistency assessment which assumes homogeneity for prior beliefs and heterogeneity for posterior beliefs with specific distributions over correct answers. It proposes to use the assumed posterior updated from observing a worker's response together with a random peer's response to the same question to calculate the reward for the worker. In the original paper, the authors consider the case in which true answers and responses are continuous and for which they assume the belief systems follow Normal distributions. %It is mainly applied to eliciting truthful subjective responses from workers.
In general, conjugate distributions are usually selected for the belief systems to facilitate belief updates. A drawback of the peer prediction approach is the existence of multiple undesired equilibria caused by the collusion among workers. The collusion allows them to exert no effort (e.g. by copying each others' answers to every question) and yet gives them higher expected payoffs than the truthful equilibrium does. Thus, subsequent work has focused on removing such equilibria \cite{jurca2007collusion,jurca2009mechanisms} or designing payment functions that penalize the ``collusion equilibria'' to make them have smaller payoffs than the truthful one \cite{kong2016putting}. However, the techniques still assume that true answers and responses are binary. 

Peer truth serum (PTS) \cite{faltings2014incentives} is an alternative when data requesters cannot find appropriate distributional assumptions for the posterior beliefs of workers. This is because PTS does not consider the posterior beliefs in the payment design. Instead, it assumes homogeneity for the prior belief of workers about correct answers for which the requesters need to provide specific distributions. PTS makes use of the assumed prior distributions and the 0/1 distance between the target worker's response and a random peer's response to the same question to calculate the target's reward. Such a payment function was shown by the authors to induce at least one ``non-truthful'' equilibrium where all workers collude with one another to always give the least likely responses to each question. Instead of using a predefined prior distribution, subsequent work~\cite{radanovic2016incentives} focused on dynamically estimating the prior using frequencies of responses from other workers to the same question to which the response of the target worker was given.

%The subsequent work has endeavoured to achieve an almost truth-telling equilibrium among large numbers of homogeneous workers based on a more realistic assumption that these workers' prior beliefs are private \cite{radanovic2015incentive}.

Bayesian truth serum (BTS) \cite{prelec2004bayesian} assumes homogeneous prior beliefs (i.e. a shared non-specific prior distribution) for crowd-workers. On the other hand, it implies that the posterior beliefs are heterogeneous by requiring additional assessments from workers about the probabilities over correct answers to each question along with their responses. BTS obtains the geometric mean of these probability estimates excluding the one from the target worker and combines it with the frequencies of collected responses to calculate the payment for the target. A weakness of BTS is that it needs a large number of workers to answer the same question in order to produce a reliable geometric mean to achieve a truthful equilibrium. Robust BTS \cite{parkes2012robust} was proposed which modified the payment function of BTS to handle the situation where only a few workers answer each question. Furthermore, a divergence-based BTS \cite{radanovic2014incentives} has been proposed to handle continuous responses (e.g. numbers). The payment function of this mechanism leverages the KL-divergence of the probability estimates over intervals that might contain the correct answer between the target worker and a random peer.

The work reviewed thus far focuses on modelling crowd-workers' beliefs or distributions on the correct answers of questions. This is different from non-incentive quality control models such as DS and GLAD which focuses on modelling the response correctness or biases given a global distribution over the correct answers. The former type of modelling emphasizes the elicitation of honest responses (i.e. workers exert efforts and truthfully report what they think to be the correct responses) which might turn out to be incorrect. The aggregation of these responses to obtain better final answers can come afterwards using the DS or GLAD models.

There are other incentive mechanisms which directly model response biases of workers (as the DS model does) for deriving payment functions that are able to achieve a truth-telling equilibrium. Unlike the DS model, they do not explicitly infer question true answers but rather aim at eliciting effort-exerted and honest responses. The first work in this regard was proposed by Dasgupta and Ghosh~\cite{dasgupta2013crowdsourced}, referred to as the \textit{DG} mechanism. They dealt with binary response options based on the one-coin DS model. They model the efforts exerted by workers as binary variables that control the switch between arbitrary guessing (i.e. zero-effort) and the workers' response correctness probabilities (i.e. effort-exerted) which were assumed to be always greater than 0.5. The payment function was designed to both recognize the response agreement between the target worker and a random peer for the same question, and penalize zero-effort (coincidence) agreement given both workers' response statistics calculated from the other questions. The authors proved that this payment function avoided a zero-effort equilibrium by making it always less appealing than a truthful equilibrium in terms of expected rewards over efforts and response correctness.

%Assuming a fixed known cost for all workers exerting effort, the payment function is designed to be able to form an \textit{effort-elicited} truthful equilibrium among the workers. In this equilibrium, investing effort to solve each question followed by truthful reporting of the answer, is the best action for the workers to maximize their expected payoffs per question with respect to their beliefs, (as well as reporting strategies and any randomness in the mechanism). 

%The above guarantee does not hold for the Dasgupta\&Ghosh (DG) mechanism when the number of response options is greater than 2. %The PTS mechanism for crowdsourcing \cite{radanovic2016incentives} is another heterogeneous mechanism, which estimates the parameters of a categorical distribution over workers' responses to questions (excluding the question being rewarded). The learned parameters, different for each worker, are then input to a payment function similar to the one used by the output agreement mechanism except that the payment amount is divided by the input parameter. The mechanism makes no assumption about the answer requester's knowledge of the belief systems of the workers. However, it requires large numbers of similar questions and a certain weak constraint on the belief update of each worker for achieving a truthful equilibrium.   

Based on the same one-coin model, Witkowski et al.~\cite{witkowski2013dwelling} additionally considers the scenario where a worker would make the decision on whether to participate in the crowdsourcing task. The probability of participation equals the worker's response correctness probability, which models the worker's self-assessment about their qualification. Correspondingly, the payment function is designed in such a way that unmotivated or unqualified workers will prefer to not participate rather than guess an answer (with zero effort), which also means that those who participate will be qualified and invest efforts in the equilibrium.

Both Dasgupta~\cite{dasgupta2013crowdsourced} and Witkowski et al.~\cite{witkowski2013dwelling} have assumed the cost induced by non-zero efforts is a constant. Within the same modelling framework, Liu and Chen~\cite{liu2016learning} proposed an extension which considers varying unknown costs randomly drawn from a distribution under non-zero efforts. Learning the cost distribution requires the workers to additionally report their costs of answering each question. The learning process is integrated with an incentivizing process, which aims to reach a truth-telling equilibrium, under a multi-armed bandit framework. The framework optimizes the trade-off between the two processes. Another hybrid mechanism that combines the DG mechanism with a multi-armed bandit framework to realize a similar goal was proposed in~\cite{liu2017sequential}. It learns the optimal choices of bonus levels at each time step for workers categorized into two peer groups that cross validate the truthfulness of each other's answer reporting behaviour. 

Based on a two-coin DS model that captures workers' biases in binary labelling, Liu and Chen~\cite{liu2017machine} leveraged binary labels generated by classification algorithms as the benchmark labels against which worker responses were compared for peer consistency assessment. The assessment results are then input to a payment function which guarantees that if the error rates of the classification algorithms on predicting the true labels converge towards zero, the function is able to achieve a truthful equilibrium.

The above mechanisms have more-or-less coped with undesirable equilibria, which yield higher expected payoffs than the truthful ones do, in their theoretical formulation of the payment functions. However, empirical evidence remains insufficient in the following two aspects \cite{faltings2014incentives,gao2014trick}. First, it is still unclear that whether the existence of these undesirable equilibria actually pose a problem to the quality of crowdsourced data in practice. Second, it remains to be seen whether these theoretically elegant truth elicitation mechanisms based on peer consistency assessment can work effectively in practice.

After any of the above incentive mechanisms is applied to the monetary payments for workers' responses, quality control methods can be further applied to the crowdsourced responses to produce more reliable estimates of the true answers. There are also unified frameworks proposed by Frongillo et al.~\cite{frongillo2015elicitation} and Ho et al.~\cite{ho2016eliciting} that integrate the above process. More specifically, Frongillo et al.~\cite{frongillo2015elicitation} combined a payment function that ensures a truth-telling equilibrium with the Bayesian statistics to allow for both truthful response elicitation and Bayesian aggregation for inferring the true answers. %Especially, the latter is done by applying the Bayes rule to the likelihood of the truthfully reported responses and the conjugate priors corresponding to the specific data type of the responses (e.g. categorical, or continuous responses). The parameters of the posterior distributions then quantify the confidence or experience of workers towards the true labels of items after seeing their respective responses.
%and are subject to the contexts in which they are situated.
%\subsection{Modelling correlations of QoA with Worker-Item-Context Interactions}
%labelling quality of workers.  In this case, modelling the latent motivation of workers allows the quality control methods to rectify the workers' behaviour (as in the example, careless behaviour) in \textit{real time} with little cost and as a result, improves the efficiency of those methods.
%To model the motivation, previous work has made explicit assumptions about how motivation of workers causes the labelling quality of workers. Then, they proposed proper \textit{incentivizing} mechanisms to intervene in the causation specified in the assumptions to improve the labelling quality. The incentivizing mechanisms are usually designed according to the characteristics of the motivation under the assumptions.
In~\cite{ho2016eliciting}, a further step was taken to optimize the multiple-choice interface with confidence shown to each worker (the interface being similar to the one employed in~\cite{shah2015approval}) along with the optimization of the Bayesian aggregation. 

\subsection{Incentive Mechanisms Based on Gamification}\label{chp2:gamification_summary}
Gamification refers to incorporating (video) game elements into various (levels of) contexts of crowdsourcing tasks in an attempt to lift the intrinsic motivations of workers to make them more engaged in answering questions. This eventually leads to improved quality of their responses. The intrinsic motivations are crowd-workers' feelings of enjoyment, playfulness, and accomplishment (e.g. through improvement of their own skills), and the welfare of the communities. Different from the payment-based incentive mechanisms discussed above which derive theoretical guarantees of expected performance of crowd-workers, the gamified incentive mechanisms have proven themselves empirically to have improved the quality of workers' responses.

\begin{figure}[h!]
\begin{tikzpicture}[scale=0.45, every node/.style={transform shape},grow'=right,level distance=2in,sibling distance=0.5in]
\tikzset{edge from parent/.style= 
            {thick, draw, edge from parent fork right},
         every tree node/.style=
            {minimum width=1in,text width=2in,align=center}}
\Large            
\Tree
[.\node[text width=2in]{Statistical Method Based Design:\\Modelling Worker Motivation \&\\Intervened Context}; [.\node[draw,rectangle,rounded corners,text width=1.5in]{Gamification}; [.\node[draw,rectangle,rounded corners,text width=1.5in]{Response-Level Context}; [.\node[draw,rectangle,rounded corners,text width=1.75in]{Score Reward Based on Accuracy Estimation}; [.\node[draw,rectangle,rounded corners,text width=1.5in]{Enjoyment Enhancement};  [.\node[,align=left]{ \cite{brenner2014people,dumitrache2013dr,harris2014beauty, he2014studying, eickhoff2012quality,saito2014skill,guillot2016towards,schlotterer2015game,moazzam2015scientific} };   ]  ] [.\node[draw,rectangle,rounded corners,text width=1.75in]{Self-Fulfilment: Knowledge \& Skill Advancement}; [.\node[draw,rectangle,rounded corners,text width=1.25in]{Quizzes};  [.\node[,align=left]{ \cite{ipeirotis2014quizz,stanculescu2016work,boyd2012besting} };   ] ]  [.\node[draw,rectangle,rounded corners,text width=1.25in]{Tutorials};  [.\node[,align=left]{ \cite{dontcheva2014combining} };   ] ]  ]   [.\node[draw,rectangle,rounded corners,text width=1.25in]{Altruism};  [.\node[,align=left]{ \cite{lee2013experiments} };   ] ]  ]  ]  [.\node[draw,rectangle,rounded corners,text width=1.5in]{Session-Level Context};  [.\node[draw,rectangle,rounded corners,text width=1.5in]{Badge Reward Based on Scores};  [.\node[draw,rectangle,rounded corners,text width=1.25in]{Enjoyment Enhancement};  [.\node[,align=left]{ \cite{guillot2016towards,mason2012tiger,itoko2014involving,massung2013using} };   ] ]  [.\node[draw,rectangle,rounded corners,text width=1.75in]{Self-Fulfilment: Knowledge \& Skill Advancement}; [.\node[draw,rectangle,rounded corners,text width=1.25in]{Quizzes};  [.\node[,align=left]{ \cite{stanculescu2016work} };   ] ] [.\node[draw,rectangle,rounded corners,text width=1.25in]{Tutorials};  [.\node[,align=left]{ \cite{dontcheva2014combining} };   ] ]  ]  [.\node[draw,rectangle,rounded corners,text width=1.25in]{Altruism};  [.\node[,align=left]{ \cite{lee2013experiments} };   ] ]  ] ] [.\node[draw,rectangle,rounded corners,text width=1.5in]{Task-Level Context}; [.\node[draw,rectangle,rounded corners,text width=1.5in]{Leaderboard Based on Scores}; [.\node[draw,rectangle,rounded corners,text width=1.5in]{Competition}; [.\node[,align=left]{ \cite{eickhoff2012quality,saito2014skill,itoko2014involving,lee2013experiments,ipeirotis2014quizz,massung2013using} };   ]  [.\node[draw,rectangle,rounded corners,text width=1.5in]{Monetary Bonus Reward};  [.\node[,align=left]{ \cite{yuan2016leaderboard} };   ] ]  ]  ]  ] ] ] ]
\end{tikzpicture}
\caption{A taxonomy of QCC papers that considered the interaction between worker context and (intrinsic) motivation. They focused on designing gamification mechanisms which, in most cases, rely on statistical estimation of worker ability.}
\label{}
\end{figure}
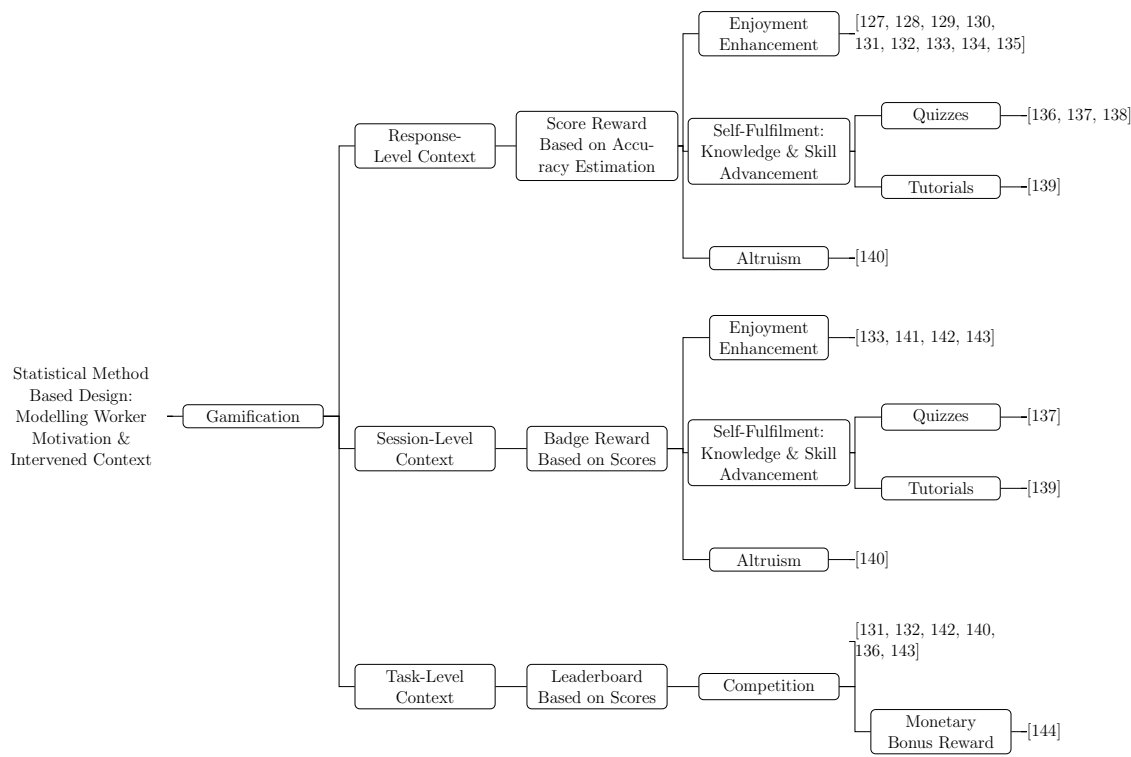

\subsection{Gamification in Response-Level Contexts}
\textit{Score feedback} (aka ``scoring'') is the most fundamental means of gamification that allocates a certain number of points to a worker depending on the quality of her response. Higher-quality responses should be rewarded with more points. This makes the scoring very much similar to dynamic monetary payments in terms of how they are allocated. They are however different with regard to the types of motivation they deal with. The former provides virtual rewards for increasing intrinsic motivations as opposed to material rewards provided by the latter for increasing extrinsic motivations. In addition, scoring often acts as the building block of other more complicated game elements such as leaderboards and level systems. 

In~\cite{brenner2014people}, a simple and fun online game was developed to help train a face recognition system to refine its classification. In the game, a crowd-worker was rewarded with certain points every time she provided correct feedback regarding an uncertain recognition result from the system. The experiments showed that the game, without paying any bonus, iteratively improved the recognition performance of the system solely based on the enjoyment/fulfilment it brought the workers. The incentivizing aspects of the above game design, including the graphics and the scoring mechanism, made the game enjoyable for workers to engage in, causing them to provide more accurate answers in general. Thus, these design aspects have appeared repeatedly in gamified crowdsourcing over different areas including medical facts elicitation \cite{dumitrache2013dr}, relevance judgement in Information Retrieval \cite{harris2014beauty, he2014studying, eickhoff2012quality}, image classification \cite{eickhoff2012quality}, video captioning \cite{saito2014skill}, language translation \cite{guillot2016towards}, and communication to the general public about culture \cite{schlotterer2015game} and science \cite{moazzam2015scientific}. 

Apart from the enjoyment and increased productivity it has brought to crowdsourcing, the scoring mechanism can also help to build more sophisticated incentive mechanisms that advance the skills and knowledge of crowd-workers. Such advancement allows the workers to produce responses of higher quality. A typical skill-development mechanism that has been boosted by the scoring mechanism from gamified crowdsourcing is the online quiz. In~\cite{ipeirotis2014quizz}, online quizzes were advertised as skill/knowledge tests for individuals in order to attract both unpaid volunteers and crowd-workers. These quizzes contain not only control questions but also target questions for which the requesters were seeking correct answers from the participants. The scoring mechanism in this case supported both the performance feedback mechanisms, which display each individual's score and the others' average score, and the all-time leader-boards, which rank the participants by their scores.
%A major issue for the quizzes is to motivate especially the large number of online volunteers to stay and engage. Therefore, several other gamification elements have been added into the quizzes including immediate feedback to an individual in terms of her performance and others' performance, individual score display, and all-time leaderboards. 
The experiment results show that such quizzes can attract a large number of participants with diverse skills over a relatively short period, and the total payment is much lower than what would have been required by AMT. 

The main idea of the above work to use gamified quizzes to attract participation of (and contributions from) workers or volunteers who seek enjoyable learning experiences, has also been adopted in~\cite{stanculescu2016work} and~\cite{boyd2012besting}. The former work leveraged the idea for engaging employees in learning about enterprise history, products and services while crowdsourcing some subjective data from them (e.g. their opinions). The latter leveraged the intrinsic fun of quiz bowl~\cite{jennings2007brainiac} for engaging online players to provide answers to questions as the labels to be used in training classifiers that can perform better automatic question-answering. 

Apart from quiz testing, tutorial training/learning is another means of stimulating workers' intrinsic needs for knowledge and skills, and has been seamlessly combined with the scoring mechanism. In~\cite{dontcheva2014combining}, a gamified crowdsourcing platform for image editing was developed which attracted large numbers of workers as they could learn skills for producing high-quality and creative images. The basic game element employed by the platform was scoring, which again also supported an all-time leaderboard. Worker satisfaction surveys were collected and showed that most of the workers appreciated the sense of achievement created by the scores when learning the image editing skills. Moreover, feedback from the requesters showed that the number of images with better quality was almost double compared to those produced by the originally novice workers.

Scoring mechanism can also help incentivize workers to make altruistic contributions to their communities. In~\cite{lee2013experiments}, the focus is on motivating workers to contribute to the construction of a new online community. The scoring mechanism in this case quantifies the amount of contribution a worker has made, and supports more advanced game elements including a badge system and an all-time leaderboard.
%, and it supports a more badge mechanism that rewards the workers based on their scores and shows them how many more scores they need to have before awarded their next badges, and an all-time contribution leaderboard that ranks the workers by their scores. 
%Moreover, there are a mission mechanism and a challenge mechanism which work closely with the other mechanisms. The mission mechanism issues missions for workers to fulfil as sessions of tutorial training. On completing each mission, a worker is awarded with some scores for her to proceed to a higher (skill) level at which she will start on a set of more advanced training missions. The challenge mechanism lists all the image editing tasks from the requesters that match a worker's current skill level. She can apply and test what she has learned by selecting and solving any of these editing tasks in which she feels confident and interested.
\subsection{Gamification in Session-Level Contexts}
\textit{Badges} are typical game elements that are awarded to people for recognizing their achievements and contributions at different levels (usually with bronze, silver and gold badges corresponding to the increasing levels). Many crowdsourcing marketplaces (e.g. CrowdFlower) have implemented their own badge systems that award workers within task-level contexts according to the numbers of tasks they have successfully completed. In a gamified task, badge awarding usually happens in session-level contexts. More specifically, when a worker completes a session/page of questions, she gains some points and whenever her total points exceed a certain threshold, a badge system is triggered to award her with the corresponding badge. Such a badge system has been integrated into the session-level contexts for various motivational purposes. They range from making laborious and tedious work (such as image annotation \cite{mason2012tiger}, proofreading \cite{itoko2014involving}, language translation \cite{guillot2016towards} and mobile application testing \cite{massung2013using}) more enjoyable, confirming one's learning progress, (e.g. on image editing \cite{dontcheva2014combining} and enterprise knowledge \cite{stanculescu2016work}), to encouraging workers' commitment to building online communities \cite{lee2013experiments}.
%Missions and challenges are two game elements

\subsection{Gamification in Task-Level Contexts}\label{chp2:gamification_final}

The most notable game element that has been utilized for gamifying task-level contexts in paid crowdsourcing is the \textit{leaderboard}. Typically, a leaderboard exists throughout the entire duration of the task and is accessible by all the workers at any time in the task. The aim is to ignite \textit{competition} amongst the workers, which motivates them to work harder to either overtake those above them in the ranking or to maintain their current rank positions. However, the past research on using all-time leaderboards to incentivize workers has yielded conflicting empirical results. In~\cite{eickhoff2012quality}, steady improvements were observed in the quality of workers' relevance judgements for documents to search queries. Quality was measured in terms of the level of agreement between workers on the same document-query pairs. In~\cite{itoko2014involving,saito2014skill}, workers were required to do proofreading. Senior workers were \textit{demotivated} by the competition brought about by the presence of a leaderboard while younger workers found it the other way round. In~\cite{lee2013experiments}, workers constantly returned to the communities as they would like to follow their status on the leaderboard, and were encouraged by doing so to make more contributions, although their quality varied significantly. In~\cite{ipeirotis2014quizz}, an all-time leaderboard was set up which provided two types of ranking: the percentage of correct
answers and the total number of correct answers submitted. The leaderboard in both cases showed positive effects on the quality of workers' answers only in the early stages of the tasks as it \textit{discouraged} the workers who came late to the tasks when other workers had already amassed a large number of points and had well-established positions on the leaderboard. A similar phenomenon has also been observed in~\cite{massung2013using,yuan2016leaderboard}. In~\cite{massung2013using}, new workers collected pro-environmental behaviour data for a mobile application. Performance was initially high before dropping for later arriving workers due to the large difference in the contribution points between the leading workers and themselves. In~\cite{yuan2016leaderboard}, workers were exposed to a task leaderboard when performing relevance judgment~\cite{lease2011overview}. The authors found that providing such a leaderboard could significantly improve the quality of responses (compared to no leaderboard) only when the workers were informed of a top-10 bonus reward with one dollar per person. They also observed that considering the number of responses made by each worker thus far in the rank calculation resulted in lower accuracy across workers (than considering the percentage of correct responses). They argued that this indicated a discouraging factor of the task leaderboard. 

To deal with the above issue, Ipeirotis and Gabrilovich~\cite{ipeirotis2014quizz} suggested the leaderboard be embedded in session-level contexts which means that there is a leaderboard dedicated to each page of a task. In this case, workers need to answer correctly much fewer questions to reach the top of a page leaderboard. As a result, workers who arrive late at a task page are less likely to be intimidated by the (page) leaderboard rankings. In~\cite{lee2013experiments}, the experiment results suggested that the leaderboard should only be ``switched on'' after a certain ``warming-up'' period for each worker, by which time she will have completed enough questions to make herself feel less disadvantaged by the late starting point.

\section{Modelling Worker Expertise and Contexts}~\label{chp2:expertise_context}
\begin{figure*}[h]
  	\scalebox{.8} 		{\begin{tikzpicture}
\node[align=center] (im) {Training\\Mechanisms};
\node[draw, diamond, right=of im] (ic) {Control};
\node[draw,right=of ic,rectangle,rounded corners] (c) {Context};
\node[draw,align=center,right=of c,rectangle,rounded corners] (efi) {Worker\\Expertise};
\node[draw,align=center,right=of efi,rectangle,rounded corners] (ei) {Worker\\Ability};
  \node[latent,align=center,right=of ei,rectangle,rounded corners] (qij) {~~~Response~~~\\~~~Correctness~~~};

%\node[draw,above=of qij,rectangle,rounded corners, xshift=1.5cm] (dj) {Question Difficulty};
%\edge{ei,dj}{qij};
\edge{ei}{qij};
\edge{efi}{ei};
\edge{c}{efi};
\edge{ic}{c};
\edge{im}{ic};

\end{tikzpicture}}
    %\caption{Motivation-Difficulty-Context Interaction}
    %\begin{subfigure}[b]{0.5\columnwidth}
  %\centering
  	%\scalebox{.8} 		{\input{models/ability_context_diagram}}
    %\caption{Expertise-Difficulty-Context Interaction}
    %\label{fig:chp1_ability_context_diagram}
  %\end{subfigure}%
  \caption{A diagram shows training mechanisms control worker contexts to improve worker expertise, which further improves the correctness their responses. %Figure~\ref{fig:chp1_ability_context_diagram} shows the \textit{expertise-difficulty-context} interaction. In this case, worker contexts can be intervened by training mechanisms to improve worker expertise and further response correctness.
  }
  \label{fig:chp1_motive_context_diagram}
\end{figure*}
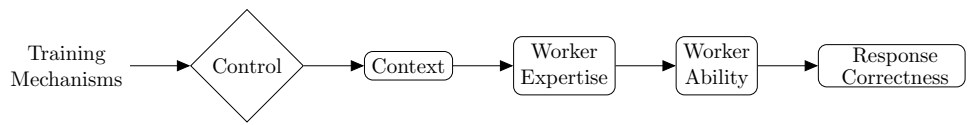
Not only can the motivation of workers be affected by intervened contexts but also their expertise can interact with the contexts in different ways. According to the QCC literature, we have found the following two types of mechanisms in which the interaction takes place:
\begin{itemize}
\item Worker expertise is improved by \textit{training mechanisms} deployed at different levels of contexts.
\item Worker expertise is leveraged by \textit{question assignment mechanisms} to control the allocation of questions into different levels of worker contexts.
%\item Expertise of a worker varies across different contexts in which this worker has been situated. For example, the worker's expertise is dynamic across different tasks she has participated, or across different pages of the same task. It is often assumed that workers exhibit similar (levels of) expertise towards similar tasks or sessions.
\end{itemize}

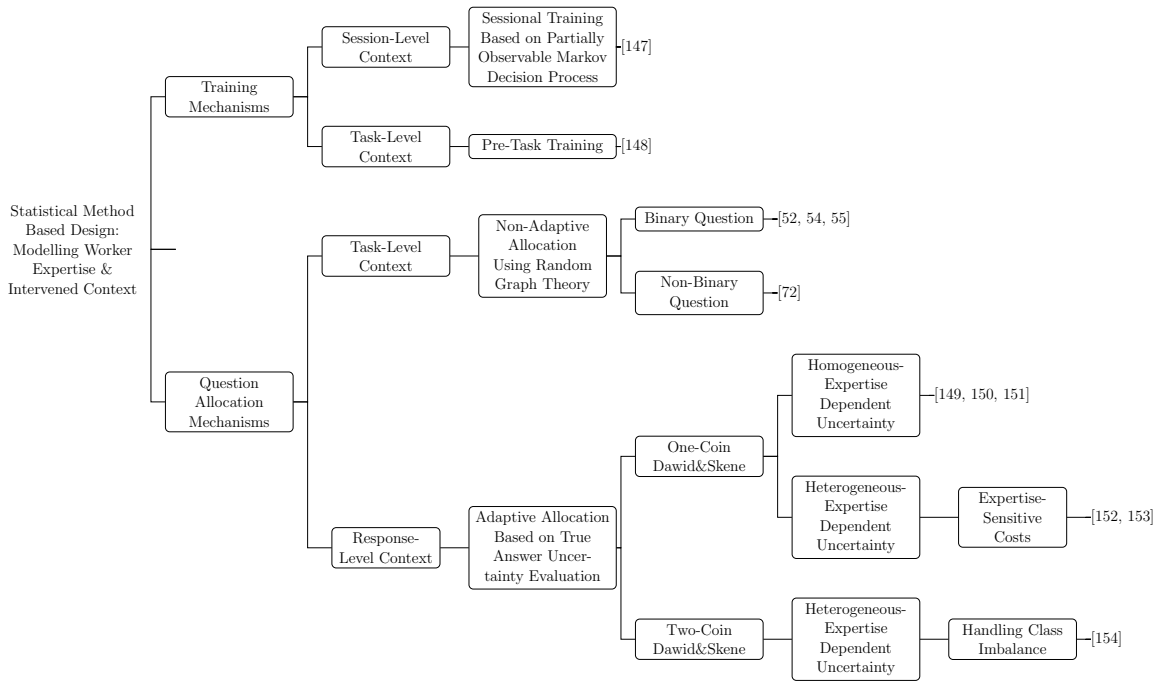
\begin{figure}[h!]
\begin{tikzpicture}[scale=0.41, every node/.style={transform shape},grow'=right,level distance=2in,sibling distance=0.5in]
\tikzset{edge from parent/.style= 
            {thick, draw, edge from parent fork right},
         every tree node/.style=
            {minimum width=1in,text width=2in,align=center}}
\Large            
\Tree
[.\node[text width=2.5in]{Statistical Method Based Design:\\Modelling Worker Expertise \&\\Intervened Context}; [.\node[draw,rectangle,rounded corners,text width=1.5in]{Training Mechanisms}; [.\node[draw,rectangle,rounded corners,text width=1.5in]{Session-Level Context}; [.\node[draw,rectangle,rounded corners,text width=1.75in]{Sessional Training Based on Partially Observable Markov Decision Process}; [.\node[,align=left]{ \cite{bragglearning} };  ] ] ] [.\node[draw,rectangle,rounded corners,text width=1.5in]{Task-Level Context}; [.\node[draw,rectangle,rounded corners,text width=1.75in]{Pre-Task Training}; [.\node[,align=left]{ \cite{gadiraju2015training} };  ] ] ] ] [.\node[draw,rectangle,rounded corners,text width=1.5in]{Question Allocation Mechanisms}; [.\node[draw,rectangle,rounded corners,text width=1.5in]{Task-Level Context}; [.\node[draw,rectangle,rounded corners,text width=1.5in]{Non-Adaptive Allocation Using Random Graph Theory}; [.\node[draw,rectangle,rounded corners,text width=1.5in]{Binary Question}; [.\node[,align=left]{ \cite{karger2011budget,karger2011iterative,karger2014budget} };  ] ] [.\node[draw,rectangle,rounded corners,text width=1.5in]{Non-Binary Question}; [.\node[,align=left]{ \cite{karger2013efficient} };  ] ] ] ]  [.\node[draw,rectangle,rounded corners,text width=1.25in]{Response-Level Context};  [.\node[draw,rectangle,rounded corners,text width=1.75in]{Adaptive Allocation Based on True Answer Uncertainty Evaluation}; [.\node[draw,rectangle,rounded corners,text width=1.5in]{One-Coin Dawid\&Skene}; [.\node[draw,rectangle,rounded corners,text width=1.5in]{Homogeneous-Expertise Dependent Uncertainty}; [.\node[,align=left]{ \cite{sheng2008get,ipeirotis2014repeated,yan2011active} };  ] ] [.\node[draw,rectangle,rounded corners,text width=1.5in]{Heterogeneous-Expertise Dependent Uncertainty}; [.\node[draw,rectangle,rounded corners,text width=1.25in]{Expertise-Sensitive Costs}; [.\node[,align=left]{ \cite{wang2013framework,wang2017cost} };  ] ] ] ]  [.\node[draw,rectangle,rounded corners,text width=1.5in]{Two-Coin Dawid\&Skene}; [.\node[draw,rectangle,rounded corners,text width=1.5in]{Heterogeneous-Expertise Dependent Uncertainty}; [.\node[draw,rectangle,rounded corners,text width=1.5in]{Handling Class Imbalance}; [.\node[,align=left]{ \cite{zhang2015active} };  ] ]  ] ]  ] ] ] ]
\end{tikzpicture}
\caption{A taxonomy of QCC papers that considered the interaction between worker expertise and context. These papers focus on designing either training mechanisms that alter contexts to improve worker expertise or question allocation mechanisms that use worker expertise to determine questions to be answered in the contexts.}
\end{figure}

\subsection{Improving Worker Expertise Using Training Mechanisms at Different Levels of Contexts}\label{chp2:worker_expertise_training}
\textit{Training mechanisms} intervene in the context to directly affect the \textit{expertise} of workers. They aim at improving workers' expertise regardless of their motivation. By default, the training of crowd-workers is performed prior to their participation in a task and aims to teach the workers basic skills and expertise required to answer the questions in the task. It has been shown in~\cite{gadiraju2015training} that the default training can significantly improve the quality of worker responses on a variety of crowdsourcing tasks. The mechanism is restricted however to a task-level context in which each worker only receives the training once throughout the entire task. As a result, even when the later performance of the workers is undesirable, they are not given a chance to be retrained to perform better. To solve this issue, Bragg et al.~\cite{bragglearning} proposed a training mechanism functioning in the session-level contexts. The mechanism models the decision-making process of whether to train a worker or collect responses from her in each of her working sessions as a partially
observable Markov decision process (POMDP)~\cite{kaelbling1998planning}. It then employs reinforcement learning to estimate the parameters of the POMDP including the expertise vectors of individual workers and the correct answers to the questions. 

\subsection{Question Allocation in Crowdsourcing}\label{chp2:question_allocation_summary}
Reducing the number of responses collected using crowdsourcing to lower costs while maintaining the prediction accuracy on question true answers has been an important QCC research topic over the years. A common crowdsourcing process involves assigning each worker a set of questions which is selected uniformly at random according to a value set up prior to the task by the data requester for the number of questions answered per worker. Each worker answers the set of assigned questions only once. Unfortunately, such a process often leads to a higher total (monetary) cost than necessary. This is because the uniformly random assignment of the questions is independent of all the informative characteristics of the workers (e.g. their domain expertise, interests, etc.) and the questions (e.g. their domain difficulty, genres, etc.), and thus fails to leverage these characteristics for more efficient performance. On the other hand, if the question assignment can be designed to be biased towards these characteristics, then the question's true answer prediction can potentially be improved with lower costs. 

\subsection{Non-Adaptive Question Allocation Based on Worker Expertise in Task-Level Contexts}
In this case, the allocation of questions happens before any worker enters the task. The total number of allocated questions equals the batch size multiplied by the number of workers (if each worker is assigned questions only once and never reused once they finish their batches). Once the task begins, workers arrive in sequence to pick up the corresponding allocated batches. Such pre-task simultaneous allocation of questions relies on designing a bipartite graph which contains two types of nodes: questions and workers, where edges between them correspond to the assignment of a question to a worker. In~\cite{karger2011budget,karger2011iterative,karger2014budget}, the authors proposed to draw a regular random bipartite graph based on the \textit{configuration model} from the random graph theory~\cite{bollobas2001random}. In the graph, the degrees of the question and the worker nodes represent how many workers to assign to each question and how many questions to assign to each worker respectively. The goal of their work is to realize a particular error rate on true answer prediction with minimum costs (i.e. minimum degree for the question nodes or equivalently, minimum number of responses\footnote{The minimum number of responses equals the minimum degree for the question nodes multiplied by the number of questions.}). The authors proved that using a regular random graph to achieve a target error rate was sufficient. This graph's actual error rate was within a constant factor of the target rate using the underlying graph (which is possibly neither regular nor random) with the best possible inference algorithm. The authors also showed that the cost incurred by each binary question to achieve a target error is the error value scaled by the inverse of the expectation of each worker's expertise. In their following work~\cite{karger2013efficient}, the authors investigated the same subjects but with non-binary questions. They derived similar results in terms of the near-optimality of the regular random graph in achieving any target error rate and the scaling effect of expertise on the cost per question.   

\subsection{Adaptive Question Allocation Based on Worker Expertise in Response-Level Contexts}~\label{chp2:question_allocation_final}
For the adaptive schemes, the question allocation happens within an ongoing task and is dependent on the current estimate of each worker's expertise based on their responses so far. Sheng et al.~\cite{sheng2008get} and Ipeirotis et al.~\cite{ipeirotis2014repeated} proposed to model one key aspect in the adaptive allocation, that is the \textit{uncertainty} of each question's true answer. In their work based on the one-coin DS model for binary questions, the uncertainty of a question depends on the expertise/ability of the workers who answered it. The higher the expertise, the lower the uncertainty will become. The authors simplified the scenario by assuming that all the workers shared the same level of expertise and were non-adversarial (i.e. their response correctness probability always greater than 0.5). They proposed a design in which at each timepoint, the question with the largest amount of uncertainty in its correct answer will be assigned to an arbitrary worker. As a result, the same question might be assigned to multiple workers. In this case, the expertise of individual workers governs the quality of their responses from which an integrated response can be derived for the question. The lower the expertise of each worker, the more responses are needed to generate an integrated response for the question. In their following work \cite{wang2013framework,wang2017cost}, the authors addressed heterogeneous worker expertise. In this case, the entropy of the probability estimates from Eq.~\ref{eqn:correct_ans} embodies the uncertainty about the true answer. The question with the highest entropy was selected for assignment at each time. The payment for each worker is proportional to their current expertise estimates. The lower the expertise, the less payment a worker will receive for a response.

In~\cite{yan2011active}, the authors modelled the uncertainty of the correct response $l_j$ for question $j$ as the squared Euclidean distance between 0.5, denoting a random response, and the correct response prediction from a binary classifier. The classifier is based on a logistic function which has global coefficients including an intercept term, and receives inputs which are the observed features of the question (i.e. $\boldsymbol{x}_j$ in Eq. \ref{eqn:convex_logistic_fun}). At each time point, a question with the greatest uncertainty (i.e. the minimum squared Euclidean distance) is selected for assignment. Each worker is also represented by a logistic function with worker-specific coefficients (i.e. $\boldsymbol{w}_i$ in Eq. \ref{eqn:convex_logistic_fun}). The selected question is then assigned to worker $i$ who is able to maximize the probability of seeing the response $r_{ij}$:
\begin{eqnarray}
P(r_{ij}|\boldsymbol{w}_i, \boldsymbol{x}_j) =\delta(\boldsymbol{w}_i, \boldsymbol{x}_j)^{ \mathbbm{1}\{r_{ij} = l_j\}}\big(1 - \delta(\boldsymbol{w}_i, \boldsymbol{x}_j)\big)^{ \mathbbm{1}\{r_{ij} \neq l_j\}}
\label{eqn:select_worker}
\end{eqnarray} where $\delta(\boldsymbol{w}_i, \boldsymbol{x}_j)=\delta_{ij}$ defined by Eq. \ref{eqn:convex_logistic_fun}.

In~\cite{zhang2015active}, the question assignment strategies were further extended to be based on two-coin models which encode biases of individual crowd-workers towards positive and negative responses. They also devised an adaptive decision boundary for determining the true answer of each question and further, degrees of the uncertainty when class imbalance exists in the true answers. In~\cite{donmez2010probabilistic}, a hidden Markov model was proposed to capture the correlation in the time-varying ability of each crowd-worker. At each time step, the workers were ranked based on estimates of their current abilities and only the top worker was assigned the question to answer. Compared to the above work, this work fails to utilize all the crowdsourcing power available.

%In~\cite{khetan2016achieving}, the authors drew inspiration from their previous work~\cite{karger2011budget,karger2011iterative,karger2014budget} which modelled question allocation as a regular random bipartite graph. Rather than construct such a graph before any worker arrives in the task, the authors in this work constructed the graphs over multiple rounds during the task. In each round, the allocation mechanism optimizes a graph for achieving a target accuracy (on true answer prediction) with respect to the degree of the question nodes. Any question with the prediction accuracy higher than the target one has its predicted true answer accepted while the rest of the questions proceed to the next round. The total number of edges in the graph represents the fraction of the remaining budget spent at the particular round. Rounds stop when all the questions have their predicted true answers accepted or the budget is exhausted. The average estimated difficulty of all the remaining questions before a particular round controls the target accuracy to be achieved at that round. Finally, the authors proved that using regular random graphs with the proposed adaptive mechanisms guarantees near-optimality in the rate of convergence to the target accuracy.

\section{Modelling Worker Motivation, Question Difficulty and Contexts}\label{chp2:worker_motive_diff_context}
In Section \ref{sec:worker_motive_context}, the response quality is modelled to be dependent on the worker motivation under different worker contexts. The \textit{heterogeneity} in question difficulty was ignored in order to simplify both the derivation of theoretical equilibrium guarantees and the practical design of the gamification techniques. When heterogeneity in question difficulty is considered, the above problems become more complicated since the response quality becomes harder to measure and estimate. 

According to the literature, both the payment-based and the gamification-based incentive mechanisms have successfully controlled the response quality by considering the question difficulty. In this case, the payment-based mechanisms are typically applied to the response-level context while the gamification-based mechanisms are applied to the session-level context.

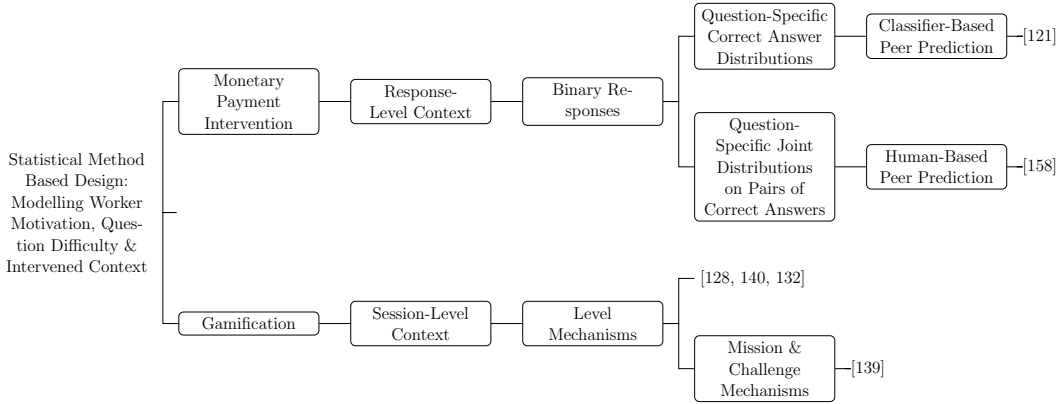
\begin{figure}[h!]
\begin{tikzpicture}[scale=0.45, every node/.style={transform shape},grow'=right,level distance=2in,sibling distance=0.5in]
\tikzset{edge from parent/.style= 
            {thick, draw, edge from parent fork right},
         every tree node/.style=
            {minimum width=1in,text width=2in,align=center}}
\Large            
\Tree
[.\node[text width=2.18in]{Statistical Method Based Design:\\Modelling Worker Motivation, Question Difficulty \&\\Intervened Context}; [.\node[draw,rectangle,rounded corners,text width=1.5in]{Monetary Payment Intervention}; [.\node[draw,rectangle,rounded corners,text width=1.5in]{Response-Level Context}; [.\node[draw,rectangle,rounded corners,text width=1.5in]{Binary Responses}; [.\node[draw,rectangle,rounded corners,text width=1.5in]{Question-Specific Correct Answer Distributions};  [.\node[draw,rectangle,rounded corners,text width=1.5in]{Classifier-Based Peer Prediction}; [.\node[,align=left]{ \cite{liu2017machine} };   ] ] ] [.\node[draw,rectangle,rounded corners,text width=1.5in]{Question-Specific Joint Distributions on Pairs of Correct Answers};  [.\node[draw,rectangle,rounded corners,text width=1.5in]{Human-Based Peer Prediction}; [.\node[,align=left]{ \cite{mandal2018peer} };   ] ] ]  ] ]  ] [.\node[draw,rectangle,rounded corners,text width=1.5in]{Gamification}; [.\node[draw,rectangle,rounded corners,text width=1.5in]{Session-Level Context}; [.\node[draw,rectangle,rounded corners,text width=1.5in]{Level\\Mechanisms}; [.\node[text width=1.5in,align=left]{ \cite{dumitrache2013dr,lee2013experiments,saito2014skill} };   ] [.\node[draw,rectangle,rounded corners,text width=1.5in]{Mission \& Challenge Mechanisms}; [.\node[,align=left]{ \cite{dontcheva2014combining} };   ] ] ] ] ]  ] ]
\end{tikzpicture}
\caption{A taxonomy of QCC papers that considered worker motivation, context and question difficulty. These papers focused on designing either monetary payment mechanisms which additionally modelled question difficulty or gamification mechanisms which increase question difficulty to challenge workers. }
\end{figure}

\subsection{Monetary Payment in Response-Level Contexts}
Only recently have payment-based incentive mechanisms started to consider the heterogeneity in question difficulty to refine their design of the response-level payment functions. They do this by increasing the payment for responses whose quality is low due to the fact that the difficulty of the questions is high rather than a lack of effort from the workers.

In the machine-aided peer prediction mechanism proposed in~\cite{liu2017machine}, the difficulty of a binary-response question is encoded by a dedicated probability distribution over its correct answers. If this distribution is (nearly) uniform a priori or a posteriori, it means the question is so difficult that its correct answer remains uncertain. In this case, the payment function was designed to achieve a truth-telling equilibrium for each question with respect to their correct answer distributions. 

In~\cite{mandal2018peer}, the proposed peer prediction mechanism was also applied to binary-response questions except that it relies solely on \textit{human} peer consistency assessment. The difficulty of a question is captured by a symmetric matrix of joint probabilities of each pair of response options (including each option with itself). Each entry represents the chance that any random pair of workers agree or disagree with one another on the correct answer for the question. The larger the summation of the diagonal entries is, the easier the question and it is the opposite for the off-diagonal entries. The payment function takes in a matrix of posterior joint probabilities given responses thus far to a question and rewards a random pair of workers according to the joint probabilities indexed by their responses to the question. The paper provided synthetic experiment results suggesting that by considering heterogeneity in question difficulty, the proposed mechanism achieved improved incentives for workers to be truthful and was less sensitive to their collusions compared to the previous mechanisms.

\subsection{Gamification in Session-Level Contexts}
The level mechanism is the most common game element that leverages differences in the difficulty of the questions for motivating the crowd-workers. The mechanism sets up different difficulty levels for the questions so that the workers progress from the easiest level to the hardest level to finish a task. In this case, proceeding to a higher level that contains more difficult questions requires the workers to exert more effort and show higher levels of expertise. In~\cite{dumitrache2013dr}, in addition to the scoring mechanism, the level mechanism controls the timing of when to change the difficulty levels of the medical documents used for fact elicitation for each worker according to the estimate of their current expertise. The mechanism was appreciated by the workers with 50\% of them praising the level progression. 

The level mechanism has played a similar role in~\cite{lee2013experiments,saito2014skill} where higher worker scores trigger higher difficulty levels in the game for the workers to play. In~\cite{dontcheva2014combining}, variants of the level mechanism were proposed, namely the \textit{mission} mechanism and the \textit{challenge} mechanism, to inspire crowd-workers to learn and develop new image editing skills and meanwhile complete the editing tasks posted by the requesters, which requires utilizing the skills they have learned. The mission mechanism issues increasingly difficult sets of questions packaged in the form of increasingly advanced training sessions for workers to improve their skills. Once a worker has successfully completed a session, she can proceed to a more sophisticated one. The challenge mechanism lists all of the image editing tasks from the requesters with difficulty levels matching the current skill level of the worker. 

\section{Modelling Worker Ability, Question Difficulty and Contexts}~\label{chp2:worker_expertise_diff_context}
A crowd-worker's ability can vary across different contexts in which this worker has been situated. For example, the ability can be dynamic across different tasks the worker has participated, or across different pages of the same task. There are statistical models that have considered the interactions among the worker ability, question difficulty and worker context. These models incorporate latent factors that represent different levels of contexts along with the worker ability and question difficulty factors.
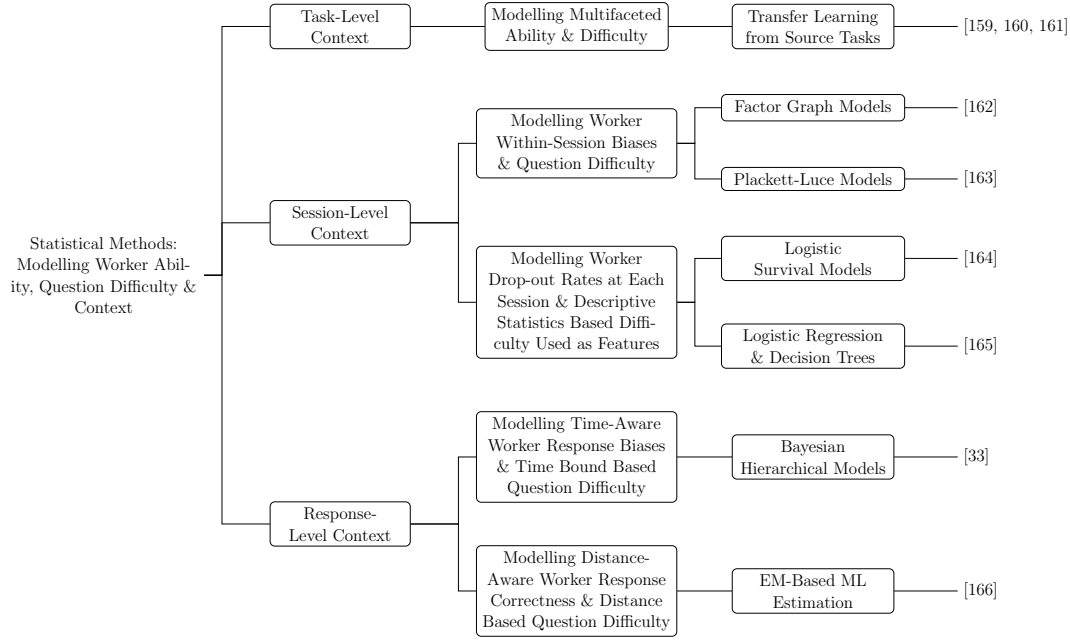
\begin{figure}[h!]
\begin{tikzpicture}[scale=0.45, every node/.style={transform shape},grow'=right,level distance=2.75in,sibling distance=0.5in]
\tikzset{edge from parent/.style= 
            {thick, draw, edge from parent fork right},
         every tree node/.style=
            {minimum width=1in,text width=2in,align=center}}
\Large            
\Tree
[.\node[text width=2.2in]{Statistical Methods:\\Modelling Worker Ability, Question Difficulty \&\\Context};  [.\node[draw,rectangle,rounded corners,text width=1.5in]{Task-Level Context};  [.\node[draw,rectangle,rounded corners,text width=2in]{Modelling Multifaceted Ability \& Difficulty};  [.\node[draw,rectangle,rounded corners,text width=1.75in]{Transfer Learning from Source Tasks};  [.\node[,align=left]{ \cite{mo2013cross,fang2013knowledge,fang2014active} };   ]  ]  ]  ] [.\node[draw,rectangle,rounded corners,text width=1.5in]{Session-Level Context}; [.\node[draw,rectangle,rounded corners,text width=2.2in]{Modelling Worker Within-Session Biases \& Question Difficulty}; [.\node[draw,rectangle,rounded corners,text width=2in]{Factor Graph Models}; [.\node[,align=left]{ \cite{zhuang2015leveraging} };   ] ]  [.\node[draw,rectangle,rounded corners,text width=2in]{Plackett-Luce Models}; [.\node[,align=left]{ \cite{zhuang2015debiasing} };   ] ] ] [.\node[draw,rectangle,rounded corners,text width=2.2in]{Modelling Worker Drop-out Rates at Each Session \& Descriptive Statistics Based Difficulty Used as Features}; [.\node[draw,rectangle,rounded corners,text width=2in]{Logistic\\Survival Models};  [.\node[,align=left]{\cite{kobren2015getting}};  ]  ] [.\node[draw,rectangle,rounded corners,text width=2in]{Logistic Regression \& Decision Trees};  [.\node[,align=left]{\cite{mao2013stop}};  ]  ] ]  ]  [.\node[draw,rectangle,rounded corners,text width=1.5in]{Response-Level Context};  [.\node[draw,rectangle,rounded corners,text width=2.2in]{Modelling Time-Aware Worker Response Biases \& Time Bound Based Question Difficulty};  [.\node[draw,rectangle,rounded corners,text width=1.75in]{Bayesian\\Hierarchical Models};  [.\node[,align=left]{\cite{venanzi2016time}};  ] ] ]  [.\node[draw,rectangle,rounded corners,text width=2.2in]{Modelling Distance-Aware Worker Response Correctness \& Distance Based Question Difficulty};  [.\node[draw,rectangle,rounded corners,text width=1.75in]{EM-Based ML Estimation};  [.\node[,align=left]{\cite{hu2016crowdsourced}};  ] ] ]  ] ] ]
\end{tikzpicture}
\caption{A taxonomy of QCC papers that considered worker ability, context and question difficulty. They focused on statistical modelling of how worker ability varies with contexts at specific levels, and the interaction between the context-aware ability and the difficulty.}
\end{figure}\subsection{Modelling the Interactions in Task-Level Contexts}~\label{chp2:worker_expertise_diff_context_task}
In this case, the context in which the interaction between the worker ability and the question difficulty takes place is the whole crowdsourcing platform. The proposed models utilize the response information from the same workers across various source tasks in which they have participated to improve the parameter estimation for a target task. This is referred as the \textit{transfer learning}~\cite{pan2010survey} in the Machine Learning literature. 

In~\cite{mo2013cross}, the authors proposed a model which encoded multi-dimensional ability and difficulty. Since this multifaceted model has a larger number of parameters, it is intrinsically more vulnerable to response sparsity. The authors thus decided to transfer worker expertise estimates from source tasks to target tasks based on estimated similarity between the tasks. This cross-task transfer was able to smooth out the unreliable expertise estimation in the target tasks. However, no transfer learning was conducted for calibrating the estimation of question difficulty in the target tasks. In contrast, in~\cite{fang2013knowledge, fang2014active}, the transfer learning helps to learn a better latent feature representation for each question in the target task from the observed features of the questions in the source tasks. The latent feature representation for each target question can be interpreted as their multi-dimensional difficulty.

\subsection{Modelling the Interactions in Session-Level Contexts}

The literature has also considered the effect that each session/page has on the interaction between worker ability and question difficulty. The corresponding models learn latent (bias) variables specific to particular (types of) sessions. Zhuang and Young~\cite{zhuang2015leveraging} did this by learning a factor graph model which encoded biases in workers' responses to questions within each session. The factor function is defined as the exponent of a linear regression over counts of different response options (within a particular session). The regression coefficients are global, meaning that they encode a latent bias structure shared across the sessions. This latent structure maps the response count distribution from a session to a bias value which \textit{offsets} each response accuracy (determined by the expertise-difficulty interaction) within that session. In another work \cite{zhuang2015debiasing}, the authors assume that a worker annotates a data item within a session either \textit{independently} from the other items or \textit{relatively} according to a ranking of the items' response correctness probabilities (determined by their difficulty). The ranking is inferred using the Plackett-Luce model \cite{plackett1975analysis}. The top-\textit{N} items in the inferred ranking are considered to be the ones that are responded correctly. The parameter \textit{N}, which is smaller than or equal to the number of questions within a session, is estimated using the ML estimation.

In~\cite{kobren2015getting}, a worker drop-out modelling framework was proposed which consists of a sequence of logistic survival models each corresponding to a particular session/page. Each model determines the probabilities of workers surviving a particular session and moving to the next. The coefficients of a model map features about workers (e.g. average response time, response accuracy over control questions, etc.), questions (e.g. difficulty, skip rate, average response time, etc.) and consecutive pairs of questions (e.g. same topic or not, their average skip rate, etc.) within the session. Similarly, Mao et al.~\cite{mao2013stop} endeavoured to predict the survival rates of the workers in their respective sequences of sessions using session-specific classification models such as logistic regression and decision trees. The side information features used by these models included the worker's dwell time on a page, the entropy of her responses from past sessions, the number of past sessions, the average response time for past sessions, etc.

\subsection{Modelling the Interactions in Response-Level Contexts}~\label{chp2:worker_expertise_diff_context_response}
We now consider how features that describe the context under which each response is made, such as response duration, order and location, can be used to improve the estimation of response quality. In~\cite{venanzi2016time}, the response time was aggregated in a Bayesian manner across the responses to each question to obtain the lower and upper bounds of an acceptable response duration for each question. The inferred bounds not only indicate the difficulty of the questions (i.e. higher upper bounds suggesting more difficult questions), but also help detect spam responses (i.e. abnormally long or short response time compared to the bounds). %In \cite{lakkaraju2016confusions} (\cite{lakkaraju2016confusions}), the group quality of workers on answering similar questions at each time step is modelled as probabilistic confusion matrices specific to every pair of worker and item clusters at that time step. For each such pair, the confusion matrix at the current time step depends on the matrix at the previous time step. The clusters of either workers or items are obtained through Bayesian mixture modelling. The ablation study shows that this temporal model significantly outperforms its variant with temporal dependency removed. 
In~\cite{hu2016crowdsourced}, the quality of a worker's response to a spatial question (e.g. labelling point of interests) is jointly determined by three factors: the worker's intrinsic ability (i.e. the probability of her being reliable\footnote{An unreliable worker randomly responds to questions and thus has a correctness probability on binary responses of 0.5.}), her (response-specific) location-aware ability, and the difficulty of the question. The location-aware ability decays as the worker's response location moves farther away from the question. The question's difficulty increases as its distance from the worker becomes larger. Shared by both the workers and the questions, the decay factor is treated as one of the model parameters over a finite set of ordinal values and estimated during the model inference. 

\section{Modelling Worker Expertise, Question Difficulty and Response Relationships}\label{chp2:worker_expertise_diff_relation}

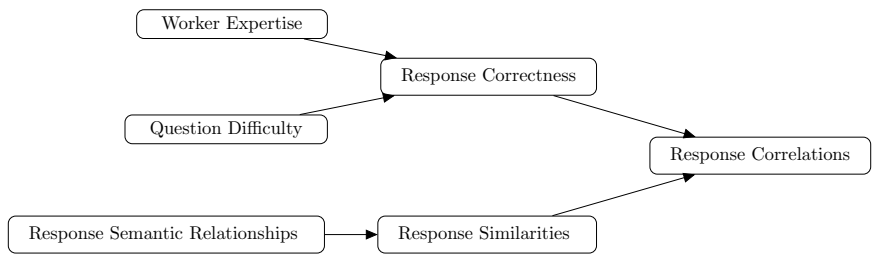
\begin{figure*}[h!]
  \centering
  	\scalebox{.7} 		{\begin{tikzpicture}
  \node[latent,rectangle,rounded corners] (rij) {~~~Response Correlations~~~~};
  \node[latent,align=center,left=of rij, rectangle,rounded corners, yshift=1.5cm] (qij) {~~~Response Correctness~~~};
   \node[latent,align=center,left=of rij,rectangle,rounded corners, yshift=-1.5cm] (cij) {~~~Response Similarities~~~~};

\node[draw,align=center,left=of qij,rectangle,rounded corners, yshift=1cm] (ei) {~~~Worker Expertise~~~};
\node[draw,align=center,left=of qij,rectangle,rounded corners, yshift=-1cm] (dj) {~~~Question Difficulty~~~};   \node[latent,align=center,left=of cij,rectangle,rounded corners] (xij) {~~~Response Semantic Relationships~~~~};
\edge{qij,cij}{rij};
\edge{ei,dj}{qij};
\edge{xij}{cij};

\end{tikzpicture}}
  %\caption{Ability-Difficulty-Response Interaction}  
  \caption{A diagram shows how worker expertise, question difficulty and response semantic relationships contribute to the correlations within responses.}
  \label{fig:chp1_ability_diff_answer_diagram}
\end{figure*}
When semantic relationships between response options are present in crowdsourcing, QCC methods measure the relationships using some \textit{distance} metric such that options with closer relationships have smaller distances. A common distance metric used by the state-of-the-art methods is the \textit{length} of (or equivalently, the number of edges in) the \textit{shortest path} between two options in a semantic structure (e.g WordNet\footnote{https://wordnet.princeton.edu/}).

\begin{figure}[h!]
\begin{tikzpicture}[scale=0.45, every node/.style={transform shape},grow'=right,level distance=2.5in,sibling distance=0.4in]
\tikzset{edge from parent/.style= 
            {thick, draw, edge from parent fork right},
         every tree node/.style=
            {minimum width=1in,text width=2in,align=center}}
\Large            
\Tree
[.\node[text width=2.2in]{Statistical Methods:\\Modelling Worker Expertise, Question Difficulty \& Response Option Relationships}; [.\node[draw,rectangle,rounded corners,text width=2in]{Modelling Worker Response Biases \& Question Difficulty}; [.\node[draw,rectangle,rounded corners,text width=2.2in]{Modelling Biases Using Pre-Computed Response Similarities from External Knowledge}; [.\node[draw,rectangle,rounded corners,text width=2in]{Observed Features about Response Options};  [.\node[draw,rectangle,rounded corners,text width=1.5in]{EM-Based ML Estimation}; [.\node[,align=left]{\cite{fang2017improving} };   ]   ] ] [.\node[draw,rectangle,rounded corners,text width=1.5in]{WordNet}; [.\node[draw,rectangle,rounded corners,text width=1.5in]{EM-Based MAP Estimation};   [.\node[,align=left]{\cite{han2016incorporating} };  ] ] ]  ] [.\node[draw,rectangle,rounded corners,text width=2.2in]{Modelling Biases Using a Latent Option-Option Confusion Matrix}; [.\node[draw,rectangle,rounded corners,text width=2.2in]{Oridinal Logistic Regressions with Expertise-Difficulty Slopes \& Off-Diagonal Entries as Intercepts}; [.\node[draw,rectangle,rounded corners,text width=1.75in]{Gibbs Sampling +\\ L-BFGS}; [.\node[,align=left]{\cite{Yuan2018LabelCategory} };  ] ] ] ]  ] ] ]
\end{tikzpicture}
\caption{A taxonomy of QCC papers that considered worker expertise, question difficulty and semantic relationships between responses. These papers focused on statistical modelling which leveraged pre-computed response similarities from external knowledge to account for response biases.}
\end{figure}
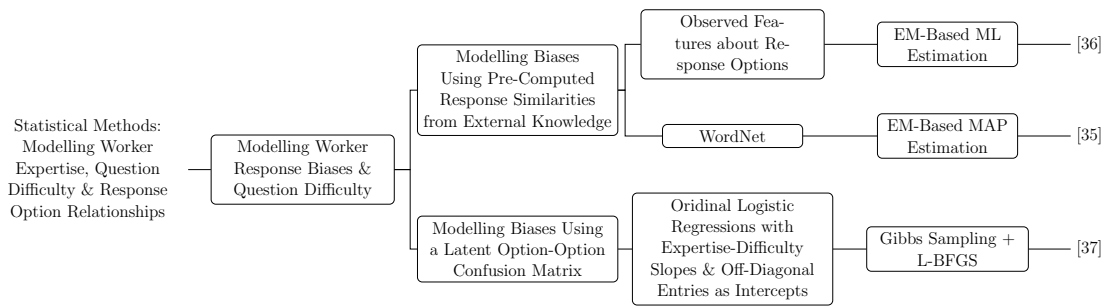

Three papers have considered leveraging semantic relationships between response options \cite{han2016incorporating,fang2017improving,Yuan2018LabelCategory} for improving the quality control of crowdsourced responses. %The core idea of both papers is to use the precomputed semantic distances to calculate response similarities which indicate possible correlations in workers' responses (both correct and incorrect responses) as shown by Figure~\ref{fig:chp1_ability_diff_answer_diagram}.
In~\cite{han2016incorporating}, a model was proposed in which the probability of each response option a worker could give to an item is conditioned on its true answer and provided by a \textit{soft-max} function. This function takes in the normalized distances between each response option and the correct response, along with the question difficulty and the worker expertise. The difference between the difficulty and the expertise is scaled by the normalized distance before computing the corresponding conditional probabilities. Due to the scaling effect, the log-odds of the probabilities are \textit{inversely proportional} to the normalized distances. The larger the distances are, the smaller the probabilities. In other words, a response option that is inherently less related to the correct answer is less likely to be selected irrespective of the worker or question.

In~\cite{fang2017improving}, the authors proposed a model which shares the same idea as~\cite{han2016incorporating} except that \textit{similarity scores} between response options are pre-computed as the inverse of the Euclidean distances between the options in terms of their \textit{observed features}. Both of these models rely on the availability of the external knowledge about the semantic relationships to pre-compute the semantic distances.

Recently, Yuan et al.~\cite{Yuan2018LabelCategory} proposed the first QCC model that directly infers the semantic relationships from responses. Their model captures the relationships using a \textit{symmetric latent} relatedness matrix. In this matrix, each off-diagonal entry is a real-valued score representing how related the response options are to one another. The authors also observed two phenomena: (i) capable workers are more likely to select options that are more (semantically) related to the correct answer and (ii) difficult questions are more likely to receive responses less related to their correct answers. The proposed model captures the above phenomena using ordinal logistic regressions. Each regression has a \textit{response-specific slope} being the expertise-difficulty difference, and the off-diagonal relatedness scores as the \textit{ordinal intercepts} specific to each option (other than the correct answer). The authors also showed that their model can elegantly incorporate the external knowledge, via a linear regression, into the prior mean of the Normal distribution which generates the off-diagonal scores. They finally showed that their model outperformed the previous two models~\cite{han2016incorporating,fang2017improving} in true answer prediction.

\section{Summary and Discussion}
Crowdsourcing has become a principle tool that allows research communities and companies to collect data for (machine learning) system development and business analysis, with significant cost savings and fast turnaround. However, cost-effective crowdsourcing is often elusive in terms of the quality of recorded responses (and their costs). The corresponding solution is the quality control methods which aim to remove or suppress low-quality responses (and possibly amplify the high-quality ones).

The two major categories of quality control methods are quality control mechanism designs and statistical inference models. Surveys exist regarding each category but their reviews were isolated without recognizing the strong connection between the categories. We view the quality control for crowdsourcing as a unified cyclic process that integrates both categories of methods (as show in Fig.~\ref{chp2:design_stats_diagram}). This viewpoint allows us to conduct a more comprehensive survey that bridges the two categories in terms of how their methods work together for crowdsourcing. To link these two categories, we proposed in the survey a framework that systematically unifies all the crowdsourcing aspects (and their important attributes) modelled by both categories to determine the response quality. Many of these aspects and attributes are, for the first time, identified from the quality control literature. Based on this framework, we proposed another graph framework that unifies the past quality control research in terms of the aspects and attributes they have exploited.

Our survey flows by following the (quality control research) graph framework and provides systematic technical insights to a wide variety of quality control methods. It also contributes, for the first time, organized taxonomies of quality control papers. Each taxonomy is characterized by the considered aspects and attributes, modelling assumptions, parameter estimation techniques, and design features of quality control methods of the same type (deemed by the graph).

According to our survey, there are several limitations existing in the current quality control research. First, even though the GLAD and DS models have been widely applied and extended in the research, these applications and extensions rarely aim at \textit{large-scale sparse} and \textit{diverse} crowdsourcing. This type of crowdsourcing is featured by large numbers of questions and crowd-workers with sparse responses across them, and the characteristics of questions vary greatly (e.g. by having diverse topics). It is prevalent in user content generation Websites where the quality of user generated content poses an issue and needs to be controlled (by estimation from responses of other users to the same content). 

In this case, workers can react very differently due to the question diversity, and their response correlations will be much weaker than small-scale (non-sparse and non-diverse) crowdsourcing which is the focus of current quality control statistical models. As for the future research directions, Bayesian hierarchical models have great opportunities to properly handle large-scale sparse and diverse crowdsourcing due to its hierarhical nature that smooths the parameter estimation. Among them, hierarchical topical models that have been successfully applied to mining large corpora with diverse topics and sparse word counts~\cite{Lim2016bayesian} can be adapted. Deep generative models~\cite{srivastava2017autoencoding} are another alternative to this purpose. They can be adapted to capture complex question and worker behavioural diversity directly from responses.

The second limitation is the lack of advanced statistical models for handling the subjectivity that might exist in crowdsourcing. Human judgment and generated content is intrinsically subjective due to personal preferences and opinions. This leads to the variation in responses (to the generated content) and current quality control methods seldom separate it from the quality of responses in their models. Although recent work~\cite{nguyen2016probabilistic,Yuan2018Distinguishing} has made some progress in this regard, more need to be done in the future. Especially, how to design measures of subjectivity in crowdsourcing remains an open question. So far, only Yuan et al.~\cite{Yuan2018Distinguishing} directly defined a question-specific subjectivity measure, which was the expected number of correct answers (for a question) with respect to underlying groups of workers. Each group embodies a particular school of thought. Other subjectivity measures can be defined specific to not only questions but also the entire tasks or individual workers. The corresponding statistical models can encode (and quantify) these measures as variables different from the response quality variables. 

Another interesting direction is to investigate the dependency between the difficulty and subjectivity of questions. According to the literature, the difficulty determines the response quality while the subjectivity results in response correlations. Yuan et al.~\cite{Yuan2018Distinguishing} made an \textit{independence} assumption about the two attributes. Future study on this subject will allow us to derive more reasonable models if the dependency does exist. More importantly, these subjectivity-aware models can be further fused with those aiming for large-scale sparse and diverse crowdsourcing as user generated contents can exhibit subjectivity. Controlling the quality of these contents requires models that separate out the subjectivity.

The third limitation is the incapability of current quality control methods in handling either \textit{highly multi-class} or \textit{highly multi-label crowdsourcing}. The former concerns crowdsourcing with many response options but there is only a single correct answer for each question. Examples include image categorization based on large numbers of ImageNet categories~\cite{khosla2011novel}, Web-page categorization based on large numbers of Wikipedia (topical) categories, and etc. The latter concerns each question having many different correct answers (from many response options). A typical example is using crowdsourcing to tag user generated contents~\cite{Jonathan2017geotagging}. Both types of crowdsourcing could exhibit response correlations resulted from \textit{semantic relationships between the response options}. Highly multi-label crowdsourcing is also a special case of partially subjective crowdsourcing which means that the subjectivity could also have contributed to the response correlation. However, the dependency between the subjectivity and the semantic relationships remains unclear and needs to investigated in the future.

Currently, only three papers have dealt with the quality control for highly multi-class crowdsourcing~\cite{han2016incorporating,fang2017improving,Yuan2018LabelCategory} and only one of them shows the potential of reconstructing the semantic relationships directly from responses~\cite{Yuan2018LabelCategory}. As for the highly multi-label crowdsourcing, an effective quality control model is still missing. According to the literature, semantic relationships between response options can be represented by an option-option matrix. This matrix is quadratic in the number of options, and therefore will become very large when the number of options is large. Thus, when estimating this matrix (to reconstruct the semantic relationships), future models need to make use of matrix factorization techniques. Such techniques are useful for lowering the complexity of the models to prevent over-fitting. The resulting factors for individual options can be modelled to interact with other factors (e.g. the expertise and difficulty factors) to determine the response quality.

The final limitation we have found regards the quality control mechanism designs, particularly for the \textit{worker payment} and the \textit{task gamification}. For the payment design, most of the current techniques rely on game-theoretic approaches. These approaches are theoretically elegant and sound but very few empirical studies have been conducted to systematically verify and compare their quality control performance in practice. Thus, studies in this regard are needed in the near future. As for the gamification design, its methodologies are much less developed compared with the methodologies for either worker payment or question allocation design. We have found from our survey that most papers on gamifying crowdsourcing tasks rely on either authors' impressions about various game elements or conventions from video games to draw the designs for their tasks. This suggests the need for empirical insights into both the individual and joint effects of the various game elements on the quality of responses in general crowdsourcing. These insights will be critical for constructing a common gamification methodology which provides empirically justified guidelines on what combinations of game elements should be used (possibly together with other mechanisms) to improve the quality of responses. 

\newpage
\section*{References}

\bibliography{bibliography}

\begin{thebibliography}{100}
\expandafter\ifx\csname url\endcsname\relax
  \def\url#1{\texttt{#1}}\fi
\expandafter\ifx\csname urlprefix\endcsname\relax\def\urlprefix{URL }\fi
\expandafter\ifx\csname href\endcsname\relax
  \def\href#1#2{#2} \def\path#1{#1}\fi

\bibitem{DBS-044}
A.~Marcus, A.~Parameswaran, Crowdsourced data management: Industry and academic
  perspectives, Foundations and Trends in Databases 6~(1-2) (2015) 1--161.
\newblock \href {http://dx.doi.org/10.1561/1900000044}
  {\path{doi:10.1561/1900000044}}.

\bibitem{zaidan2011crowdsourcing}
O.~F. Zaidan, C.~Callison-Burch, Crowdsourcing translation: Professional
  quality from non-professionals, in: Proceedings of the 49th Annual Meeting of
  the Association for Computational Linguistics: Human Language
  Technologies-Volume 1, Association for Computational Linguistics, 2011, pp.
  1220--1229.

\bibitem{liu2012crowdsourcing}
D.~Liu, R.~G. Bias, M.~Lease, R.~Kuipers, Crowdsourcing for usability testing,
  in: Proceedings of the Association for Information Science and Technology,
  Vol.~49, Wiley Online Library, 2012, pp. 1--10.

\bibitem{stol2014two}
K.-J. Stol, B.~Fitzgerald, Two's company, three's a crowd: a case study of
  crowdsourcing software development, in: Proceedings of the 36th International
  Conference on Software Engineering, ACM, 2014, pp. 187--198.

\bibitem{ipeirotis2010analyzing}
P.~G. Ipeirotis, Analyzing the amazon mechanical turk marketplace, XRDS:
  Crossroads, the ACM Magazine for Students 17~(2) (2010) 16--21.

\bibitem{kumar2011learning}
A.~Kumar, M.~Lease, Learning to rank from a noisy crowd, in: Proceedings of the
  34th International ACM SIGIR conference on Research and Development in
  Information Retrieval, ACM, 2011, pp. 1221--1222.

\bibitem{ambati2012active}
V.~Ambati, Active learning and crowdsourcing for machine translation in low
  resource scenarios, Ph.D. thesis (2012).

\bibitem{brew2010interaction}
A.~Brew, D.~Greene, P.~Cunningham, the interaction between supervised learning
  and crowdsourcing, in: NIPS workshop on computational social science and the
  wisdom of crowds, 2010.

\bibitem{deng2013fine}
J.~Deng, J.~Krause, L.~Fei-Fei, Fine-grained crowdsourcing for fine-grained
  recognition, in: Proceedings of the 2013 IEEE Conference on Computer Vision
  and Pattern Recognition, IEEE Computer Society, 2013, pp. 580--587.

\bibitem{cheng2015flock}
J.~Cheng, M.~S. Bernstein, Flock: Hybrid crowd-machine learning classifiers,
  in: Proceedings of the 18th ACM Conference on Computer Supported Cooperative
  Work \& Social Computing, ACM, 2015, pp. 600--611.

\bibitem{eickhoff2013increasing}
C.~Eickhoff, A.~P. de~Vries, Increasing cheat robustness of crowdsourcing
  tasks, Information Retrieval 16~(2) (2013) 121--137.

\bibitem{surowiecki2005wisdom}
J.~Surowiecki, the Wisdom of Crowds, Anchor, 2005.

\bibitem{allahbakhsh2013quality}
M.~Allahbakhsh, B.~Benatallah, A.~Ignjatovic, H.~R. Motahari-Nezhad,
  E.~Bertino, S.~Dustdar, Quality control in crowdsourcing systems: Issues and
  directions, IEEE Internet Computing 17~(2) (2013) 76--81.

\bibitem{chittilappilly2016survey}
A.~I. Chittilappilly, L.~Chen, S.~Amer-Yahia, A survey of general-purpose
  crowdsourcing techniques, IEEE Transactions on Knowledge and Data Engineering
  28~(9) (2016) 2246--2266.

\bibitem{daniel2018quality}
F.~Daniel, P.~Kucherbaev, C.~Cappiello, B.~Benatallah, M.~Allahbakhsh, Quality
  control in crowdsourcing: A survey of quality attributes, assessment
  techniques, and assurance actions, ACM Computing Surveys 51~(1) (2018) 7.

\bibitem{muhammadi2015unified}
J.~Muhammadi, H.~R. Rabiee, A.~Hosseini, A unified statistical framework for
  crowd labeling, Knowledge and Information Systems 45~(2) (2015) 271--294.

\bibitem{zhang2016learning}
J.~Zhang, X.~Wu, V.~S. Sheng, Learning from crowdsourced labeled data: a
  survey, Artificial Intelligence Review 46~(4) (2016) 543--576.

\bibitem{zheng2017truth}
Y.~Zheng, G.~Li, Y.~Li, C.~Shan, R.~Cheng, Truth inference in crowdsourcing: is
  the problem solved?, Proceedings of the VLDB Endowment 10~(5) (2017)
  541--552.

\bibitem{kittur2013future}
A.~Kittur, J.~V. Nickerson, M.~Bernstein, E.~Gerber, A.~Shaw, J.~Zimmerman,
  M.~Lease, J.~Horton, the future of crowd work, in: Proceedings of the 2013
  Conference on Computer Supported Cooperative Work, ACM, 2013, pp. 1301--1318.

\bibitem{yin2014survey}
X.~Yin, W.~Liu, Y.~Wang, C.~Yang, L.~Lu, What? how? where? a survey of
  crowdsourcing, in: Frontier and Future Development of Information Technology
  in Medicine and Education, Springer, 2014, pp. 221--232.

\bibitem{geiger2014personalized}
D.~Geiger, M.~Schader, Personalized task recommendation in crowdsourcing
  information systems - current state of the art, Decision Support Systems 65
  (2014) 3--16.

\bibitem{luz2015survey}
N.~Luz, N.~Silva, P.~Novais, A survey of task-oriented crowdsourcing,
  Artificial Intelligence Review 44~(2) (2015) 187--213.

\bibitem{lease2013crowdsourcing}
M.~Lease, E.~Yilmaz, Crowdsourcing for information retrieval: introduction to
  the special issue, Information Retrieval 16~(2) (2013) 91--100.

\bibitem{mao2015survey}
K.~Mao, L.~Capra, M.~Harman, Y.~Jia, A survey of the use of crowdsourcing in
  software engineering, Journal of Systems and Software 126 (2017) 57 -- 84.

\bibitem{xintong2014brief}
G.~Xintong, W.~Hongzhi, Y.~Song, G.~Hong, Brief survey of crowdsourcing for
  data mining, Expert Systems with Applications 41~(17) (2014) 7987--7994.

\bibitem{ranard2014crowdsourcing}
B.~L. Ranard, Y.~P. Ha, Z.~F. Meisel, D.~A. Asch, S.~S. Hill, L.~B. Becker,
  A.~K. Seymour, R.~M. Merchant, Crowdsourcing - harnessing the masses to
  advance health and medicine, a systematic review, Journal of General Internal
  Medicine 29~(1) (2014) 187--203.

\bibitem{gomes2012crowdsourcing}
C.~Gomes, D.~Schneider, K.~Moraes, J.~d.~Souza, Crowdsourcing for music: Survey
  and taxonomy, in: 2012 IEEE International Conference on Systems, Man, and
  Cybernetics, 2012, pp. 832--839.

\bibitem{heipke2010crowdsourcing}
C.~Heipke, Crowdsourcing geospatial data, ISPRS Journal of Photogrammetry and
  Remote Sensing 65~(6) (2010) 550--557.

\bibitem{frongillo2015elicitation}
R.~M. Frongillo, Y.~Chen, I.~A. Kash, Elicitation for aggregation, in:
  Proceedings of the Twenty-Ninth AAAI Conference on Artificial Intelligence,
  AAAI Press, 2015, pp. 900--906.

\bibitem{ho2016eliciting}
C.-J. Ho, R.~Frongillo, Y.~Chen, Eliciting categorical data for optimal
  aggregation, in: Advances in Neural Information Processing Systems, Curran
  Associates, Inc., 2016, pp. 2450--2458.

\bibitem{Yuan2018Distinguishing}
Y.~Jin, M.~Carman, Y.~Zhu, W.~Buntine, Distinguishing question subjectivity
  from difficulty for improved crowdsourcing, in: J.~Zhu, I.~Takeuchi (Eds.),
  Proceedings of The 10th Asian Conference on Machine Learning, Vol.~95 of
  Proceedings of Machine Learning Research, PMLR, 2018, pp. 192--207.

\bibitem{jung2014predicting}
H.~J. Jung, Y.~Park, M.~Lease, Predicting next label quality: A time-series
  model of crowdwork, in: Second AAAI Conference on Human Computation and
  Crowdsourcing, 2014.

\bibitem{venanzi2016time}
M.~Venanzi, J.~Guiver, P.~Kohli, N.~R. Jennings, Time-sensitive bayesian
  information aggregation for crowdsourcing systems, Journal of Artificial
  Intelligence Research 56 (2016) 517--545.

\bibitem{jung2015modeling}
H.~J. Jung, M.~Lease, Modeling temporal crowd work quality with limited
  supervision, in: Third AAAI Conference on Human Computation and
  Crowdsourcing, 2015.

\bibitem{han2016incorporating}
T.~Han, H.~Sun, Y.~Song, Y.~Fang, X.~Liu, Incorporating external knowledge into
  crowd intelligence for more specific knowledge acquisition, in: Proceedings
  of the Twenty-Fifth International Joint Conference on Artificial
  Intelligence, AAAI Press, 2016, pp. 1541--1547.

\bibitem{fang2017improving}
Y.-L. Fang, H.-L. Sun, P.-P. Chen, T.~Deng, Improving the quality of
  crowdsourced image labeling via label similarity, Journal of Computer Science
  and Technology 32~(5) (2017) 877--889.

\bibitem{Yuan2018LabelCategory}
Y.~Jin, L.~Du, Y.~Zhu, M.~Carman, Leveraging label category relationships in
  multi-class crowdsourcing, in: Proceedings of the 22nd Pacific-Asia
  Conference on Advances in Knowledge Discovery and Data Mining, Springer
  International Publishing, 2018, pp. 128--140.

\bibitem{Loni:2013}
B.~Loni, M.~Menendez, M.~Georgescu, L.~Galli, C.~Massari, I.~S. Altingovde,
  D.~Martinenghi, M.~Melenhorst, R.~Vliegendhart, M.~Larson, Fashion-focused
  creative commons social dataset, in: Proceedings of the 4th ACM Multimedia
  Systems Conference, 2013, pp. 72--77.

\bibitem{Loni:2014}
B.~Loni, L.~Y. Cheung, M.~Riegler, A.~Bozzon, L.~Gottlieb, M.~Larson, Fashion
  10000: An enriched social image dataset for fashion and clothing, in:
  Proceedings of the 5th ACM Multimedia Systems Conference, 2014, pp. 41--46.

\bibitem{deng2009imagenet}
J.~Deng, W.~Dong, R.~Socher, L.~Li, K.~Li, L.~Fei-Fei, Imagenet: A large-scale
  hierarchical image database, in: 2009 IEEE Conference on Computer Vision and
  Pattern Recognition, 2009, pp. 248--255.

\bibitem{khosla2011novel}
A.~Khosla, N.~Jayadevaprakash, B.~Yao, F.-f. Li, L.: Novel dataset for
  fine-grained image categorization, in: First Workshop on Fine-Grained Visual
  Categorization, CVPR (2011, Citeseer.

\bibitem{dawid1979maximum}
A.~P. Dawid, A.~M. Skene, Maximum likelihood estimation of observer error-rates
  using the em algorithm, Applied Statistics (1979) 20--28.

\bibitem{snow2008cheap}
R.~Snow, B.~O'Connor, D.~Jurafsky, A.~Y. Ng, Cheap and fast - but is it good?:
  evaluating non-expert annotations for natural language tasks, in: Proceedings
  of the Conference on Empirical Methods in Natural Language Processing,
  Association for Computational Linguistics, 2008, pp. 254--263.

\bibitem{tang2011semi}
W.~Tang, M.~Lease, Semi-supervised consensus labeling for crowdsourcing, in:
  SIGIR 2011 Workshop on Crowdsourcing for Information Retrieval, 2011, pp.
  1--6.

\bibitem{zhang2014spectral}
Y.~Zhang, X.~Chen, D.~Zhou, M.~I. Jordan, Spectral methods meet em: A provably
  optimal algorithm for crowdsourcing, in: Advances in Neural Information
  Processing Systems, 2014, pp. 1260--1268.

\bibitem{zhang2016spectral}
Y.~Zhang, X.~Chen, D.~Zhou, M.~Jordan, Spectral methods meet em: A provably
  optimal algorithm for crowdsourcing, Journal of Machine Learning Research
  17~(1) (2016) 3537--3580.

\bibitem{gao2013minimax}
C.~Gao, D.~Zhou, Minimax optimal convergence rates for estimating ground truth
  from crowdsourced labels, arXiv preprint arXiv:1310.5764.

\bibitem{carpenter2011hierarchical}
B.~Carpenter, A hierarchical bayesian model of crowdsourced relevance coding.

\bibitem{kim2012bayesian}
H.-C. Kim, Z.~Ghahramani, Bayesian classifier combination, in: Artificial
  Intelligence and Statistics, 2012, pp. 619--627.

\bibitem{simpson2013dynamic}
E.~Simpson, S.~J. Roberts, I.~Psorakis, A.~Smith, Dynamic bayesian combination
  of multiple imperfect classifiers., Decision Making and Imperfection 474
  1--35.

\bibitem{ghosh2011moderates}
A.~Ghosh, S.~Kale, P.~McAfee, Who moderates the moderators?: crowdsourcing
  abuse detection in user-generated content, in: Proceedings of the 12th ACM
  Conference on Electronic Commerce, ACM, 2011, pp. 167--176.

\bibitem{karger2011budget}
D.~R. Karger, S.~Oh, D.~Shah, Budget-optimal crowdsourcing using low-rank
  matrix approximations, in: 2011 49th Annual Allerton Conference on
  Communication, Control, and Computing (Allerton), IEEE, 2011, pp. 284--291.

\bibitem{dalvi2013aggregating}
N.~Dalvi, A.~Dasgupta, R.~Kumar, V.~Rastogi, Aggregating crowdsourced binary
  ratings, in: Proceedings of the 22nd International Conference on World Wide
  Web, ACM, 2013, pp. 285--294.

\bibitem{karger2011iterative}
D.~R. Karger, S.~Oh, D.~Shah, Iterative learning for reliable crowdsourcing
  systems, in: Advances in Neural Information Processing Systems, 2011, pp.
  1953--1961.

\bibitem{karger2014budget}
D.~R. Karger, S.~Oh, Shah, Budget-optimal task allocation for reliable
  crowdsourcing systems, Operations Research 62~(1) (2014) 1--24.

\bibitem{liu2012variational}
Q.~Liu, J.~Peng, A.~T. Ihler, Variational inference for crowdsourcing, in:
  Advances in Neural Information Processing Systems, 2012, pp. 692--700.

\bibitem{ok2016optimality}
J.~Ok, S.~Oh, J.~Shin, Y.~Yi, Optimality of belief propagation for crowdsourced
  classification, in: International Conference on Machine Learning, 2016, pp.
  535--544.

\bibitem{bonald2017minimax}
T.~Bonald, R.~Combes, A minimax optimal algorithm for crowdsourcing, in:
  Advances in Neural Information Processing Systems, 2017.

\bibitem{venanzi2014community}
M.~Venanzi, J.~Guiver, G.~Kazai, P.~Kohli, M.~Shokouhi, Community-based
  bayesian aggregation models for crowdsourcing, in: Proceedings of the 23rd
  International Conference on World Wide Web, ACM, 2014, pp. 155--164.

\bibitem{moreno2015bayesian}
P.~G. Moreno, A.~Artes-Rodriguez, Y.~W. Teh, F.~Perez-Cruz, Bayesian
  nonparametric crowdsourcing, Journal of Machine Learning Research.

\bibitem{liu2012truelabel+}
C.~Liu, Y.-M. Wang, Truelabel+ confusions: a spectrum of probabilistic models
  in analyzing multiple ratings, in: Proceedings of the 29th International
  Coference on International Conference on Machine Learning, Omnipress, 2012,
  pp. 17--24.

\bibitem{kamar2015identifying}
E.~Kamar, A.~Kapoor, Horvitz, Identifying and accounting for task-dependent
  bias in crowdsourcing, in: Third AAAI Conference on Human Computation and
  Crowdsourcing, 2015.

\bibitem{aydin2014crowdsourcing}
B.~I. Aydin, Y.~S. Yilmaz, Y.~Li, Q.~Li, J.~Gao, M.~Demirbas, Crowdsourcing for
  multiple-choice question answering., in: Proceedings of the 28th AAAI
  Conference on Artificial Intelligence, 2014, pp. 2946--2953.

\bibitem{yan2015opportunities}
R.~Yan, Y.~Song, C.-T. Li, M.~Zhang, X.~Hu, Opportunities or risks to reduce
  labor in crowdsourcing translation? characterizing cost versus quality via a
  pagerank-hits hybrid model, in: Proceedings of the 24th International
  Conference on Artificial Intelligence, AAAI Press, 2015, pp. 1025--1032.

\bibitem{kajimura2015quality}
S.~Kajimura, Y.~Baba, H.~Kajino, H.~Kashima, Quality control for crowdsourced
  poi collection, in: Pacific-Asia Conference on Knowledge Discovery and Data
  Mining, Springer, 2015, pp. 255--267.

\bibitem{sunahase2017pairwise}
T.~Sunahase, Y.~Baba, H.~Kashima, Pairwise hits: Quality estimation from
  pairwise comparisons in creator-evaluator crowdsourcing process., in:
  Proceedings of the 31st AAAI Conference on Artificial Intelligence, 2017, pp.
  977--984.

\bibitem{li2016survey}
Y.~Li, J.~Gao, C.~Meng, Q.~Li, L.~Su, B.~Zhao, W.~Fan, J.~Han, A survey on
  truth discovery, Acm Sigkdd Explorations Newsletter 17~(2) (2016) 1--16.

\bibitem{ouyang2015debiasing}
R.~W. Ouyang, L.~Kaplan, P.~Martin, A.~Toniolo, M.~Srivastava, T.~J. Norman,
  Debiasing crowdsourced quantitative characteristics in local businesses and
  services, in: Proceedings of the 14th International Conference on Information
  Processing in Sensor Networks, ACM, 2015, pp. 190--201.

\bibitem{ouyang2016aggregating}
R.~W. Ouyang, L.~M. Kaplan, A.~Toniolo, M.~Srivastava, T.~J. Norman,
  Aggregating crowdsourced quantitative claims: Additive and multiplicative
  models, IEEE Transactions on Knowledge and Data Engineering 28~(7) (2016)
  1621--1634.

\bibitem{raykar2010learning}
V.~C. Raykar, S.~Yu, L.~H. Zhao, G.~H. Valadez, C.~Florin, L.~Bogoni, L.~Moy,
  Learning from crowds, Journal of Machine Learning Research 11~(Apr) (2010)
  1297--1322.

\bibitem{gao2016exact}
C.~Gao, Y.~Lu, D.~Zhou, Exact exponent in optimal rates for crowdsourcing, in:
  International Conference on Machine Learning, 2016, pp. 603--611.

\bibitem{karger2013efficient}
D.~R. Karger, S.~Oh, D.~Shah, Efficient crowdsourcing for multi-class labeling,
  ACM SIGMETRICS Performance Evaluation Review 41~(1) (2013) 81--92.

\bibitem{kleinberg1999authoritative}
J.~M. Kleinberg, Authoritative sources in a hyperlinked environment, Journal of
  the ACM (JACM) 46~(5) (1999) 604--632.

\bibitem{whitehill2009whose}
J.~Whitehill, T.-f. Wu, J.~Bergsma, J.~R. Movellan, P.~L. Ruvolo, Whose vote
  should count more: Optimal integration of labels from labelers of unknown
  expertise, in: Advances in Neural Information Processing Systems, 2009, pp.
  2035--2043.

\bibitem{bachrach2012grade}
Y.~Bachrach, T.~Graepel, T.~Minka, J.~Guiver, How to grade a test without
  knowing the answers---a bayesian graphical model for adaptive crowdsourcing
  and aptitude testing, arXiv preprint arXiv:1206.6386.

\bibitem{zhou2012learning}
D.~Zhou, S.~Basu, Y.~Mao, J.~C. Platt, Learning from the wisdom of crowds by
  minimax entropy, in: Advances in Neural Information Processing Systems, 2012,
  pp. 2195--2203.

\bibitem{zhou2015regularized}
D.~Zhou, Q.~Liu, J.~C. Platt, C.~Meek, N.~B. Shah, Regularized minimax
  conditional entropy for crowdsourcing, arXiv preprint arXiv:1503.07240.

\bibitem{ruvolo2010exploiting}
P.~Ruvolo, J.~Whitehill, J.~R. Movellan, Exploiting structure in crowdsourcing
  tasks via latent factor models, Tech. rep. (2010).

\bibitem{ruvolo2013exploiting}
P.~Ruvolo, J.~Whitehill, J.~Movellan, Exploiting commonality and interaction
  effects in crowdsourcing tasks using latent factor models, in: Neural
  Information Processing Systems. Workshop on Crowdsourcing: Theory, Algorithms
  and Applications, 2013.

\bibitem{kajino2012convex}
H.~Kajino, Y.~Tsuboi, H.~Kashima, A convex formulation for learning from
  crowds., in: Proceedings of the 26th AAAI Conference on Artificial
  Intelligence, 2012.

\bibitem{ma2015faitcrowd}
F.~Ma, Y.~Li, Q.~Li, M.~Qiu, J.~Gao, S.~Zhi, L.~Su, B.~Zhao, H.~Ji, J.~Han,
  Faitcrowd: Fine grained truth discovery for crowdsourced data aggregation,
  in: Proceedings of the 21th ACM SIGKDD International Conference on Knowledge
  Discovery and Data Mining, ACM, 2015, pp. 745--754.

\bibitem{welinder2010multidimensional}
P.~Welinder, S.~Branson, P.~Perona, S.~J. Belongie, the multidimensional wisdom
  of crowds, in: Advances in Neural Information Processing Systems, Curran
  Associates, Inc., 2010, pp. 2424--2432.

\bibitem{wauthier2011bayesian}
F.~L. Wauthier, M.~I. Jordan, Bayesian bias mitigation for crowdsourcing, in:
  Advances in Neural Information Processing Systems, 2011, pp. 1800--1808.

\bibitem{Yuan2017sideinfo}
Y.~Jin, M.~Carman, D.~Kim, L.~Xie, Leveraging side information to improve label
  quality control in crowd-sourcing, in: Proceedings of the 5th AAAI Conference
  on Human Computation and Crowdsourcing (HCOMP), 2017, pp. 79--88.

\bibitem{yin2017aggregating}
L.~Yin, J.~Han, W.~Zhang, Y.~Yu, Aggregating crowd wisdoms with label-aware
  autoencoders, in: Proceedings of the 26th International Joint Conference on
  Artificial Intelligence, AAAI Press, 2017, pp. 1325--1331.

\bibitem{atarashi2018semi}
K.~Atarashi, S.~Oyama, M.~Kurihara, Semi-supervised learning from crowds using
  deep generative models, in: Proceedings of the 32nd AAAI Conference on
  Artificial Intelligence, AAAI, 2018.

\bibitem{gaunt2016training}
A.~Gaunt, D.~Borsa, Y.~Bachrach, Training deep neural nets to aggregate
  crowdsourced responses, in: Proceedings of the 32nd Conference on Uncertainty
  in Artificial Intelligence. AUAI Press, 2016, p. 242251.

\bibitem{lakshminarayanan2013inferring}
B.~Lakshminarayanan, Y.~W. Teh, Inferring ground truth from multi-annotator
  ordinal data: a probabilistic approach, arXiv preprint arXiv:1305.0015.

\bibitem{metrikov2013modification}
P.~Metrikov, J.~Wu, J.~Anderton, V.~Pavlu, J.~A. Aslam, A modification of
  lambdamart to handle noisy crowdsourced assessments, in: Proceedings of the
  2013 Conference on the Theory of Information Retrieval, ACM, 2013, p.~31.

\bibitem{jung2012inferring}
H.~J. Jung, M.~Lease, Inferring missing relevance judgments from crowd workers
  via probabilistic matrix factorization, in: Proceedings of the 35th
  International ACM SIGIR Conference on Research and Development in Information
  Retrieval, ACM, 2012, pp. 1095--1096.

\bibitem{jung2012improving}
H.~J. Jung, Lease, Improving quality of crowdsourced labels via probabilistic
  matrix factorization, in: Proceedings of the 4th Human Computation Workshop
  at AAAI, 2012, pp. 101--106.

\bibitem{jung2013crowdsourced}
H.~J. Jung, M.~Lease, Crowdsourced task routing via matrix factorization, arXiv
  preprint arXiv:1310.5142.

\bibitem{jung2014quality}
H.~J. Jung, Quality assurance in crowdsourcing via matrix factorization based
  task routing, in: Proceedings of the 23rd International Conference on World
  Wide Web, ACM, 2014, pp. 3--8.

\bibitem{zhou2014aggregating}
D.~Zhou, Q.~Liu, J.~Platt, C.~Meek, Aggregating ordinal labels from crowds by
  minimax conditional entropy, in: Proceedings of the 31st International
  Conference on International Conference on Machine Learning, 2014, pp.
  262--270.

\bibitem{griffiths2011indian}
T.~L. Griffiths, Z.~Ghahramani, the indian buffet process: An introduction and
  review, Journal of Machine Learning Research 12~(Apr) (2011) 1185--1224.

\bibitem{blei2003latent}
D.~M. Blei, A.~Y. Ng, M.~I. Jordan, Latent dirichlet allocation, Journal of
  Machine Learning Research 3~(Jan) (2003) 993--1022.

\bibitem{lecun2015deep}
Y.~LeCun, Y.~Bengio, G.~Hinton, Deep learning, Nature 521~(7553) (2015) 436.

\bibitem{kingma2013auto}
D.~P. Kingma, M.~Welling, Auto-encoding variational bayes, in: Proceedings of
  the 2nd International Conference on Learning Representations, 2013.

\bibitem{metrikov2015aggregation}
P.~Metrikov, V.~Pavlu, J.~A. Aslam, Aggregation of crowdsourced ordinal
  assessments and integration with learning to rank: A latent trait model, in:
  Proceedings of the 24th ACM International on Conference on Information and
  Knowledge Management, ACM, 2015, pp. 1391--1400.

\bibitem{tian2012learning}
Y.~Tian, J.~Zhu, Learning from crowds in the presence of schools of thought,
  in: Proceedings of the 18th ACM SIGKDD International Conference on Knowledge
  discovery and Data Mining, ACM, 2012, pp. 226--234.

\bibitem{nguyen2016probabilistic}
A.~T. Nguyen, M.~Halpern, B.~C. Wallace, M.~Lease, Probabilistic modeling for
  crowdsourcing partially-subjective ratings, in: Fourth AAAI Conference on
  Human Computation and Crowdsourcing, 2016.

\bibitem{antoniak1974mixtures}
C.~E. Antoniak, Mixtures of dirichlet processes with applications to bayesian
  nonparametric problems, the Annals of Statistics (1974) 1152--1174.

\bibitem{ryan2000intrinsic}
R.~M. Ryan, E.~L. Deci, Intrinsic and extrinsic motivations: Classic
  definitions and new directions, Contemporary Educational Psychology 25~(1)
  (2000) 54--67.

\bibitem{shah2015approval}
N.~Shah, D.~Zhou, Y.~Peres, Approval voting and incentives in crowdsourcing,
  in: Proceedings of the 32nd International Conference on Machine Learning,
  2015, pp. 10--19.

\bibitem{shah2015double}
N.~B. Shah, D.~Zhou, Double or nothing: Multiplicative incentive mechanisms for
  crowdsourcing, in: Advances in Neural Information Processing Systems, 2015,
  pp. 1--9.

\bibitem{shah2016no}
N.~Shah, D.~Zhou, No oops, you won't do it again: mechanisms for
  self-correction in crowdsourcing, in: Proceedings of the 33rd International
  Conference on Machine Learning, 2016, pp. 1--10.

\bibitem{von2008designing}
L.~Von~Ahn, L.~Dabbish, Designing games with a purpose, Communications of the
  ACM 51~(8) (2008) 58--67.

\bibitem{waggoner2014output}
B.~Waggoner, Y.~Chen, Output agreement mechanisms and common knowledge, in:
  Second AAAI Conference on Human Computation and Crowdsourcing, 2014.

\bibitem{miller2005eliciting}
N.~Miller, P.~Resnick, R.~Zeckhauser, Eliciting informative feedback: the
  peer-prediction method, Management Science 51~(9) (2005) 1359--1373.

\bibitem{jurca2007collusion}
R.~Jurca, B.~Faltings, Collusion-resistant, incentive-compatible feedback
  payments, in: Proceedings of the 8th ACM Conference on Electronic Commerce,
  ACM, 2007, pp. 200--209.

\bibitem{jurca2009mechanisms}
R.~Jurca, B.~Faltings, et~al., Mechanisms for making crowds truthful, Journal
  of Artificial Intelligence Research 34~(1) (2009) 209.

\bibitem{kong2016putting}
Y.~Kong, K.~Ligett, G.~Schoenebeck, Putting peer prediction under the micro
  (economic) scope and making truth-telling focal, in: International Conference
  on Web and Internet Economics, Springer, 2016, pp. 251--264.

\bibitem{prelec2004bayesian}
D.~Prelec, A bayesian truth serum for subjective data, science 306~(5695)
  (2004) 462--466.

\bibitem{parkes2012robust}
D.~C. Parkes, J.~Witkowski, A robust bayesian truth serum for small
  populations, in: Proceedings of the 26th AAAI Conference on Artificial
  Intelligence, Association for the Advancement of Artificial Intelligence,
  2012.

\bibitem{faltings2014incentives}
B.~Faltings, R.~Jurca, P.~Pu, B.~D. Tran, Incentives to counter bias in human
  computation, in: Second AAAI conference on human computation and
  crowdsourcing, 2014.

\bibitem{radanovic2016incentives}
G.~Radanovic, B.~Faltings, R.~Jurca, Incentives for effort in crowdsourcing
  using the peer truth serum, ACM Transactions on Intelligent Systems and
  Technology (TIST) 7~(4) (2016) 48.

\bibitem{dasgupta2013crowdsourced}
A.~Dasgupta, A.~Ghosh, Crowdsourced judgement elicitation with endogenous
  proficiency, in: Proceedings of the 22nd International Conference on World
  Wide Web, ACM, 2013, pp. 319--330.

\bibitem{witkowski2013dwelling}
J.~Witkowski, Y.~Bachrach, P.~Key, D.~C. Parkes, Dwelling on the negative:
  Incentivizing effort in peer prediction, in: First AAAI Conference on Human
  Computation and Crowdsourcing, 2013.

\bibitem{liu2016learning}
Y.~Liu, Y.~Chen, Learning to incentivize: eliciting effort via output
  agreement, Proceedings of the 25th International Joint Conference on
  Artificial Intelligence.

\bibitem{liu2017sequential}
Y.~Liu, Chen, Sequential peer prediction: Learning to elicit effort using
  posted prices., in: Proceedings of the 31st AAAI Conference on Artificial
  Intelligence, 2017, pp. 607--613.

\bibitem{liu2017machine}
Y.~Liu, Chen, Machine-learning aided peer prediction, in: Proceedings of the
  2017 ACM Conference on Economics and Computation, ACM, 2017, pp. 63--80.

\bibitem{gao2014trick}
X.~A. Gao, A.~Mao, Y.~Chen, R.~P. Adams, Trick or treat: putting peer
  prediction to the test, in: Proceedings of the 5th ACM conference on
  Economics and Computation, ACM, 2014, pp. 507--524.

\bibitem{singer2013pricing}
Y.~Singer, M.~Mittal, Pricing mechanisms for crowdsourcing markets, in:
  Proceedings of the 22nd International Conference on World Wide Web, ACM,
  2013, pp. 1157--1166.

\bibitem{ho2015incentivizing}
C.-J. Ho, A.~Slivkins, S.~Suri, J.~W. Vaughan, Incentivizing high quality
  crowdwork, in: Proceedings of the 24th International Conference on World Wide
  Web, 2015, pp. 419--429.

\bibitem{faltings2017game}
B.~Faltings, G.~Radanovic, Game theory for data science: Eliciting truthful
  information, Synthesis Lectures on Artificial Intelligence and Machine
  Learning 11~(2) (2017) 1--151.

\bibitem{radanovic2014incentives}
G.~Radanovic, B.~Faltings, Incentives for truthful information elicitation of
  continuous signals, in: Proceedings of the 28th AAAI Conference on Artificial
  Intelligence, no. EPFL-CONF-215878, 2014, pp. 770--776.

\bibitem{brenner2014people}
M.~Brenner, N.~Mirza, E.~Izquierdo, People recognition using gamified ambiguous
  feedback, in: Proceedings of the 1st International Workshop on Gamification
  for Information Retrieval, ACM, 2014, pp. 22--26.

\bibitem{dumitrache2013dr}
A.~Dumitrache, L.~Aroyo, C.~Welty, R.-J. Sips, A.~Levas, Dr. detective:
  combining gamication techniques and crowdsourcing to create a gold standard
  in medical text, in: Proceedings of the 1st International Conference on
  Crowdsourcing the Semantic Web, 2013, pp. 16--31.

\bibitem{harris2014beauty}
C.~G. Harris, the beauty contest revisited: Measuring consensus rankings of
  relevance using a game, in: Proceedings of the 1st International Workshop on
  Gamification for Information Retrieval, ACM, 2014, pp. 17--21.

\bibitem{he2014studying}
J.~He, M.~Bron, L.~Azzopardi, A.~de~Vries, Studying user browsing behavior
  through gamified search tasks, in: Proceedings of the 1st International
  Workshop on Gamification for Information Retrieval, ACM, 2014, pp. 49--52.

\bibitem{eickhoff2012quality}
C.~Eickhoff, C.~G. Harris, A.~P. de~Vries, P.~Srinivasan, Quality through flow
  and immersion: gamifying crowdsourced relevance assessments, in: Proceedings
  of the 35th ACM SIGIR Conference on Research and Development in Information
  Retrieval, ACM, 2012, pp. 871--880.

\bibitem{saito2014skill}
S.~Saito, T.~Watanabe, M.~Kobayashi, H.~Takagi, Skill development framework for
  micro-tasking, in: International Conference on Universal Access in
  Human-Computer Interaction, Springer, 2014, pp. 400--409.

\bibitem{guillot2016towards}
L.~Guillot, Q.~Bragard, R.~Smith, A.~Ventresque, Towards a gamified system to
  improve translation for online meetings, in: the 3rd International Workshop
  on Gamification for Information Retrieval, CEUR, 2016.

\bibitem{schlotterer2015game}
J.~Schlotterer, C.~Seifert, L.~Wagner, M.~Granitzer, A game with a purpose to
  access europe's cultural treasure., in: the 2nd International Workshop
  Gamification for Information Retrieval, 2015.

\bibitem{moazzam2015scientific}
W.~Moazzam, M.~Riegler, S.~Sen, M.~Nygaard, Scientific hangman: Gamifying
  scientific evidence for general public., in: the 2nd International Workshop
  Gamification for Information Retrieval, 2015.

\bibitem{ipeirotis2014quizz}
P.~G. Ipeirotis, E.~Gabrilovich, Quizz: Targeted crowdsourcing with a billion
  (potential) users, in: Proceedings of the 23rd International Conference on
  World Wide Web, ACM, 2014, pp. 143--154.

\bibitem{stanculescu2016work}
L.~C. Stanculescu, A.~Bozzon, R.-J. Sips, G.-J. Houben, Work and play: An
  experiment in enterprise gamification, in: Proceedings of the 19th ACM
  Conference on Computer-Supported Cooperative Work \& Social Computing, ACM,
  2016, pp. 346--358.

\bibitem{boyd2012besting}
J.~Boyd-Graber, B.~Satinoff, H.~He, H.~Daume~III, Besting the quiz master:
  Crowdsourcing incremental classification games, in: Proceedings of the 2012
  Joint Conference on Empirical Methods in Natural Language Processing and
  Computational Natural Language Learning, Association for Computational
  Linguistics, 2012, pp. 1290--1301.

\bibitem{dontcheva2014combining}
M.~Dontcheva, R.~R. Morris, J.~R. Brandt, E.~M. Gerber, Combining crowdsourcing
  and learning to improve engagement and performance, in: Proceedings of the
  32nd annual ACM conference on Human factors in computing systems, ACM, 2014,
  pp. 3379--3388.

\bibitem{lee2013experiments}
T.~Y. Lee, C.~Dugan, W.~Geyer, T.~Ratchford, J.~C. Rasmussen, N.~S. Shami,
  S.~Lupushor, Experiments on motivational feedback for crowdsourced workers.,
  in: the 7th International AAAI Conference on Web and Social Media, 2013, pp.
  341--350.

\bibitem{mason2012tiger}
A.~D. Mason, G.~Michalakidis, P.~J. Krause, Tiger nation: Empowering citizen
  scientists, in: the 2012 6th IEEE International Conference on Digital
  Ecosystems and Technologies, IEEE, 2012, pp. 1--5.

\bibitem{itoko2014involving}
T.~Itoko, S.~Arita, M.~Kobayashi, H.~Takagi, Involving senior workers in
  crowdsourced proofreading, in: International Conference on Universal Access
  in Human-Computer Interaction, Springer, 2014, pp. 106--117.

\bibitem{massung2013using}
E.~Massung, D.~Coyle, K.~F. Cater, M.~Jay, C.~Preist, Using crowdsourcing to
  support pro-environmental community activism, in: Proceedings of the SIGCHI
  Conference on Human Factors in Computing Systems, ACM, 2013, pp. 371--380.

\bibitem{yuan2016leaderboard}
Y.~Jin, M.~Carman, X.~Lexing, A little competition never hurt anyone's
  relevance assessments, in: Proceedings of the Third International Workshop on
  Gamification for Information Retrieval (GamifIR), Vol. 1642, CEUR Workshop
  Proceedings, 2016, pp. 29--36.

\bibitem{jennings2007brainiac}
K.~Jennings, Brainiac: adventures in the curious, competitive, compulsive world
  of trivia buffs, Villard Books, 2007.

\bibitem{lease2011overview}
M.~Lease, G.~Kazai, Overview of the {TREC} 2011 crowdsourcing track, in:
  Proceedings of the Text Retrieval Conference, 2011.

\bibitem{bragglearning}
J.~Bragg, W.~EDU, D.~S. Weld, Learning on the job: Optimal instruction for
  crowdsourcing, in: ICML Workshop on Crowdsourcing and Machine Learning, 2015.

\bibitem{gadiraju2015training}
U.~Gadiraju, B.~Fetahu, R.~Kawase, Training workers for improving performance
  in crowdsourcing microtasks, in: Design for Teaching and Learning in a
  Networked World, Springer, 2015, pp. 100--114.

\bibitem{sheng2008get}
V.~S. Sheng, F.~Provost, P.~G. Ipeirotis, Get another label? improving data
  quality and data mining using multiple, noisy labelers, in: Proceedings of
  the 14th ACM SIGKDD International Conference on Knowledge Discovery and Data
  Mining, ACM, 2008, pp. 614--622.

\bibitem{ipeirotis2014repeated}
P.~G. Ipeirotis, F.~Provost, V.~S. Sheng, J.~Wang, Repeated labeling using
  multiple noisy labelers, Data Mining and Knowledge Discovery 28~(2) (2014)
  402--441.

\bibitem{yan2011active}
Y.~Yan, R.~Rosales, G.~Fung, J.~G. Dy, Active learning from crowds, in:
  Proceedings of the 28th International Conference on Machine Learning,
  Omnipress, 2011, pp. 1161--1168.

\bibitem{wang2013framework}
J.~Wang, P.~Ipeirotis, A framework for quality assurance in crowdsourcing.

\bibitem{wang2017cost}
J.~Wang, P.~G. Ipeirotis, F.~Provost, Cost-effective quality assurance in crowd
  labeling, Information Systems Research 28~(1) (2017) 137--158.

\bibitem{zhang2015active}
J.~Zhang, X.~Wu, V.~S. Shengs, Active learning with imbalanced multiple noisy
  labeling, IEEE Transactions on Cybernetics 45~(5) (2015) 1095--1107.

\bibitem{kaelbling1998planning}
L.~P. Kaelbling, M.~L. Littman, A.~R. Cassandra, Planning and acting in
  partially observable stochastic domains, Artificial Intelligence 101~(1-2)
  (1998) 99--134.

\bibitem{bollobas2001random}
B.~Bollob{\'a}s, Random Graphs, no.~73, Cambridge University Press, 2001.

\bibitem{donmez2010probabilistic}
P.~Donmez, J.~Carbonell, J.~Schneider, A probabilistic framework to learn from
  multiple annotators with time-varying accuracy, in: Proceedings of the 2010
  SIAM International Conference on Data Mining, SIAM, 2010, pp. 826--837.

\bibitem{mandal2018peer}
D.~Mandal, M.~Leifer, D.~C. Parkes, G.~Pickard, V.~Shnayder, Peer prediction
  with heterogeneous tasks, arXiv preprint arXiv:1612.00928.

\bibitem{mo2013cross}
K.~Mo, E.~Zhong, Q.~Yang, Cross-task crowdsourcing, in: Proceedings of the 19th
  ACM SIGKDD International Conference on Knowledge Discovery and Data Mining,
  ACM, 2013, pp. 677--685.

\bibitem{fang2013knowledge}
M.~Fang, J.~Yin, X.~Zhu, Knowledge transfer for multi-labeler active learning,
  in: Joint European Conference on Machine Learning and Knowledge Discovery in
  Databases, Springer, 2013, pp. 273--288.

\bibitem{fang2014active}
M.~Fang, J.~Yin, D.~Tao, Active learning for crowdsourcing using knowledge
  transfer., in: Proceedings of the 28th AAAI Conference on Artificial
  Intelligence, 2014, pp. 1809--1815.

\bibitem{zhuang2015leveraging}
H.~Zhuang, J.~Young, Leveraging in-batch annotation bias for crowdsourced
  active learning, in: Proceedings of the 8th ACM International Conference on
  Web Search and Data Mining, ACM, 2015, pp. 243--252.

\bibitem{zhuang2015debiasing}
H.~Zhuang, A.~Parameswaran, D.~Roth, J.~Han, Debiasing crowdsourced batches,
  in: Proceedings of the 21th ACM SIGKDD International Conference on Knowledge
  Discovery and Data Mining, ACM, 2015, pp. 1593--1602.

\bibitem{kobren2015getting}
A.~Kobren, C.~H. Tan, P.~Ipeirotis, E.~Gabrilovich, Getting more for less:
  Optimized crowdsourcing with dynamic tasks and goals, in: Proceedings of the
  24th International Conference on World Wide Web, 2015, pp. 592--602.

\bibitem{mao2013stop}
A.~Mao, E.~Kamar, E.~Horvitz, Why stop now? predicting worker engagement in
  online crowdsourcing, in: First AAAI Conference on Human Computation and
  Crowdsourcing, 2013.

\bibitem{hu2016crowdsourced}
H.~Hu, Y.~Zheng, Z.~Bao, G.~Li, J.~Feng, R.~Cheng, Crowdsourced poi labelling:
  Location-aware result inference and task assignment, in: Proceedings of the
  IEEE 32nd International Conference on Data Engineering, IEEE, 2016, pp.
  61--72.

\bibitem{pan2010survey}
S.~J. Pan, Q.~Yang, A survey on transfer learning, IEEE Transactions on
  Knowledge and Data Engineering 22~(10) (2010) 1345--1359.

\bibitem{plackett1975analysis}
R.~L. Plackett, The analysis of permutations, Applied Statistics (1975)
  193--202.

\bibitem{Lim2016bayesian}
K.~W. Lim, W.~Buntine, C.~Chen, L.~Du, Nonparametric bayesian topic modelling
  with the hierarchical pitman-yor processes, Int. J. Approx. Reasoning 78~(C)
  (2016) 172--191.

\bibitem{srivastava2017autoencoding}
A.~Srivastava, C.~Sutton, Autoencoding variational inference for topic models,
  arXiv preprint arXiv:1703.01488.

\bibitem{Jonathan2017geotagging}
C.~Jonathan, M.~F. Mokbel, A demonstration of stella: A crowdsourcing-based
  geotagging framework, Proc. VLDB Endow. 10~(12) (2017) 1969--1972.

\end{thebibliography}

\end{document}